\documentclass[prep]{svprep}
\usepackage{mydef3}

\begin{document}

\title{\begin{center}{Vertex Algebras in Higher Dimensions and
	Globally Conformal Invariant Quantum Field Theory}\end{center}}
\titlerunning{Vertex Algebras in Higher Dimensions}

\author{\begin{center}{\rm Nikolay M. Nikolov}\end{center}}
\institute{\begin{center}{Institute for Nuclear Research and Nuclear Energy \\
Tsarigradsko Chaussee 72, 1784 Sofia, Bulgaria, \\
e-mail: mitov@inrne.bas.bg}\end{center}}
\authorrunning{Vertex Algebras in Higher Dimensions}

\date{\begin{center}{\today}\end{center}}
\communicated{}

\maketitle
${}$ \hfill {\parbox{230pt}{{\small \begin{abstract}
We propose an extension of the definition of vertex algebras
in arbitrary space--time dimensions together with their basic
structure theory.
An one-to-one correspondence between these vertex algebras
and axiomatic quantum field theory (QFT) with global conformal
invariance (GCI) is constructed.
\end{abstract}}}} \hfill ${}$

\tableofcontents

\newpage

\msection{Introduction and Notations}{sec:1}

The axiomatic QFT was proposed and accepted by
the physics community about 50 years ago as a collection of mathematically
precise structures and their properties which any QFT should posses.
Despite the fact that no four dimensional nontrivial model of the axiomatic
QFT has been found so far,
the long time efforts in this directions have led to several general results
such as the Bargmann--Hall--Wightman (BHW) theorem about analytic properties of
correlation (i.~e. Wightman) functions, the TCP and the spin and statistic
theorems.
A basic structure in the axiomatic approach is the Poincar\'{e} symmetry.
Right from the beginning the question of extending the space--time symmetry
to the conformal one has been posed.
It was shown in the article~\cite{NT 01} that the condition of GCI,
i.~e. group conformal invariance, in the frame
of the axiomatic QFT leads to the rationality of all correlation functions
in any number $D$ of space--time dimensions.
This result can be viewed as an extension of the above mentioned BHW theorem.
Since the Wightman functions carry the full information of the theory
this result shows that the QFT with GCI is essentially algebraic.
This gives new insight to the problem of constructing nonfree QFT models
in higher dimensions.

In 2 dimensional conformal QFT the theory of vertex algebras
is based on simple axiomatic conditions with a straightforward physical
interpretation \cite{Kac}.
One of them is the axiom of locality stating that the commutators or
anticommutators of the fields vanish when multiplied by a sufficiently
large power of the coordinate difference.
This axiom has a natural extension to higher dimensions by replacing
the coordinate difference with the space--time interval and this is
a consequence of GCI in the
axiomatic QFT~--~this is a form of the \textit{Huygens principle} in QFT
called in \cite{NT 01} (see Remark~3.1) strong locality.
On the other hand, the rationality of correlation functions in a
QFT with GCI allows to define a precise state--field correspondence
and an expansion of fields as formal power series in their coordinates
\(z = \left( z^1,\, \dots ,\, z^D \right)\)
and the inverse square interval
\(\frac{\raisebox{1pt}{$1$}}{\raisebox{-3pt}{$z^{\, 2}$}}
\)\gvspc{-5pt}
(\(z^{\, 2} = z \spr z := \left( z^1 \right)^2 + \dots + \left( z^D \right)^2\)).
This provides the second axiomatic structure for the vertex algebras.
The coordinates ``$z$''
define a chart in the \textit{complex compactified Minkowski space}
containing the entire real compact space and they are useful for connecting
the vertex algebra approach with the axiomatic QFT with GCI
(see Sect.~\ref{sec:9},
they are introduced for \(D = 4\) in \cite{Tod 86} and for general $D$, in
\cite{NT 02} Sect.~2.2).
The existence of the last connection motivates our approach from
physical point of view~--~giving examples of such
vertex algebras one would actually obtain models of the Wightman axioms.
Physically, one could regard the vertex algebras as providing a realization of
the observable field algebra in higher dimensional conformal QFT.
The proposed construction of vertex algebras allows to give
a precise definition of the notion of \textit{their representation}
which would realize the charged sectors in accord with Haag's program in
the algebraic QFT~\cite{Haag}.

There is a more general definition of vertex algebras in higher dimensions
proposed by Borcherds \cite{Borcherds} which allows arbitrary type of
singularities occurring in the correlation functions.
From the point of view of GCI the only type of singularities arising
is the light cone type~\cite{NT 01}.

\vspace{5pt}

The paper is organized as follows.

In Sect.~\ref{sec:2} we give the basic definitions and prove the
existence of operator product expansions (Theorem~\ref{th:2.1}).
The vertex algebra fields are denoted by
$Y \left( a,\, z \right)$ as in the \textit{chiral}
two--dimensional conformal QFT (chiral CFT),
depending on the state $a$ and being formal power series in $z$ including
negative powers of $z^{\, 2}\,$.
A convenient basis for such series is provided by the \textit{harmonic}
decomposition of the polynomials in $z$, which we will briefly recall
bellow.
The operator product expansion of two fields $Y \left( a,\, z \right)$
and $Y \left( b,\, z \right)$ is described in terms of infinite series
of ``products''
\(Y \left( a,\, z \right)_{\left\{ n,\, m,\, \sigma \right\}}
Y \left( b,\, z \right)\) labeled by integers which generalize
the analogous products
\(Y \left( a,\, t \right)_{\left( n \right)}
Y \left( b,\, t \right)\) in the chiral CFT.
The $\left\{ 0,0,1 \right\}$--product in our notations
is the natural candidate for the normal product in higher--dimensional
vertex algebras.
In Sect.~\ref{sec:3} we obtain an analogue (Theorem~\ref{th:2.4})
of the (corollary) of the Reeh--Schlider theorem~--~the separating
property of the vacuum~\cite{Jost 65}.
It is also shown that the state--field correspondence exhausts the
class of translation invariant local fields
(i.~e. the Borchers class, Proposition~\ref{pr:2.5}).
We also obtain
generalizations
of some basic formulas for the vertex algebras from the chiral CFT.
In Sect.~\ref{sec:4} we prove a higher dimensional analogue
(Theorem~\ref{th:2.8})
of the Kac existence theorem (\cite{Kac}, Theorem~4.5)
which provides examples of vertex algebras (at least the free ones).
In this section we also find a higher dimensional analogue of the
associativity identity
``\(Y \left( a,\, z \right) Y \left( b,\, w \right)=
Y \left( Y \left( a,\, z-w \right) b,\, w \right)\)''
(Theorem~\ref{th:2.11}).

In Sect.~\ref{sec:5} we present the free field examples of
higher dimensional vertex algebras and also a more general
construction based on Lie superalgebras of formal distributions.
In Sect.~\ref{sec:6} we introduce some constructions with vertex
algebras including the basic categorical notions, tensor product
and representations of vertex algebras.
Sects.~\ref{sec:7} and \ref{sec:8} are devoted to the incorporation
of the conformal symmetry in higher dimensions and the Hermitean structure
(needed for the passage to the GCI~QFT) within the vertex algebras.

In Sect.~\ref{sec:9} we give an one--to--one correspondence between
vertex algebras with additional conformal and Hermitean structure,
and the GCI~QFT.
Thus the free GCI~QFT models provide examples for the vertex algebras
with additional structure introduced in the previous sections.

\subsection*{Notations}
The $z$-- and $w$--variables as
$z\,$, $z_1\,$, $z_2\,$, $w$ etc. will \textbf{always} denote $D$ component
variables:
\beq\label{eqnew1.1a}
z \, = \, \left( z^1,\, \dots ,\, z^D \right) \, , \quad
z_k \, = \, \left( z^1_k,\, \dots ,\, z^D_k \right) \, , \quad
w \, = \, \left( w^1,\, \dots ,\, w^D \right)
\, . \qquad
\eeq
We fix the standard scalar product:
\beq\label{eqnew1.1}
z_1 \spr z_2 \, = \, \Su_{\mu \, = \, 1}^D z_1^{\mu} \, z_2^{\mu}
\, , \quad
z^{\, 2} \, \equiv \, z \spr z
\, . \qquad
\eeq
\(\N \equiv \left\{ 1,\, 2,\, \dots \hspace{1pt} \right\}\,\),
\(\Z \equiv \left\{ 0,\, \pm \hspace{1pt} 1,\, \dots \hspace{1pt} \right\}\,\).

For a complex vector space $\Lin\,$, $\Lin \Brk{z}$
stands for the space of polynomials with coefficients in $\Lin$
(i.~e., \(\Lin \Brk{z} \equiv \Lin \otimes \C \Brk{z}\,\)).
Similarly, $\Lin \Bbrk{z}$ is the space of formal power series in $z$
with coefficients in $\Lin\,$.
We introduce the formal derivatives on $\Lin \Brk{z}$ and
$\Lin \Bbrk{z}\,$:
\beq\label{eqnew1.3n}
\di_{z}
\, \equiv \,
\left( \di_{z^1},\, \dots ,\, \di_{z^D} \right) \equiv
\left( \frac{\textstyle \di}{\textstyle \di z^1},\, \dots ,\,
\frac{\textstyle \di}{\textstyle \di z^D} \right)
\, , \qquad
\eeq
as well as the Euler and Laplace operators:
\beq\label{eqnew1.4n}
z \spr \di_{z}
\, \equiv \,
\Su_{\mu \, = \, 1}^D z^{\mu} \, \di_{z^{\mu}}
\, , \quad
\di_z^{\, 2}
\, \equiv \,
\di_{z} \spr \di_{z} \equiv
\Su_{\mu \, = \, 1}^D \left( \di_{z^{\mu}} \right)^2
\, . \qquad
\eeq
Then $\di_z$ obeys the Leibniz rule and
the homogeneous polynomials of degree $n$ are
characterized by the Euler equation
\(\lb z \spr \di_{z} \, p \rb \lb z \rb =
n \, p \left( z \right)\,\).
A harmonic polynomial
\(p \left( z \right)\in \Lin \Brk{z}\)
is such
that the Laplace equation
\(\lb \di_{z}^{\, 2} \, p \rb \lb z \rb =  0\)
is satisfied.
\spc

The basic fact about the existence of harmonic
decomposition can be stated as follows:
\textit{if}
$p \left( z \right) \in \Lin \Brk{z}$
\textit{is a homogeneous polynomial of degree} $n$
(\(\mathit{deg} \, p = n\))
\textit{then there exists a unique
decomposition}
\beq\label{harmdec}
p \left( z \right) =
\Su_{k \, = \, 0}^{
\left\llbracket \frac{n}{2} \right\rrbracket}
\left( z^{\, 2} \right)^k
h_k \left( z \right)
, \ \
\di_{z}^{\, 2} h_k \lb z \rb
= 0
, \ \
z \spr \di_{z} h_k \lb z \rb
= \lb n - 2 k \rb h_k \left( z \right)
,
\eeq
\textit{where} $\left\llbracket a \right\rrbracket$
\textit{stands for the integer part of the real number} $a\,$.
\spc

The \textit{proof} is based on induction in
$n = \mathit{deg} \, p\,$:
if
$\di_{z}^{\, 2} \, p \lb z \rb$
has by the inductive assumption a unique decomposition
\(\di_{z}^{\, 2} p \lb z \rb
= \Su_{k \, = \, 0}^{
\left[ \! \left[ \frac{n}{2} \right] \! \right] - 1}
\left( z^{\, 2} \right)^k
h_k' \left( z \right)\,\),
\(\mathit{deg} \, h_k' = n-2-2k\,\),
then the difference
$$
h_0 \left( z \right) :=
p \left( z \right) -
\Su_{k \, = \, 0}^{
\left[ \! \left[ \frac{n}{2} \right] \! \right] - 1}
\left(
4 \left( k+1 \right)
\lb
n - k +
\raisebox{-0.8pt}{\Large \(\frac{D-4}{2}\)}
\rb
\right)^{-1}
\left( z^{\, 2} \right)^{k+1}
h_k' \left( z \right)
$$
is verified to be a harmonic homogeneous polynomial by a
straightforward computation.
Thus we obtain that $h_k \left( z \right)$
\(= \, \lb 4 \, k \lb n - k + 1 +
\raisebox{-1pt}{\Large \(\frac{D-4}{2}\)}
\rb \rb^{-1}
h_{k-1}' \left( z \right)\)
for \(k>0\,\).\qed
\spc

In such a way if we denote by $\Lin_m \Brk{z}$ the subspace
of homogeneous polynomials of $\Lin \Brk{z}$ of
degree $m$ and by $\Lin^{\text{harm}}_m \Brk{z}$
the subspace of $\Lin_m \Brk{z}$
of the harmonic polynomials
(\(\Lin_m \Brk{z} \equiv
\Lin^{\text{harm}}_m \Brk{z} \equiv \{0\}\) for
\(m < 0\))
then we have the decomposition
\beqa\label{1.7}
\hspace{-30pt}
\Lin_m \Brk{z} \, = && \hspace{-2pt}
\Lin^{\text{harm}}_m \Brk{z} \, \oplus \,
z^{\, 2} \, \Lin_{m-2} \Brk{z}
\, ,
\vspe
\\ \label{1.8}
\hspace{-30pt}
\har_m^D \equiv
\har_m :=
\text{\textit{dim}} \, \C^{\text{harm}}_m \Brk{z}
= && \hspace{-2pt}
\text{\textit{dim}} \, \C_m \Brk{z}
-
\text{\textit{dim}} \, \C_{m-2} \Brk{z}
=
\nn \hspace{-30pt} = && \hspace{-2pt}
\txtsmall{\left(\hspace{-3pt}\begin{array}{c} m+D-1 \\ D-1
\end{array}\hspace{-2pt}\right)}
\hspace{-2pt}
- \!\!
\hspace{-1.5pt}
\txtsmall{\left(\hspace{-3pt}\begin{array}{c} m+D-3 \\ D-1
\end{array}\hspace{-2pt}\right)}
\,
\eeqa
(recall that
\(\left( 1-q \right)^{-D} =
\su_{m\, = \,0}^{\infty}
\left( \text{\textit{dim}} \, \C_m \Brk{z} \right) \, q^m\,\)).
The space
\(\C^{\text{harm}}_m \Brk{z}\) carries an irreducible representation
of the complex orthogonal group \(\mathit{SO} \left( D; \C \right)\)
for every \(m = 0, 1, \dots\).
Note that \(\har_0^1 = \har_1^1 = \har_0^2 = 1\) and
\(\har_{m+1}^1 = 0\,\), \(\har_m^2 = 2\) for \(m \geqslant 1\,\);
\(\har_m^3 = 2m+1\)
and for \(D \geqslant 4\,\):
\beq\label{ad1.9}
\har_{m \hspace{1pt} - \hspace{0.3pt}
\raisebox{-0.5pt}{{\footnotesize \(\frac{\,
\text{\raisebox{0pt}{$D$}}}{2}\)}} \hspace{1.5pt}
+ \hspace{0.8pt} 1}^D \, = \,
\frac{\text{\raisebox{-1pt}{$2m$}}}{\,\text{\raisebox{-0.5pt}{$\left( D \! - \! 2 \right) !$}}} \
\left( m +
\raisebox{-0.8pt}{\large {\(\frac{\, \text{\raisebox{0pt}{$D$}}}{
\, \text{\raisebox{-1pt}{$2$}}}\)}} \hspace{0.3pt}
- 2 \right) \dots
\left( m -
\raisebox{-0.8pt}{{\large \(\frac{\, \text{\raisebox{0pt}{$D$}}}{
\, \text{\raisebox{-1pt}{$2$}}}\)}} \hspace{0.3pt}
+ 2\right)
\, , \qquad
\eeq
so that \(\har_{m \hspace{1pt} - \hspace{0.3pt}
\text{{\footnotesize \raisebox{-0.5pt}{\(\frac{\,
\text{\raisebox{0pt}{$D$}}}{\, 2^{\vspe}}\)}}} \hspace{1pt}
+ \hspace{0.8pt} 1}^D\)
is a polynomial in $m$ of degree $D-2$ for \(D \geqslant 2\,\),
which is even for $D$ even, and odd for $D$ odd.
For \(D=4\,\): \(\har_m = \left( m+1 \right)^2\,\).

Let us fix for every \(m = 0,\, 1,\, \dots\) a basis in
\(\C^{\text{harm}}_m \Brk{z}\):
\beq\label{1.9}
\left\{ h^{\left( m \right)}_{\sigma} \left( z \right) \, : \,
\sigma \, = \, 1,\, \dots ,\, \har_m \right\}
, \ \ h^{\left( 0 \right)}_1 \left( z \right) \equiv 1
, \ \ h^{\left( m \right)}_{\sigma} \left( z \right) \equiv 0
\ \, \text{iff} \ \, m \, < \, 0
\, .
\eeq
Then for every \(a \left( z \right) \in \Lin \Bbrk{z}\) we have
a unique representation:
\beq\label{1.10}
a \left( z \right) \, = \,
\Su_{n \, = \, 0}^{\infty} \,
\Su_{m \, = \, 0}^{\infty} \,
\Su_{\sigma \, = \, 1}^{\har_m} \,
a_{\,\left\{ n,\, m,\, \sigma \right\}} \,
\left( z^{\, 2} \right)^n \, h^{\left( m \right)}_{\sigma}
\left( z \right)
\, , \quad
a_{\,\left\{ n,\, m,\, \sigma \right\}} \, \in \, \Lin
. \
\eeq
For every \(n,\, m = 0,\, 1,\,  \dots\) and \(\sigma = 1,\, \dots ,\, \har_m\)
there exists a unique homogeneous polynomial
$P_{\left\{ n,\, m,\, \sigma \right\}} \left( z \right)$ of degree \(2n+m\) such that:
\beq\label{ad1.11}
P_{\left\{ n,\, m,\, \sigma \right\}} \left( \di_{z} \right) \,
a \left( z \right)
\!\left.\vspe\right|_{\,
\raisebox{1pt}{\small \(z = 0\)}}
\, = \,
a_{\left\{ n,\, m,\, \sigma \right\}}
\, \qquad
\eeq
for any \(a \left( z \right)\) (\ref{1.10}).
In the special case of \(m=0\) (\(\har_m=1\)):
\beq\label{polynomial}
P_{\left\{ n,\, 0,\, 1 \right\}} \left(  z\right)
\, = \,
K_n \, \left( z^2 \right)^n
\, , \qquad
\eeq
where
\(K_n := \frac{\textstyle \left( D-2 \right) !!}{\textstyle 2^n \, n! \,
\left( 2n+D-2 \right) !!}\) and
$k\hspace{1pt}!!$ $:=$ $k$ $\left( k-2 \right) \dots$
$\left( k - 2\left\llbracket \frac{k}{2} \right\rrbracket \right)$.
In general, $P_{\left\{ n,\, m,\, \sigma \right\}} \left( z \right)$
could be proven
to be proportional to
\(\left( z^2 \right)^n h^{\left( m \right)}_{\sigma} \left( z \right)\)
under the additional assumption of orthogonality of
$h^{\left( m \right)}_{\sigma}$ but we will not need its explicit form.

Denote by $\Lin \Bbrk{z,\, \frc{1}{z^{\, 2}}}$
the vector space of all formal series:
\beq\label{1.14}
a \left( z \right) \, = \,
\Su_{n \, \in \, \Z} \,
\Su_{m \, = \, 0}^{\infty} \,
\Su_{\sigma \, = \, 1}^{\har_m} \,
a_{\,\left\{ n,\, m,\, \sigma \right\}} \,
\left( z^{\, 2} \right)^n \, h^{\left( m \right)}_{\sigma}
\left( z \right)
\, , \quad
a_{\,\left\{ n,\, m,\, \sigma \right\}} \, \in \, \Lin
. \
\eeq
The subspace of $\Lin \Bbrk{z,\, \frc{1}{z^{\, 2}}}$
of \textit{finite} series~(\ref{1.14}) will be denoted by
$\Lin \Brk{z,\, \frc{1}{z^{\, 2}}}\,$; the subspace
of formal series~(\ref{1.14})
whose sum in $n$
is bounded from below:~\(\Lin \Bbrk{z}_{\,\lscw{z}{\, 2}{}}\)~--~i.~e.
the localization of
$\Lin \Bbrk{z}$ with respect to the multiplicative
system
$\left\{ \left( z^{\, 2} \right)^n \right\}_{\, n \, \in \, \N}\,$.
Thus \(a \left( z \right) \in \Lin \Bbrk{z}_{\,\lscw{z}{\, 2}{}}\)
iff \(\left( z^{\, 2} \right)^N a \left( z \right) \in \Lin \Bbrk{z}\)
for sufficiently large $N\,$, which we will briefly write as
\(N \mgrt 0\,\).

The spaces $\Lin \Bbrk{z}$ and
\(\Lin \Bbrk{z}_{\,\lscw{z}{\, 2}{}}^{}\) are
\(\C \Bbrk{z}\) and
\(\C \Bbrk{z}_{\,\lscw{z}{\, 2}{}}\) modules, respectively, with derivations
\(\left\{ \di_{z^{\mu}} \right\}_{\mu = 1,\, \dots ,\, D}\,\).
For
$\Lin \Bbrk{z,\, \frc{1}{z^{\, 2}}}$ we have a structure
of a $\C \Brk{z,\, \frc{1}{z^{\, 2}}}$ module with derivations
$\di_{z^{\mu}}\,$.
To define this structure we use the isomorphism
\beqa\label{1.21}
&
\Lin \Bbrk{z,\, \frc{1}{z^{\, 2}}} \, \cong \, &
\nn & \, \cong \,
\raisebox{3pt}{\(
\left(\raisebox{9pt}{\hspace{-2pt}}\right.
\Lin \Bbrk{z}_{\,\lscw{z}{\, 2}{}} \oplus \hspace{0.6pt}
\Lin \Bbrk{z}_{\,\lscw{z}{\, 2}{}}
\left.\raisebox{9pt}{\hspace{-2pt}}\right)
\)}
\hspace{-2pt}\left/\raisebox{9pt}{${}$}\right.\hspace{-3pt}
\raisebox{-3pt}{\(
\left\{\raisebox{9pt}{\hspace{-2pt}}\right.
\left( \hspace{0.5pt} c \left( z \right),\, - J \left[ c \left( z \right)
\hspace{0.5pt} \right] \right)
\, : \,
c \left( z \right) \, \in \, \Lin \blrk{z,\, \frc{1}{z^{\, 2}}}
\left.\raisebox{9pt}{\hspace{-2pt}}\right\}
\)}
\, , & \qquad
\eeqa
where we set $\Lin \blrk{z,\, \frc{1}{z^{\, 2}}}$
to be the space of all formal series~(\ref{1.14}) with finite sum
in $n$ but possibly infinite in $m\,$
(being thus a $\C \Brk{z,\, \frc{1}{z^{\, 2}}}$--module),
and
\beq
\label{1.22}
J \left[ a \left( z \right) \right] :=
a \left( w \right)
\!\left.\vspe\right|_{\,
\raisebox{1pt}{\small \(w =
\frac{\raisebox{0.7pt}{\small $z$}}{\raisebox{-1pt}{\small $\lscw{z}{2}{}$}}\)}}
\equiv
\Su_{
n \in \Z
} \,
\Su_{
m = 0
}^{
\infty
} \,
\Su_{
\sigma = 1
}^{
\har_m
} \,
a_{\,\left\{ -n-m,\, m,\, \sigma \right\}} \,
\left( z^{\, 2} \right)^n \, h^{\left( m \right)}_{\sigma}
\left( z \right)
\, ,
\eeq
for $a \left( z \right)$ in Eq.~(\ref{1.14}).
The so defined $J$ is an involutive automorphism of
$\Lin \Brk{z,\, \frc{1}{z^{\, 2}}}$
(\(J^2 \, = \, \ID\,\)
and \(J \left[ f a \right] = J \left[ f \right]
J \left[ \hspace{0.5pt} a \hspace{0.3pt} \right]\)
for \(f \in \C \Brk{z,\, \frc{1}{z^{\, 2}}}\,\),
\(a \in \Lin \Brk{z,\, \frc{1}{z^{\, 2}}}\,\)).
The isomorphism~(\ref{1.21}) is generated by the map
\beq\label{1.23}
\Lin \Bbrk{z}_{\,\lscw{z}{\, 2}{}} \oplus \hspace{0.6pt}
\Lin \Bbrk{z}_{\,\lscw{z}{\, 2}{}} \, \ni \,
\left( a \left( z \right),\, b \left( z \right) \right)
\, \longmapsto \,
a \left( z \right) + J \left[ b \left( z \right) \right]
\, \in \,
\Lin \Bbrk{z,\, \frc{1}{z^{\, 2}}}
\, .
\eeq
Then the actions of
\(f \left( z \right) \in \C \Brk{z,\, \frc{1}{z^{\, 2}}}\)
and $\di_{z^{\mu}}$ on
$\Lin \Bbrk{z,\, \frc{1}{z^{\, 2}}}$
are generated,
in view of the isomorphism~(\ref{1.21}),
by:
\beqa\label{1.24}
&
f \left( z \right) \hspace{0.5pt}
\left( a \left( z \right),\, b \left( z \right) \right) \,
\, := \,
\left( f \left( z \right) \, a \left( z \right),\,
J \left[ \hspace{0.5pt} f \left( z \right) \hspace{0.5pt} \right]
\, b \left( z \right) \right)
\, , & \vspe
\\ \label{1.25} &
\di_{z^{\mu}} \hspace{0.5pt}
\left( a \left( z \right),\, b \left( z \right) \right) \,
\, := \,
\left( \di_{z^{\mu}} \, a \left( z \right),\,
\delta_{z^{\mu}}
\, b \left( z \right) \right)
\, , &
\\ \label{1.26} &
\text{where} \quad
\delta_{z^{\mu}} \, := \,
J \, \di_{z^{\mu}} \, J \, \equiv \,
z^{\, 2} \, \di_{z^{\mu}} -
2 \, z^{\mu}
\, z \spr \di_{z}
\, . & \qquad \vspe
\eeqa
The derivations
\(\left\{ \di_{z^{\mu}},\, \delta_{z^{\nu}} \right\}_{\,
\mu,\, \nu \, = \, 1,\, \dots ,\, D}\)
give rise to an action of the (complex) conformal Lie algebra
$\mathit{so}\left( D+2;\, \C \right)$ on
$\Lin \Bbrk{z,\, \frc{1}{z^{\, 2}}}\,$.

To obtain the explicit form of the actions of
$\C \Brk{z,\, \frc{1}{z^{\, 2}}}$ and $\di_{z^{\mu}}$
on $\Lin \Bbrk{z,\, \frc{1}{z^{\, 2}}}$ let us note first that
for a homogeneous harmonic polynomial $h_m \left( z \right)$
of degree $m$ (\(h_m \in \Lin^{\text{harm}}_m \Brk{z}\,\)),
the polynomials:
$$
\di_{z^{\mu}} \, h_m \left( z \right)
\quad \text{and} \quad
z^{\mu} \, h_m \left( z \right) -
\frac{1}{
\raisebox{-3.5pt}{\(
2 \left(
m +
\raisebox{-0.3pt}{{\large \(\frac{\raisebox{0pt}{{\footnotesize \(D\)}}}{\raisebox{-1pt}{{\footnotesize \(2\)}}}\)}}
- 1
\right)
\)}
} \ \,
z^{\, 2} \ \di_{z^{\mu}} \, h_m \left( z \right)
$$
are harmonic and homogeneous of degrees $m-1$ and $m+1\,$,
respectively.
Therefore, there exist constants
$A_{\mu\,\sigma_1\,\sigma_2}^{\left( m \right)}$ and
$B_{\mu\,\sigma_1\,\sigma_2}^{\left( m \right)}$ such that
\beqa\label{actions}
&
\di_{z^{\mu}} h^{\left( m \right)}_{\sigma_1} \! \left( z \right)
=
\mathop{\raisebox{0pt}{\large \(\sum\)}}\limits_{
\sigma_2 \, = \, 1}^{\har_{m-1}} \!
A_{\mu\,\sigma_1\sigma_2}^{\left( m \right)}
h^{\left( m-1 \right)}_{\sigma_2} \left( z \right)
\, , \quad
z^{\mu} h^{\left( m \right)}_{\sigma_1} \! \left( z \right)
=
\mathop{\raisebox{0pt}{\large \(\sum\)}}\limits_{
\sigma_2 \, = \, 1}^{\har_{m+1}} \!
B_{\mu\,\sigma_1\sigma_2}^{\left( m \right)}
h^{\left( m+1 \right)}_{\sigma_2} \left( z \right)
+
& \nn & + \
\frac{\raisebox{1pt}{$1$}}{
\raisebox{-3.5pt}{\(
2 \left(
m +
\raisebox{-0.3pt}{{\large \(\frac{\raisebox{0pt}{{\footnotesize \(D\)}}}{\raisebox{-1pt}{{\footnotesize \(2\)}}}\)}}
- 1
\right)
\)}
} \ \, z^{\, 2}
\mathop{\raisebox{0pt}{\large \(\sum\)}}\limits_{
\sigma_2 \, = \, 1}^{\har_{m-1}} \!
A_{\mu\,\sigma_1\sigma_2}^{\left( m \right)}
h^{\left( m-1 \right)}_{\sigma_2} \left( z \right)
. & \hspace{-10pt}
\eeqa
Using this equations one can obtain the explicit form of the actions of
$z^{\mu}$ and of the derivations $\di_{z^{\mu}}$ on a general
series \(a \left( z \right) \in \Lin \Bbrk{z, \lfrc{1}{z^{\, 2}}}\,\).
The coefficients
$A_{\mu\,\sigma_1\,\sigma_2}^{\left( m \right)}$ and
$B_{\mu\,\sigma_1\,\sigma_2}^{\left( m \right)}$ define
intertwining operators
$A^{\left( m \right)}$~:
$\C^{\text{harm}}_1 \Brk{z}$ $\otimes$ $\C^{\text{harm}}_m \Brk{z}$
$\rightarrow$ $\C^{\text{harm}}_{m-1} \Brk{z}$ and
$B^{\left( m \right)}:$
$\C^{\text{harm}}_1 \Brk{z}$ $\otimes$ $\C^{\text{harm}}_m \Brk{z}$
$\rightarrow$ $\C^{\text{harm}}_{m+1} \Brk{z}$
as $\mathit{SO} \left( D;\, \C \right)$ representations.

In the same way one can define the spaces
\\ ${}$ \hfill
\(\Lin \Brk{z_1, \lfrc{1}{z^{\, 2}_1}
\hspace{1pt} ; \dots ; \hspace{1pt} z_n, \lfrc{1}{z^{\, 2}_n}}, \
\Lin \Bbrk{z_1, \lfrc{1}{z^{\, 2}_1}\hspace{1pt} ; \hspace{1pt}
\dots ; \hspace{1pt} z_n, \lfrc{1}{z^{\, 2}_n}} \ \text{and} \
\Lin \Bbrk{z_1,\, \dots ,\, z_n}_{\,\lscw{z}{\, 2}{1} \dots \, \lscw{z}{\, 2}{n}}
\mgvspc{15pt}\mgvspc{-11pt}\) \hfill ${}$ \\
(the last symbol stands for the localization of
\(\Lin \Bbrkl{z_1,\, \dots ,}\) \(\Bbrkr{z_n}\) with respect to the multiplicative
system
\(\left\{ \left( z^{\, 2}_1 \dots \, z^{\, 2}_n \right)^n \right\}_{\,
n \in \N_{\vspe}^{}}\vspe\)).
Note that
\beqa\label{add2.1}
\Lin \Brk{z_1, \lfrc{1}{z^{\, 2}_1}
\hspace{1pt} ; \dots ; \hspace{1pt} z_n, \lfrc{1}{z^{\, 2}_n}}
\, = && \hspace{-4pt}
\left( \Lin \Brk{z_1, \lfrc{1}{z^{\, 2}_1}
\hspace{1pt} ; \dots ; \hspace{1pt} z_{n-1}, \lfrc{1}{z^{\, 2}_{n-1}}} \right)
\Brk{z_n, \lfrc{1}{z^{\, 2}_n}}
\, , \quad
\\ \label{add2.2}
\Lin \Bbrk{z_1, \lfrc{1}{z^{\, 2}_1}
\hspace{1pt} ; \dots ; \hspace{1pt} z_n, \lfrc{1}{z^{\, 2}_n}}
\, = && \hspace{-4pt}
\left( \Lin \Bbrk{z_1, \lfrc{1}{z^{\, 2}_1}
\hspace{1pt} ; \dots ; \hspace{1pt} z_{n-1}, \lfrc{1}{z^{\, 2}_{n-1}}} \right)
\Bbrk{z_n, \lfrc{1}{z^{\, 2}_n}}
\, . \qquad
\eeqa

It is important that the
\(\C \Bbrk{z}_{\,\lscw{z}{\, 2}{}}\)--module
\(\Lin \Bbrk{z}_{\,\lscw{z}{\, 2}{}}\) has no
``zero divisors'', i.~e. if
\(f \left( z \right) a \left( z \right) = 0\)
for \(f \left( z \right) \in \C \Bbrk{z}_{\,\lscw{z}{\, 2}{}}\)
and \(a \left( z \right) \in \Lin \Bbrk{z}_{\,\lscw{z}{\, 2}{}}\)
then \(f \left( z \right) = 0\) or \(a \left( z \right) = 0\,\).
This is not the case for the \(\C \Brk{z}_{\,\lscw{z}{\, 2}{}}\)--module
$\Lin \Bbrk{z, \lfrc{1}{z^{\, 2}}}$ as it is seen by the following example.

\begin{mexample}\label{ex:2.1}
Let \(c = \left( c^1,\, \dots ,\, c^D \right) \in \C^D\)
be a complex vector such that $c^{\, 2}$ $\equiv$ $c \spr c$ $=$ $1\,$.
The polynomial $\left( z-c \right)^2$ is invertible in
$\C \Bbrk{z}$ and let $t \left( z \right)$ be its inverse.
Then
\(\left( z-c \right)^2
\left( t \left( z \right) -
\frac{\textstyle 1}{\textstyle z^{\, 2}} \,
J \left[ \hspace{1pt} t \left( z \right) \hspace{1pt} \right] \right)
= 0\) and
\(0 \neq t \left( z \right) -
\frac{\textstyle 1}{\textstyle z^{\, 2}} \,
J \left[ \hspace{1pt} t \left( z \right) \hspace{1pt} \right] \in
\C \Bbrk{z, \lfrc{1}{z^{\, 2}}}\,\).
\end{mexample}

\msection{Vertex Algebra Definition and Operator Product Expansion}{sec:2}

In the next three sections we will use the notation \(z_{kl} := z_k - z_l\)
as an abbreviation of the polynomial \(z_k - z_l \in \C \Brk{z_k,\, z_l}\)
but not as new variable.

\begin{mdefinition}\label{def:2.1}
Let \(\VA = \VA_0 \oplus \VA_1\) be a
$\Z_2$--graded complex vector space
(i.~e., a superspace) and
\(\mathit{End}\, \VA =
\left( \mathit{End}\, \VA \right)_0 \oplus
\left( \mathit{End}\, \VA \right)_1\,\)
be the corresponding Lie superalgebra
with bracket
\(\left[ \hspace{0.6pt} \_ \hspace{2pt} ,
\hspace{0pt} \_ \hspace{2pt} \right]\).
Then $\VA$ is said to be a
\textbf{vertex algebra over} $\C^D$ if it is
equipped with a parity preserving linear map
\(\VA \rightarrow \left( \mathit{End} \, \VA \right)
\Bbrk{z,\, \frc{1}{z^{\, 2}}} :
a \mapsto Y \left( a,\, z \right)\,\),
endomorphisms
\(\Trn_{\mu} \in \left( \mathit{End} \, \VA \right)_0\)
for \(\mu = 1,\, \dots ,\, D\) called
\textbf{translation endomorphisms}
and an element \(\varvac \in \VA_0\) called \textbf{vacuum}
such that for every
\(a,\, a_1,\, a_2,\, b \in \VA\,\):
\begin{mlist}
\item
\(\left( z^{\, 2} \right)^{N} \,Y \left( a,\, z \right) \, b
\in \VA \Bbrk{z}\) for \(N \mgrt 0\)
(\(\Leftrightarrow \ Y \left( a,\, z \right) \, b
\in \VA \Bbrk{z}_{\,\lscw{z}{\, 2}{}}\,\));
\vspace{-3pt}
\item
\(\left( z_{12}^{\, 2} \right)^N \,
\left[ Y \left( a_1,\, z_1 \right),\, Y \left( a_2,\, z_2 \right) \right]
\, = \, 0\) for \(N \mgrt 0\)
(\(z_{12} := z_1 - z_2\,\));
\item
\(\left[ \hspace{0.6pt} \Trn_{\mu} \hspace{0.6pt} ,\,
Y \left( a,\, z \right) \right] = \di_{z^{\mu}}
\, Y \left( a,\, z \right)\) for \(\mu = 1,\, \dots ,\, D\,\);
\item
\(Y \left( a,\, z \right) \, \varvac \in \VA \Bbrk{z}\) and
\(Y \left( a,\, z \right) \, \varvac
\!\left.\vspe\right|_{\,
\raisebox{1pt}{\small \(z = 0\)}} = a\,\);
\item
\(\Trn_{\mu} \, \varvac = 0\)
for \(\mu = 1,\, \dots D\,\);
\(Y \left(\raisebox{9pt}{\hspace{-2pt}}\right. \hspace{1pt} \varvac,\, z
\left.\raisebox{9pt}{\hspace{-2pt}}\right) = \ID\,\).
\end{mlist}
\end{mdefinition}

The map \(a \mapsto Y \left( a,\, z \right)\) is represented
as a formal series by:
\beq\label{2.1}
Y \left( a,\, z \right) \hspace{-1pt} = \hspace{-3pt}
\Su_{n \, \in \, \Z} \hspace{1pt}
\Su_{m \, = \, 0}^{\infty} \hspace{1pt}
\Su_{\sigma \, = \, 1}^{\har_m} \hspace{1pt}
a_{\,\left\{ n,\, m,\, \sigma \right\}} \hspace{1pt}
\left( z^{\, 2} \right)^n h^{\left( m \right)}_{\sigma}
\left( z \right)
\hspace{1pt} , \ \,
a_{\,\left\{ n,\, m,\, \sigma \right\}} \hspace{1pt} \in \hspace{1pt}
\mathit{End} \hspace{1pt} \VA
\eeq
and \(Y \left( a,\, z \right) b\) is understood as the series
\(\Su_{n \, \in \, \Z}
\Su_{m \, = \, 0}^{\infty}
\Su_{\sigma \, = \, 1}^{\har_m}
a_{\,\left\{ n,\, m,\, \sigma \right\}}b \,
\left( z^{\, 2} \right)^n h^{\left( m \right)}_{\sigma}
\left( z \right)\) $\in$ $\VA \Bbrk{z,\, \frc{1}{z^{\, 2}}}\,$.
For every \(a,\, b \in \VA\,\):
\beq\label{2.2}
a_{\left\{ n,\, m,\, \sigma \right\}} b \, = \,
P_{\left\{ n+N,\, m,\, \sigma \right\}} \left( \di_z \right) \,
\left( z^{\, 2} \right)^N
Y \left( a,\, z \right) \, b
\left.\vspe\right|_{\,
\raisebox{1pt}{\small \(z = 0\)}}
\quad \text{for} \quad
N \mgrt 0
\,
\eeq
($P_{\left\{ n,\, m ,\, \sigma \right\}} \left( z \right)$
are defined by Eq.~(\ref{ad1.11})).
The product
\(Y \left( a_1,\, z_1 \right) \dots Y \left( a_N,\, z_N \right)\)
will be presented by the series
\\ ${}$ \hfill \(
Y \left( a_1,\, z_1 \right) \, \dots \, Y \left( a_N,\, z_N \right) \, =
\mgvspc{15pt}\) \hfill ${}$ \\ ${}$ \hfill \(
=
\Su_{n_1 \, \in \, \Z} \,
\Su_{m_1 \, = \, 0}^{\infty} \,
\Su_{\sigma_1 \, = \, 1}^{\har_{m_1}} \,
\dots
\Su_{n_N \, \in \, \Z} \,
\Su_{m_N \, = \, 0}^{\infty} \,
\Su_{\sigma_N \, = \, 1}^{\har_{m_N}} \,
a_{1\,\left\{ n_1,\, m_1,\, \sigma_1 \right\}} \, \dots
a_{N\,\left\{ n_N,\, m_N,\, \sigma_N \right\}} \, \times
\mgvspc{15pt}\) \hfill ${}$ \\ ${}$ \hfill \(
\times \,
\left( z^{\, 2}_1 \right)^{n_1} \dots \, \left( z^{\, 2}_N \right)^{n_N}
h^{\left( m_1 \right)}_{\sigma_1} \left( z_1 \right) \, \dots \,
h^{\left( m_N \right)}_{\sigma_N} \left( z_N \right)
\mgvspc{15pt}\mgvspc{-11pt}\) \hfill ${}$ \\
belonging to
\(\left( \mathit{End} \, \VA \right) \Bbrk{z_1, \lfrc{1}{z^{\, 2}_1}
\hspace{1pt} ; \dots ; \hspace{1pt} z_N, \lfrc{1}{z^{\, 2}_N}}\,\).

\begin{mdefinition}\label{def:2.2}
Let \(\VA\) be a superspace.
An element
\\ ${}$ \hfill
$u \left( z_1,\, \dots ,\, z_n \right)$ $\in$
\(\left( \mathit{End} \, \VA \right)
\Bbrk{z_1,\frc{1}{z_1^{\, 2}}\hspace{0.5pt};\dots;
\hspace{0.5pt}z_n,\frc{1}{z_n^{\, 2}}}\mgvspc{12pt}\mgvspc{-10pt}\)
\hfill ${}$ \\
is said to be a \textbf{field} if for every \(a \in \VA\,\):
\(u \left( z_1,\, \dots ,\, z_n \right) a \in
\VA \Bbrk{z_1,\, \dots ,\, z_n}_{\,
\lscw{z}{\, 2}{1} \dots \hspace{1pt} \lscw{z}{\, 2}{n}}\)
(i.~e., if
\(\left( z_1^{\, 2} \dots z_n^{\, 2} \right)^{N_a}\)
\(u \left( z_1,\, \dots ,\, z_n \right) \, a \in
\VA \Bbrk{z_1,\, \dots ,\, z_n}\) for \(N_a \mgrt 0\,\)).
\end{mdefinition}

Thus in the case of a vertex algebra $\VA\,$,
$Y \left( a,\, z \right)$ are fields for every \(a \in \VA\,\),
in accord with Definition~\ref{def:2.1} (\textit{a}).
If $u \left( z_1,\, \dots ,\, z_n \right)$ is a field then
we can define $u \left( z,\, \dots ,\, z \right)$ by setting:
\beq\label{2.3}
u \hspace{-1pt} \left( z, \dots , z \right)
a
:= \hspace{-1pt}
\frac{1}{\left( z^{\, 2} \right)^{n \hspace{0.5pt} N_a}}
\left[ \left( z_1^{\, 2} \hspace{-2pt}\dots z_n^{\, 2} \right)^{N_a}
\hspace{-2pt}
u \hspace{-1pt} \left( z_1, \dots , z_n \right)
a
\left.\vspe\!\right|_{\,
\raisebox{1pt}{\small \(z_1 = \dots = z_n = z\)}}
\vspe\!\right]
\hspace{-1pt} ,
\eeq
for
\(a \in \VA\) and \(N_a \mgrt 0\,\),
which does not depend on \(N_a \in \N\,\).
Clearly, if $u \left( z_1,\, \dots ,\, z_n \right)$ is a
field then $u \left( z,\, \dots ,\, z \right)$ and
\(\di_{z_k^{\mu}} \, u \left( z_1,\, \dots ,\, z_n \right)\)
are fields too.

\begin{mdefinition}\label{def:2.3}
Let $a \left( z \right)$ and $b \left( z \right)$ be
two fields on a superspace $\VA$,
$a^{\left( 0 \right)} \left( z \right)$,
$b^{\left( 0 \right)}  \left( z \right)$ and
$a^{\left( 1 \right)}  \left( z \right)$,
$b^{\left( 1 \right)}  \left( z \right)$
be their even and odd parts, respectively
(i.~e. if $a \left( z \right)$ has an expansion of type (\ref{1.14})
with \(a_{\left\{ n,\, m,\, \sigma \right\}} \in \mathit{End} \, \VA\)
then $a^{\left( 0,\, 1 \right)} \left( z \right)$ is the formal series with
coefficients \(a_{\left\{ n,\, m,\, \sigma \right\}}^{\left( 0,\, 1 \right)}
\in \left( \mathit{End} \, \VA \right)_{0,\, 1}\mgvspc{10pt}\),
\(a_{\left\{ n,\, m,\, \sigma \right\}} =
a_{\left\{ n,\, m,\, \sigma \right\}}^{\left( 0 \right)} +
a_{\left\{ n,\, m,\, \sigma \right\}}^{\left( 1 \right)}\)).
The fields $a \left( z \right)$ and $b \left( z \right)$
are said to be
\textbf{mutually local}
if
\(\left( z_{12}^{\, 2} \right)^N\mgvspc{11pt} \hspace{-3pt}
\left[ a^{\left( \varepsilon_1 \right)}  \left( z_1 \right),\hspace{1pt}
b^{\left( \varepsilon_2 \right)}  \left( z_2 \right) \right]\)
$=$ $0$
for \(N \mgrt 0\) and \(\varepsilon_1,\, \varepsilon_2
= 0,\, 1\,\).

\textit{Then}
\(\left( z_{12}^{\, 2} \right)^N
a \left( z_1 \right)  b \left( z_2 \right)\gvspc{11pt}\)
\textit{is a field for}
\(N \mgrt 0\,\).
Indeed, if \(N \mgrt 0\)
then for
all \(v \in \VA\):
\(\left( z_{12}^{\, 2} \right)^N \,
a \left( z_1 \right)  b \left( z_2 \right) \, v
=
\left( z_{12}^{\, 2} \right)^N
\left[\right.
b^{\left( 0 \right)} \left( z_2 \right)
a^{\left( 0 \right)} \left( z_1 \right)
+ \hspace{1pt}
b^{\left( 0 \right)} \left( z_2 \right)
a^{\left( 1 \right)} \left( z_1 \right)
+ \hspace{1pt}
b^{\left( 1 \right)}\hspace{-1pt} \left( z_2 \right)
a^{\left( 0 \right)}\hspace{-1pt} \left( z_1 \right)
- \hspace{1pt}
b^{\left( 1 \right)}\hspace{-1pt} \left( z_2 \right)
a^{\left( 1 \right)}\hspace{-1pt} \left( z_1 \right)
\left.\hspace{-1pt}\right] \hspace{1pt} v
\),
in accord with locality, so that
for
\(M \mgrt 0\,\):
\(\left( z_1^{\, 2} z_2^{\, 2} \right)^M \hspace{-2pt}
\left( z_{12}^{\, 2} \right)^{N_{}^{}} \hspace{-2pt}
a \left( z_1 \right) b \left( z_2 \right)
v \in
\left( \VA \Bbrk{z_1}_{\,\lscw{z}{\, 2}{1}}
\hspace{-1pt}\right) \Bbrk{z_2}
\hspace{1pt} \cap
\left( \VA \Bbrk{z_2}_{\,\lscw{z}{\, 2}{2}}
\hspace{-1pt}\right) \Bbrk{z_1}
\equiv
\VA \Bbrk{z_1,\, z_2}^{\vspe}\).
\end{mdefinition}

\begin{mtheorem}\label{th:2.1}
Let $a \left( z \right)$ and $b\left( z \right)$
be mutually local fields on a superspace $\VA\,$.
Then for \(N \mgrt 0\) and every
\(M \in \N
\,\), \(v \in \VA\,\),
there exists a unique decomposition:
\beqa\label{2.4}
\left( z_{12}^{\, 2} \right)^N \hspace{-1pt}
a \left( z_1 \right) b \left( z_2 \right)
\hspace{1pt} v
\, = && \!\!\!\!
\Su_{\raisebox{0pt}{\scriptsize \(
\begin{array}{c}
n, m = 0, 1, \dots ;
\\
2n + m \leqslant M
\end{array}\)}}
\,
\Su_{\sigma \, = \, 1}^{\har_m} \
\theta^{\left( N,\, M \right)}_{\left\{ n,\, m,\, \sigma \right\}}
\left( z_2 \right)
\, v
\, \left( z_{12}^{\, 2} \right)^n \,
h^{\left( m \right)}_{\sigma} \left( z_{12} \right)
\, + \,
\nn && \!\!\!\!
+ \,
\Su_{\mu_1,\, \dots ,\, \mu_{\text{\tiny \(M \!\! + \!\! 1\)}} \, = \, 1}^D \,
z_{12}^{\mu_1} \dots z_{12}^{\mu_{\text{\tiny \(M \!\! + \!\! 1\)}}} \,
\psi_{\mu_1 \dots\, \mu_{\text{\tiny \(M \!\! + \!\! 1\)}}}^{\left( N,\, M \right)}
\left( z_1,\, z_2 \right)
\, v
\, , \quad
\eeqa
where
\(\theta^{\left( N,\, M \right)}_{\left\{ n,\, m,\, \sigma \right\}}
\left( z \right)\) and
$\psi_{\mu_1 \dots\, \mu_{\text{\tiny \(M \!\! + \!\! 1\)}}}^{\left( N,\, M \right)} \left( z_1,\, z_2 \right)$
are fields.
The fields \(
a \left( z \right)_{\left\{ n,\, m,\, \sigma \right\}}
b \left( z \right)\) $:=$
\(\theta^{\left( N,\, M \right)}_{\left\{ n+N,\, m,\, \sigma \right\}}
\left( z \right)\)
do not depend on $N$ and $M$\gvspc{10pt}
and are determined by
\beq\label{2.5}
a \left( z \right)_{\left\{ n,\, m,\, \sigma \right\}}
\hspace{-1pt} b \left( z \right)
\hspace{1pt} v
=
P_{\left\{ n+N,\, m ,\, \sigma \right\}} \left( \di_{z_1} \right)
\left( z_{12}^{\, 2} \right)^N
a \left( z_1 \right) b \left( z_2 \right)
\hspace{1pt} v
\hspace{-1pt}\left.\vspe\right|_{\,
\raisebox{1pt}{\small \(z_1 = z_2 = z\)}}
\,
\eeq
for sufficiently large $N\,$, independent on
\(v \in \VA\,\),
\(n \in \Z\,\), \(m = 0,\, 1,\, \dots\) and
\(\sigma = 1,\, \dots ,\, \har_m\)
($P_{\left\{ n+N,\, m,\, \sigma \right\}} \left( z \right)\mgvspc{-5pt}$
are the polynomials introduced by~(\ref{ad1.11})).
If $c \left( z \right)$ is another field which is
local with respect to $a \left( z \right)$ and $b \left( z \right)$
then every field
\(a \left( z \right)_{\left\{ n, m, \sigma \hspace{-1pt} \right\}}
\hspace{-2pt} b \left( z \right)\)
is~also~local~with~respect~to~$c \left( z \right)\,$.
\end{mtheorem}

We will prove first two lemmas.

\begin{mlemma}\label{lm:2.2}
Let \(f \left( z_1,\, z_2 \right) \in
\VA \Bbrk{z_1,\, z_2}_{\,\lscw{z}{\, 2}{1} \lscw{z}{\, 2}{2}}\,\).
Then for every \(M \in \N\)
there exists a unique decomposition
\beqa\label{2.6}
f \left( z_1,\, z_2 \right)
\, = && \!\!\!\!
\Su_{\raisebox{0pt}{\scriptsize \(
\begin{array}{c}
n, m = 0, 1, \dots ;
\\
2n + m \leqslant M
\end{array}\)}}
\,
\Su_{\sigma \, = \, 1}^{\har_m} \ \,
g_{\left\{ n,\, m,\, \sigma \right\}}
\left( z_2 \right)
\, \left( z_{12}^{\, 2} \right)^n \,
h^{\left( m \right)}_{\sigma} \left( z_{12} \right)
\, + \,
\nn && \!\!\!\!
+ \,
\Su_{\mu_1,\, \dots ,\, \mu_{\text{\tiny \(M \!\! + \!\! 1\)}} \, = \, 1}^D \,
z_{12}^{\mu_1} \dots z_{12}^{\mu_{\text{\tiny \(M \!\! + \!\! 1\)}}} \,
g_{\mu_1 \dots\, \mu_{\text{\tiny \(M \!\! + \!\! 1\)}}}^{\left( M \right)}
\left( z_1,\, z_2 \right)
\, , \quad
\eeqa
where
\(g_{\left\{ n,\, m,\, \sigma \right\}}
\left( z \right) \in
\VA \Bbrk{z}_{\,\lscw{z}{\, 2}{}}\) and
\(g_{\mu_1 \dots\, \mu_{\text{\tiny \(M \!\! + \!\! 1\)}}}^{
\left( M \right)} \left( z_1,\, z_2 \right) \in
\VA \Bbrk{z_1,\, z_2}_{\,\lscw{z}{\, 2}{1} \lscw{z}{\, 2}{2}}\,\).
Moreover, if
\(f \left( z_1,\, z_2 \right) \hspace{-1pt} \in
\VA \Bbrk{z_1,\, z_2}\)
then
\(g_{\left\{ n,\, m,\, \sigma \right\}}
\left( z \right) \hspace{-1pt} \in
\VA \Bbrk{z}\) and
\(g_{\mu_1 \dots\, \mu_{\text{\tiny \(M \!\! + \!\! 1\)}}}^{
\left( M \right)} \left( z_1,\, z_2 \right) \hspace{-1pt} \in
\VA \Bbrk{z_1,\, z_2}\mgvspc{11pt}\).
\end{mlemma}

\begin{proof}
The uniqueness of the decomposition~(\ref{2.6}) follows from the equality
$$
g_{\left\{ n,\, m,\, \sigma \right\}} \left( z \right)
\, = \,
P_{\left\{ n,\, m,\, \sigma \right\}} \left( \di_{z_1} \right) \,
f \left( z_1,\, z_2 \right)
\!\left.\vspe\right|_{\,
\raisebox{1pt}{\small \(z_1 = z_2 = z\)}}
\, \qquad
$$
in accord with Eq.~(\ref{ad1.11}), so that if
\(f \left( z_1,\, z_2 \right) \in
\VA \Bbrk{z_1,\, z_2}\)
then
\(\forall \, g_{\left\{ n,\, m,\, \sigma \right\}}
\left( z \right) \in
\VA \Bbrk{z}\) as well as
\(\forall \, g_{\mu_1 \dots\, \mu_{\text{\tiny \(M \!\! + \!\! 1\)}}}^{
\left( M \right)} \left( z_1,\, z_2 \right) \in
\VA \Bbrk{z_1,\, z_2}\,\).
One proves the existence first when
\(f \left( z_1, z_2 \right) \in
\VA \Bbrk{z_1, z_2}\mgvspc{11pt}\) by the change of variables
\(\left( z_1, z_2 \right) \mapsto
\left( z_{12} = z_1 - z_2,\, z_2 \right)\,\).
In the general case:
\(f \left( z_1,\, z_2 \right)
= \left( z_1^{\, 2} z_2^{\, 2} \right)^{- \hspace{0.5pt} N}
\phi \left( z_1,\, z_2 \right)\mgvspc{11pt}\)
for \(N \mgrt 0\) and
\(\phi \left( z_1,\, z_2 \right) \in
\VA \Bbrk{z_1,\, z_2}\mgvspc{11pt}\).
Then it is sufficient to prove that there exists the decomposition
$$
f \left( z_1,\, z_2 \right) \, = \,
f \left( z_2,\, z_2 \right) +
\Su_{\mu \, = \, 1}^D \,
z_{12}^{\mu} \
g_{\mu}^{\left( 1 \right)}
\left( z_1,\, z_2 \right)
\, , \qquad
$$
where
\(g_{\mu}^{\lb 1 \rb} \lb z_1,\, z_2 \rb \in
\VA \Bbrk{z_1,\, z_2}_{\, \lscw{z}{\, 2}{1} \lscw{z}{\, 2}{2}}\,\).
The existence of such a decomposition follows, on the other hand, from
$$
\hspace{-3pt}
\begin{array}{c}
f \left( z_1,\, z_2 \right) - f \left( z_2,\, z_2 \right)
= \\ =
\left( z_1^{\, 2} z_2^{\, 2} \right)^{- \hspace{0.5pt} N}
\hspace{-4pt}
\left(\!\vspe\right.
\phi \left( z_1,\, z_2 \right)
\hspace{0.5pt} - \hspace{0.5pt}
\phi \left( z_2,\, z_2 \right)
\left.\vspe\!\right)
\hspace{0.5pt} + \hspace{1pt}
\phi \left( z_2,\, z_2 \right)
\left(\!\vspe\right.
\left( z_1^{\, 2} z_2^{\, 2} \right)^{- \hspace{0.5pt} N}
\hspace{-3pt} - \hspace{0.5pt}
\left( z_2^{\, 2}\right)^{- \hspace{0.5pt} 2 \hspace{0.5pt} N}
\left.\vspe\!\right)
,
\vspace{1pt} \mgvspc{14pt} \\
\left( z_1^{\, 2} z_2^{\, 2} \right)^{- \hspace{0.5pt} N}
\hspace{-3pt} - \hspace{0.5pt}
\left( z_2^{\, 2}\right)^{- \hspace{0.5pt} 2 \hspace{0.5pt} N}
= \mgvspc{14pt} \mgvspc{-5pt} \\ =
\left( z_1^{\, 2} \right)^{- \hspace{0.3pt} N}
\left( z_2^{\, 2} \right)^{- \hspace{0.3pt} 2 \hspace{0.3pt} N}
\left( \Su_{\mu \, = \, 1}^{D}
z_{12}^{\mu} \left( z_1^{\mu} + z_2^{\mu} \right) \right)
\left( \Su_{k \, = \, 0}^{N-1}
\left( z_1^{\, 2} \right)^{k}
\left( z_2^{\, 2} \right)^{N-k-1} \right)
.\text{\qed}
\end{array}
$$
\end{proof}

\begin{mlemma}\label{lm:2.3}
For every \(M \in \N\) and
\(P \left( z \right) \in \C \Brk{z}\) there exist
\(N \in \N\) and
\(Q \left( z,\, w \right) \in \C \Brk{z,\, w}\) such that:
\beq\label{2.9}
\left( z^{\, 2} \right)^N \, P \left( \di_z \right)
\, = \,
Q \left( z,\, \di_z \right) \, \left( z^{\, 2} \right)^M
\, , \qquad
\eeq
where the equation is understood as an operator equality and
$Q \left( z, \di_z \right)$
stands for the polynomial $Q \left( z, w \right)$
with each monomial
\(z^{\mu_1} \!\dots z^{\mu_k}\)
\(w^{\nu_1} \!\dots w^{\nu_l}\) replaced by
\(z^{\mu_1} \!\dots z^{\mu_k}\)
\(\di_{z^{\nu_1}} \!\dots\) \(\di_{z^{\nu_l}}\,\).
\end{mlemma}

\begin{proof}
Apply induction in $\mathit{deg} \, P$.
If \(\mathit{deg} \, P = 0\) then \(N = M\) and \(Q = P\,\).
If \(\mathit{deg} \, P > 0\) then
\(P \left( z \right) = \Su_{\mu \, = \, 1}^D z^{\mu}
\, P_{\mu} \left( z \right) + P_0 \left( z \right)\,\)
where \(\mathit{deg} \, P_{\mu} \left( z \right) <
\mathit{deg} \, P \left( z \right)\) for
\(\mu = 0,\, \dots ,\, D\,\).
By induction:
for every \(\mu = 0,\, \dots ,\, D\) there exist
\(N_{\mu} \in \N\) and \(Q_{\mu}' \left( z,\, w \right)
\in \C \Brk{z,\, w}\) such that
\(\left( z^{\, 2} \right)^{N_{\mu}} P \left( \di_z \right)
=
Q_{\mu}' \left( z,\, \di_z \right)
\left( z^{\, 2} \right)^{M+1}\mgvspc{11pt}\).
Then let \(N = \mathit{max} \,
\left\{\mgvspc{9pt}\hspace{-3pt}\right. N_{\mu} :
\mu = 0,\, \dots ,\)
\(D \left.\mgvspc{9pt}\hspace{-3pt}\right\}\) so that
\(\left( z^{\, 2} \right)^N P \left( \di_z \right)
=
Q_{\mu}'' \left( z,\, \di_z \right)
\left( z^{\, 2} \right)^{M+1}\gvspc{11pt}\).
Thus:
$$
\begin{array}{c}
\left( z^{\, 2} \right)^N P \left( \di_z \right)
= \\ =
\left(
\Su_{\mu \, = \, 1}
\left(\vspe\!\!\right.
Q_{\mu}'' \left( z,\, \di_z \right) \di_{z^{\mu}} \,
z^{\, 2}
-
2 \left( M+1 \right)
Q_{\mu}'' \left( z,\, \di_z \right) z^{\mu}
\left.\vspe\!\!\right)
+
Q_{0}'' \left( z,\, \di_z \right) z^{\, 2}
\right)
\left( z^{\, 2} \right)^M
\! .
\end{array}
$$
${}$ \hfill \QED
\end{proof}

\textit{Proof of Theorem~\ref{th:2.1}.}
The first part of the theorem follows from Lemma~\ref{lm:2.2}.
For the last statement
we have to prove that:
\beq\label{2.10}
\left( z_{12}^{\, 2} \right)^N \,
\left[ c \left( z_1 \right) ,\,
a \left( z_2 \right)_{\left\{ n,\, m,\, \sigma \right\}}
b \left( z_2 \right) \right]
\, v
\, = \, 0
\ \ \text{for} \ \ N \mgrt 0 \ \ \text{and} \ \
v \, \in \, \VA
\, .
\eeq
Because of the equality
\\ ${}$ \hfill
\(
\left( a \left( z \right)_{\left\{ n,\, m,\, \sigma \right\}}
b \left( z \right) \right)^{\left( \varepsilon \right)} =
\Su_{\varepsilon_1 \, = \, 0,\, 1 \, \text{\textit{mod}} \, 2} \,
a^{\left( \varepsilon + \varepsilon_1 \right)}
\left( z \right)_{\left\{ n,\, m,\, \sigma \right\}}
b^{\left( \varepsilon_1 \right)} \left( z \right)
\)
\hfill ${}$ \\
(\gvspc{11pt}using the notations of Definition~\ref{def:2.3})
it is sufficient to consider the case when the fields
$a \left( z \right)\,$, $b \left( z \right)$ and
$c \left( z \right)$ have fixed parities \(p_a,\,
p_b,\, p_c \in \Z_2\,\), respectively.
Then we have
\beqa\label{new2.8.2}
&& \hspace{-4pt}
\left( z_{12}^{\, 2} \right)^{N+M} \,
c \left( z_1 \right) \,
a \left( z_2 \right)_{\left\{ n,\, m,\, \sigma \right\}}
b \left( z_2 \right)
\, v
=
\vspace{4pt} \nn && \hspace{-10pt} =
\left( z_{12}^{\, 2} \right)^{N}
\left( z_{13}^{\, 2} \right)^{M}
P_{\left\{ n+M,\, m,\, \sigma \right\}} \left( \di_{z_2} \right)
\left( z_{23}^{\, 2} \right)^{M}
c \left( z_1 \right)
a \left( z_2 \right)
b \left( z_3 \right)
\, v
\!\left.\vspe\right|_{\,
\raisebox{1pt}{\small \(z_3 = z_2\)}}
\hspace{-5pt} =
\vspace{3pt} \nn && \hspace{-10pt} =
Q \left( z_{12},\, \di_{z_2} \right)
\left( z_{12}^{\, 2} \right)^{M}
\left( z_{13}^{\, 2} \right)^{M}
\left( z_{23}^{\, 2} \right)^{M}
c \left( z_1 \right)
a \left( z_2 \right)
b \left( z_3 \right)
\, v
\!\left.\vspe\right|_{\,
\raisebox{1pt}{\small \(z_3 = z_2\)}}
\hspace{-5pt} =
\vspace{3pt} \nn && \hspace{-10pt} =
\left( -1 \right)^{p_ap_c + p_bp_c}
Q \left( z_{12},\, \di_{z_2} \right)
\left( z_{12}^{\, 2} \right)^{M} \!
\left( z_{13}^{\, 2} \right)^{M} \!
\left( z_{23}^{\, 2} \right)^{M}
a \left( z_2 \right)
b \left( z_3 \right)
c \left( z_1 \right)
\, v
\!\left.\vspe\right|_{\,
\raisebox{1pt}{\small \(z_3 = z_2\)}}
\hspace{-5pt} =
\vspace{3pt} \hspace{-10pt} \nn && \hspace{-10pt} =
\left( -1 \right)^{p_ap_c + p_bp_c}
\left( z_{12}^{\, 2} \right)^{N}
\left( z_{13}^{\, 2} \right)^{M} \, \times \,
\vspace{3pt} \nn && \hspace{-10pt} \hspace{12pt} \times \,
P_{\left\{ n+M,\, m,\, \sigma \right\}} \left( \di_{z_2} \right)
\left( z_{23}^{\, 2} \right)^{M}
a \left( z_2 \right)
b \left( z_3 \right)
c \left( z_1 \right)
\, v
\!\left.\vspe\right|_{\,
\raisebox{1pt}{\small \(z_3 = z_2\)}}
\hspace{-5pt} =
\vspace{3pt} \nn && \hspace{-10pt} =
\left( -1 \right)^{\lb p_a + p_b \rb \hspace{1pt} p_c} \,
\left( z_{12}^{\, 2} \right)^{N+M} \,
a \left( z_2 \right)_{\left\{ n,\, m,\, \sigma \right\}}
b \left( z_2 \right) \,
c \left( z_1 \right)
\, v
,
\hspace{-10pt}
\eeqa
for sufficiently large $M$ and $N\,$, independent on \(v \in \VA\,\),
in accord with Eq.~(\ref{2.5}) and
Lemma~\ref{lm:2.3}.\qed
\spc

The $\left\{ 0,0,1 \right\}$--product is the natural candidate for
the notion of normal product in vertex algebras which generalizes
the corresponding one from the chiral CFT:
\beq\label{normpr.1}
: \hspace{-1pt}
Y \left( a,\, z \right) Y \left( b,\, z \right)
\hspace{-1pt} : \ := \
Y \left( a,\, z \right)_{\left\{ 0,\, 0,\, 1 \right\}} Y \left( b,\, z \right)
\, . \qquad
\eeq
As a consequence of Eqs.~(\ref{2.5}) and (\ref{polynomial}) it
could be expressed as:
\beq\label{mormpr.2}
\hspace{2pt}
: \hspace{-1pt}
Y \left( a,\, z \right) Y \left( b,\, z \right)
\hspace{-1pt} : v
=
K_N
\left( \di_{z_1}^{\, 2} \right)^N
\left[ \left( z_{12}^{\, 2} \right)^N
Y \left( a,\, z_1 \right) Y \left( b,\, z_2 \right) \right]
v
\hspace{-1pt}\left.\vspe\right|_{\,
\raisebox{1pt}{\small \(z_1 = z_2 = z\)}}
\hspace{-10pt}
\eeq
for \(N \mgrt 0\) and every \(v \in \VA\).

\msection{Consequences of the Existence of a Vacuum and of
Translation Invariance}{sec:3}

There is a vertex algebra analog of (the corollary of) the
\textit{Reeh--Schlider} theorem~--~the separating property of
the vacuum~\cite{Jost 65}.

\begin{mtheorem}\label{th:2.4}
Let $\VA$ be a vertex algebra and
$u \left( z \right)$ be a field on $\VA$
which is mutually local with respect to all fields
$Y \left( a,\, z \right)\,$, \(a \in \VA\,\).
Then if \(u \left( z \right) \, \varvac = 0\)
it follows that \(u \left( z \right) = 0\,\).
\end{mtheorem}

\begin{proof}
Because of locality we have for
every \(a \in \VA\) and
\(N_a \mgrt 0\,\):
\\ ${}$ \hfill
\(
\left( z_{12}^{\, 2} \right)^{N_a}
u \left( z_1 \right) Y \left( a,\, z_2 \right) \, \varvac
=
\left( z_{12}^{\, 2} \right)^{N_a}
Y \left( a,\, z_2 \right) u \left( z_1 \right) \, \varvac
\mgvspc{15pt}\mgvspc{-8pt}\),
\hfill ${}$ \\
thus obtaining
\(
\left( z_{12}^{\, 2} \right)^{N_a}
u \left( z_1 \right) Y \left( a,\, z_2 \right) \, \varvac
= 0
\mgvspc{11pt}\).
Then we can set \(z_2 = 0\) and divide by
\(\left( z_{12}^{\, 2} \right)^{N_a} =
\left( z_1^{\, 2} \right)^{N_a}\mgvspc{11pt}\)
because it multiplies an element of
\(\VA \Bbrk{z_1}_{\,\lscw{z}{\, 2}{1}}\,\)
(in the $\left( \C \Bbrk{z}_{\,\lscw{z}{\, 2}{}} \right)$--module
$\VA \Bbrk{z}_{\,\lscw{z}{\, 2}{}}\mgvspc{11pt}$
there are no zero divisors).
Thus we obtain that \(u \left( z \right) a = 0\)\gvspc{12.4pt}
for every
\(a \in \VA\,\).\qed
\end{proof}

The following proposition shows that the system of fields
\(\left\{ Y \left( a,\, z \right) : a \in \VA \right\}\)
is a maximal system of translation invariant local fields.

\begin{mproposition}\label{pr:2.5}
Let $\VA$ be a vertex algebra and
$u \left( z \right)$ be a field on $\VA$
which is mutually local with respect to all fields
$Y \left( a,\, z \right)\,$, \(a \in \VA\,\).
Then the following conditions are equivalent:
\begin{mlist}
\item
\(\left[ \Trn_{\mu},\, u \left( z \right) \right] =
\di_{z^{\mu}} \, u \left( z \right)\)
for
\(\mu = 1,\, \dots ,\, D\)
and
\(u \left( z \right) \, \varvac \in \VA \Bbrk{z}\,\),
as for \(z=0\),
\(u \left( z \right) \, \varvac
\!\left.\vspe\right|_{\,
\raisebox{1pt}{\small \(z = 0\)}}\) $=$ $c\mgvspc{13pt}$;
\vspace{-6pt}
\item
\(u \left( z \right) \, \varvac
= \exp \left( {\Trn \hspace{-1pt} \spr \hspace{1pt} z} \right)
\, c\,\), where
\(\Trn \hspace{-1pt} \spr \hspace{1pt} z
\equiv \Su_{\mu \, = \, 1}^D \, \Trn_{\mu} \, z^{\mu}\)
and
\(\exp \left( {\Trn \hspace{-1pt} \spr \hspace{1pt} z} \right)\)
$=$
\linebreak
\(\Su_{n \, = \, 0}^{\infty} \, \raisebox{-0.3pt}{\Large \(\frac{1}{n!}\)}\)
$\left( {\Trn \spr z} \right)^n$
$\in$ $\left( \mathit{End} \, \VA \right) \Bbrk{z}\,$;
\item
\(u \left( z \right) = Y \left( c,\, z \right)\,\).
\end{mlist}
\end{mproposition}

\begin{proof}
\(\left( a \right) \Rightarrow \left( b \right)\,\).
The equality \(u \left( z \right) \hspace{1pt} \varvac =
\exp \left( {\Trn \hspace{-1pt} \spr \hspace{1pt} z} \right) \hspace{1pt} c\)
appears as the unique solution of
the equations
\(\di_{z^{\mu}} \left( u \left( z \right) \hspace{1pt} \varvac \right) =
\Trn_{\mu} \left( u \left( z \right)
\hspace{1pt} \varvac \right)\)\gvspc{10pt}
for \(\mu = 1,\, \dots ,\, D\) with initial condition
\(u \left( z \right) \hspace{1pt} \varvac
\!\left.\vspe\right|_{\,
\raisebox{1pt}{\small \(z = 0\)}} = c\,\).
Indeed, if
\\ ${}$ \hfill
\(u \left( z \right) \, \varvac =
\Su_{n \, = \, 0}^{\infty} \, \Su_{\mu_1,\, \dots ,\, \mu_n \, = \, 1}^{D}
\, c^{\left( n \right)}_{\mu_1 \hspace{1pt} \dots \hspace{2pt} \mu_n}
\, z^{\mu_1} \dots z^{\mu_n}\,\),
\hspace{0.5cm}
\(c^{\left( n \right)}_{\mu_1 \hspace{1pt} \dots \hspace{2pt} \mu_n} \in \VA\,\),
\hfill ${}$ \\
then \(c^{\left( 0 \right)} = c\)
and
\(c^{\left( n \right)}_{\mu_1 \hspace{1pt} \dots \hspace{2pt} \mu_n} =
\raisebox{-0.3pt}{\Large \(\frac{1}{n}\)} \ \Trn_{\mu_1} \,
c^{\left( n-1 \right)}_{\mu_2 \hspace{1pt} \dots \hspace{2pt} \mu_n} =
\dots =
\raisebox{-0.3pt}{\Large \(\frac{1}{n!}\)} \
\Trn_{\mu_1} \dots \Trn_{\mu_n}\, c\mgvspc{13pt}\)
for \(n > 1\,\).
\\
\(\left( b \right) \Rightarrow \left( c \right)\mgvspc{11pt}\).
By Definition~\ref{def:2.1} (\textit{c}) and~(\textit{d}),
and the implication
\(\left( a \right) \Rightarrow \left( b \right)\) above we have
\(Y \left( c,\, z \right) \hspace{1pt} \varvac =
\exp \left( {\Trn \hspace{-1pt} \spr \hspace{1pt} z} \right) \hspace{1pt} c\,\).
Then
\(\left(\vspe\! u \left( z \right) - Y \left( c,\, z \right) \right)
\hspace{1pt} \varvac = 0\) and by Theorem~\ref{th:2.4}
we conclude that \(u \left( z \right) = Y \left( c,\, z \right)\,\).
\\
\(\left( c \right) \Rightarrow \left( a \right)\mgvspc{11pt}\).
This is a part of Definition~\ref{def:2.1}
(conditions~(\textit{c}) and~(\textit{d})).\qed
\end{proof}

\begin{mcorollary}\label{cr:2.9}
Let $\VA$ be a vertex algebra. Then
for all \(a \in \VA\)
and
\(\mu \, = \, 1,\, \dots ,\, D\,\):
\beq\label{add2.14.1}
Y \left( \Trn_{\mu} \, a,\, z \right) \, = \,
\di_{z^{\mu}} \, Y \left( a,\, z \right)
\, \equiv \,
\left[ \Trn_{\mu},\, Y \left( a,\, z \right) \right]
\, . \qquad
\eeq
\end{mcorollary}

\begin{proof}
Eq.~(\ref{add2.14.1}) follows from the equality
\\ ${}$ \hfill
\(Y \left( \Trn_{\mu} \, a,\, z \right) \, \varvac
\restr{z = 0} = \Trn_{\mu} \, a =
\left[ \Trn_{\mu},\, Y \left( a,\, z \right) \right] \, \varvac
\restr{z = 0}
\mgvspc{14pt}\mgvspc{-8pt}\)
\hfill ${}$ \\
and Proposition~\ref{pr:2.5}.\qed
\end{proof}

\begin{mproposition}\label{cr:2.6}
Let $\VA$ be a vertex algebra.
Then for all \(a,\, b \in \VA\) and
\(n \in \Z\,\),
\(m = 0,\, 1,\, \dots\,\),
\(\sigma \, = \, 1,\, \dots ,\, \har_m\):
\beq\label{n2.8}
Y \left( a,\, z \right)_{\left\{ n,\, m,\, \sigma \right\}}
Y \left( b,\, z \right)
\, = \,
Y \left(  a_{\left\{ n,\, m,\, \sigma \right\}}b,\, z \right)
\, , \qquad
\eeq
and for \(n \geqslant 0\,\):
\beq\label{n2.9}
Y \left( a,\, z \right)_{\left\{ n,\, m,\, \sigma \right\}}
Y \left( b,\, z \right)
\, = \,
\left(\vspe\!
P_{\left\{ n,\, m,\, \sigma \right\}} \left( \di_{z} \right) \,
Y \left(  a,\, z \right) \right){\!}_{\left\{ 0,\, 0,\, 1 \right\}}
Y \left( b,\, z \right)
\, . \quad
\eeq
\end{mproposition}

\begin{proof}
To prove Eq.~(\ref{n2.8}) we will basically use Eq.~(\ref{2.5}).
First we have for \(N \mgrt 0\)
and all \(v \in \VA\,\), \(\mu = 1,\, \dots ,\, D\,\):
\beqa\label{new2.11.3}
&& \hspace{-10pt}
\left[
\Trn_{\mu},\,
Y \left( a,\, z \right)_{\left\{ n,\, m,\, \sigma \right\}}
Y \left( b,\, z \right)
\right]
\, v
\, = \,
\nn && \hspace{-10pt}
= \,
P_{\left\{ n+N,\, m,\, \sigma \right\}} \left( \di_{z_1} \right) \,
\left( z_{12}^{\, 2} \right)^{N} \,
\left[
\Trn_{\mu},\,
Y \left( a,\, z_1 \right)
Y \left( b,\, z_2 \right)
\right]
\, v
\!\left.\vspe\right|_{\,
\raisebox{1pt}{\small \(z_1 = z_2 = z\)}}
\, = \,
\nn && \hspace{-10pt}
= \,
P_{\left\{ n+N,\, m,\, \sigma \right\}} \left( \di_{z_1} \right) \,
\left( z_{12}^{\, 2} \right)^{N} \,
\times \nn && \hspace{-10pt} \hspace{12pt} \times \hspace{1pt}
\left(\vspe\hspace{-2.5pt}
\di_{z_1^{\mu}} \,
Y \left( a,\, z_1 \right) \,
Y \left( b,\, z_2 \right)
+
Y \left( a,\, z_1 \right) \,
\di_{z_2^{\mu}} \,
Y \left( b,\, z_2 \right)
\right)
\, v
\!\left.\vspe\right|_{\,
\raisebox{1pt}{\small \(z_1 = z_2 = z\)}}
\, = \,
\hspace{-10pt}
\nn && \hspace{-10pt}
= \,
P_{\left\{ n+N,\, m,\, \sigma \right\}} \left( \di_{z_1} \right) \,
\left(
\di_{z_1^{\mu}} + \di_{z_2^{\mu}}
\right) \,
\left( z_{12}^{\, 2} \right)^{N} \,
Y \left( a,\, z_1 \right) \,
Y \left( b,\, z_2 \right)
\, v
\!\left.\vspe\right|_{\,
\raisebox{1pt}{\small \(z_1 = z_2 = z\)}}
\, = \,
\nn && \hspace{-10pt}
= \,
\di_{z^{\mu}} \,
\left[\vspe\hspace{-2pt}
P_{\left\{ n+N,\, m,\, \sigma \right\}} \left( \di_{z_1} \right) \,
\left( z_{12}^{\, 2} \right)^{N} \,
Y \left( a,\, z_1 \right) \,
Y \left( b,\, z_2 \right)
\, v
\!\left.\vspe\right|_{\,
\raisebox{1pt}{\small \(z_1 = z_2 = z\)}}
\right]
\, = \,
\nn && \hspace{-10pt}
= \,
\di_{z^{\mu}} \,
\left(\vspe\hspace{-2pt}
Y \left( a,\, z \right)_{\left\{ n,\, m,\, \sigma \right\}}
Y \left( b,\, z \right)
\, v
\right)
\, . \qquad
\eeqa
On the other hand, for \(N \mgrt 0\,\):
\beqa\label{new2.11.4}
&
Y \left( a,\, z \right)_{\left\{ n,\, m,\, \sigma \right\}}
Y \left( b,\, z \right)
\, \varvac
\!\left.\vspe\right|_{\,
\raisebox{1pt}{\small \(z = 0\)}}
\, = \, & \nn & \, = \,
\left[\vspe\hspace{-2pt}
P_{\left\{ n+N,\, m,\, \sigma \right\}} \left( \di_{z_1} \right) \,
\left( z_{12}^{\, 2} \right)^{N} \,
Y \left( a,\, z_1 \right) \,
Y \left( b,\, z_2 \right)
\, \varvac
\!\left.\vspe\right|_{\,
\raisebox{1pt}{\small \(z_1 = z_2 = z\)}}
\right]_{\,
\raisebox{1pt}{\small \(z = 0\)}}
\, . & \qquad
\eeqa
But
\(
P_{\left\{ n+N,\, m,\, \sigma \right\}} \left( \di_{z_1} \right) \,
\left( z_{12}^{\, 2} \right)^{N} \,
Y \left( a,\, z_1 \right) \,
Y \left( b,\, z_2 \right)
\, \varvac \in \VA \Bbrk{z_1,\, z_2}\mgvspc{-5pt}\),
so that the consecutive restrictions \(z_1 = z_2 = z\) and
\(z = 0\) are equivalent to the restrictions:
first \(z_2 = 0\) and then \(z_1 = 0\,\).
In such a way we obtain that
\\ ${}$ \hfill
\(Y \left( a,\, z \right)_{\left\{ n,\, m,\, \sigma \right\}}
Y \left( b,\, z \right)
\, \varvac
\!\left.\vspe\right|_{\,
\raisebox{1pt}{\small \(z = 0\)}} =
a_{\left\{ n,\, m,\, \sigma \right\}} b
\mgvspc{14pt}\mgvspc{-8pt}\).
\hfill ${}$ \\
Combining these two results we conclude by
Proposition~\ref{pr:2.5} that Eq.~(\ref{n2.8}) holds.

The proof of~(\ref{n2.9}) uses Eqs.~(\ref{n2.8}), (\ref{2.2})
and some additional properties of the polynomials
$P_{\left\{ n,\, m,\, \sigma \right\}} \left( z \right)$.
We will not prove (\ref{n2.9}) since we will not use it further.\qed
\end{proof}

\begin{mcorollary}\label{cr:2.7}
Let
\(u \left( z \right)  =
\Su_{n \, \in \, \Z} \,
\Su_{m \, = \, 0}^{\infty} \,
\Su_{\sigma \, = \, 1}^{\har_m} \,
u_{\,\left\{ n,\, m,\, \sigma \right\}} \,
\left( z^{\, 2} \right)^n \, h^{\left( m \right)}_{\sigma}
\left( z \right)\) and
\(v \left( z \right)
\)
be two mutually local fields on a superspace $\VA$
and \(\varvac \in \VA_0\) be such that
\(u \left( z \right) \hspace{1pt} \varvac\) and
\(v \left( z \right) \hspace{1pt} \varvac\)
belong to
$\in$ $\VA \Bbrk{z}\mgvspc{10pt}$.
Then
\beq\label{n2.10}
u \left( z \right)_{\left\{ n,\, m,\, \sigma \right\}}
v \left( z \right)
\, \varvac
\!\left.\vspe\right|_{\,
\raisebox{1pt}{\small \(z = 0\)}}
\, = \,
u_{\left\{ n,\, m,\, \sigma \right\}}
\left(
v \left( z \right)
\hspace{1pt} \varvac
\!\left.\vspe\right|_{\,
\raisebox{1pt}{\small \(z = 0\)}} \right)
\, . \qquad
\eeq
\end{mcorollary}

\begin{proof}
This can be derived as in the proof of
Proposition~\ref{cr:2.6}
(the derivation of~(\ref{new2.11.4})).\qed
\end{proof}

\msection{Existence Theorem. Analytic Continuations}{sec:4}

The next theorem allows one to construct a vertex algebra
from a system of mutually local and ``translation covariant''
fields which give rise to
the entire space by acting on the vacuum.

\begin{mtheorem}\label{th:2.8}
{\rm (``Existence Theorem''.)} \\
Let
\(u^{\alpha} \left( z \right)  =
\Su_{n \, \in \, \Z} \,
\Su_{m \, = \, 0}^{\infty} \,
\Su_{\sigma \, = \, 1}^{\har_m} \,
u_{\,\left\{ n,\, m,\, \sigma \right\}}^{\alpha} \,
\left( z^{\, 2} \right)^n \, h^{\left( m \right)}_{\sigma}
\left( z \right)\) for \(\alpha \in \mathcal{A}\)
be a system of mutually local fields on a superspace
\(\VA\,\).
Let \(\varvac \in \VA_0\) and
\(\Trn_{\mu} \in \left( \mathit{End} \, \VA \right)_0\)
be such that
\(\Trn_{\mu} \varvac = 0\)
for \(\mu = 1,\, \dots ,\, D\mgvspc{10pt}\)
and:
\begin{mlist}
\item
\(\left[ \Trn_{\mu}, u^{\alpha} \left( z \right) \right] =
\di_{z^{\mu}} u^{\alpha} \left( z \right)\)
and
\(u^{\alpha} \left( z \right) \, \varvac \in \VA \Bbrk{z}\)
for all \(\alpha \in \mathcal{A}\),
\(\mu = 1, \dots , D\mgvspc{10pt}\);
\item
the set of all elements
\(u^{\alpha_1}_{\left\{ n_1,\, m_1,\, \sigma_1 \right\}}
\dots
u^{\alpha_N}_{\left\{ n_N,\, m_N,\, \sigma_N \right\}} \, \varvac
\)
for \(N = 0, 1, \dots ,\)
\(\alpha_k \in \mathcal{A}\,\), \(n_k \in \Z\,\), \(n_N \geqslant 0\,\),
\(m_k = 0, 1, \dots\,\),
\(\sigma_k = 1,\, \dots ,\, \har_{m_k}\)
(\(k = 1,\, \dots ,\,  N\mgvspc{10pt}\)),
spans the space $\VA\mgvspc{9pt}$.
\end{mlist}
Then
there exists a unique structure of a vertex algebra with vacuum
$\varvac$ and translation endomorphisms $\Trn_{\mu}$ on $\VA$
such that
\beq\label{n2.12}
Y \left( u^{\alpha},\, z \right) \, = \,
u^{\alpha} \left( z \right) \quad \text{\textit{for}} \quad
u^{\alpha} \, := \,
u^{\alpha} \left( z \right) \, \varvac
\!\left.\vspe\right|_{\,
\raisebox{1pt}{\small \(z = 0\)}}
\, , \quad
\alpha \, \in \, \mathcal{A}
\, . \qquad
\eeq
The operators $Y \left( a,\, z \right)$
are determined for the vectors of the set in
the above condition~{\rm (}\textit{b}{\rm )} by:
\beqa\label{n2.13}
&
Y
\left(
u^{\alpha_1}_{\left\{ n_1,\, m_1,\, \sigma_1 \right\}}
\dots \,
u^{\alpha_N}_{\left\{ n_N,\, m_N,\, \sigma_N \right\}} \, \varvac,\, z
\right)
\, = \,
& \nn & \, = \,
u^{\alpha_1}
\left( z \right)_{\left\{ n_1,\, m_1,\, \sigma_1 \right\}}
\left(\mgvspc{12pt}\hspace{-3pt}\right.
\dots \
u^{\alpha_{N-1}}
\left( z \right)_{\left\{ n_{N-1},\, m_{N-1},\, \sigma_{N-1} \right\}}
\left(\mgvspc{9pt}\hspace{-3pt}\right.
P_{\left\{ n_{N},\, m_{N},\, \sigma_{N} \right\}} \left( \di_{z} \right) \,
\times \hspace{-20pt} & \nn & \times \,
u^{\alpha_{N}} \left( z \right)
\left.\mgvspc{9pt}\hspace{-3pt}\right) \dots
\left.\mgvspc{12pt}\hspace{-3pt}\right)
\, . & \qquad
\eeqa
\end{mtheorem}

\begin{proof}
Set \(Y \left(\raisebox{8pt}{\hspace{-2pt}}\right. \varvac,\, z
\left.\raisebox{8pt}{\hspace{-2pt}}\right) = \ID\) and
take Eq.~(\ref{n2.13}) as a definition for the operators
$Y \left( a,\, z \right)$ restricting to a subsystem of the set
displayed in condition~(\textit{b}) which
contains $\varvac$ and
forms a basis of $\VA\,$.
By Theorem~\ref{th:2.1} we obtain a system of mutually local
fields.
The conditions~(\textit{c}) and (\textit{d})
of Definition~\ref{def:2.1}
can be proven by induction in $N$ for the fields~(\ref{n2.13})
following the argument of the first part of the proof of
Proposition~\ref{cr:2.6}
(the computations of~(\ref{new2.11.3}) and~(\ref{new2.11.4})).
The uniqueness follows from Proposition~\ref{pr:2.5}.\qed
\end{proof}

Now
we will find an analogue of the analytic
continuation of products of Wightman fields acting on the vacuum.

Let $\Alg$ be a ring and $\Lin$ be an $\Alg$--module.
Then $\Lin \Bbrk{z}_{\,\lscw{z}{\, 2}{}}$ is an
$\left( R \Bbrk{z}_{\,\lscw{z}{\, 2}{}} \right)$--module with derivations
$\di_{z^{\mu}}$ for \(\mu = 1,\, \dots D\,\).
Moreover, if the $\Alg$--module $\Lin$ has no zero divisors
then this is also true for the
$\left( R \Bbrk{z}_{\,\lscw{z}{\, 2}{}} \right)$--module
$\Lin \Bbrk{z}_{\,\lscw{z}{\, 2}{}}\,$.

From this simple fact it follows by induction that
\beq\label{3.1.1}
\Lin \Bbrk{z_1}_{\,\lscw{z}{\, 2}{1}}
\dots \, \Bbrk{z_n}_{\,\lscw{z}{\, 2}{n}}
\, := \,
\left( \Lin \Bbrk{z_1}_{\,\lscw{z}{\, 2}{1}}
\dots \right) \Bbrk{z_n}_{\,\lscw{z}{\, 2}{n}}
\, \qquad
\eeq
is a
\(\left( \C \Bbrk{z_1}_{\,\lscw{z}{\, 2}{1}}
\dots \Bbrk{z_n}_{\,\lscw{z}{\, 2}{n}} \right)\)--module\gvspc{-10pt}
with derivations $\di_{z_k^{\mu}}$
(\(k = 1,\, \dots ,\) $n$,
\(\mu = 1,\, \dots ,\) $D$), which has no zero divisors.
Note that
\beq\label{3.1.2}
\hspace{3pt}
\Lin \Bbrk{z_1,\, \dots ,\, z_n}_{\,\lscw{z}{\, 2}{1}
\hspace{0pt}\dots\hspace{2pt}\lscw{z}{\, 2}{n}}
\hspace{1pt} \varsubsetneqq \hspace{1pt}
\Lin \Bbrk{z_1}_{\,\lscw{z}{\, 2}{1}}
\dots \, \Bbrk{z_n}_{\,\lscw{z}{\, 2}{n}}
\hspace{1pt} \varsubsetneqq \hspace{1pt}
\Lin \Bbrk{z_1,\, \frc{1}{\lscw{z}{\, 2}{1}}\,
;\hspace{1pt}\dots\hspace{1pt}; \,
z_n,\, \frc{1}{\lscw{z}{\, 2}{n}}}
\hspace{1pt} .
\hspace{-20pt}
\eeq

\vspace{0.1in}

It follows from the definition of vertex algebra (Def.~\ref{def:2.1})
that in a vertex algebra $\VA\,$, for all
\(a_1,\, \dots ,\, a_n,\, b \in \VA\,\):
\beq\label{new2.15.x}
Y \left( a_1,\, z_1 \right) \dots \hspace{1pt}
Y \left( a_n,\, z_n \right) \hspace{1pt} b
\, \in \,
\VA \Bbrk{z_1}_{\,\lscw{z}{\, 2}{1}}
\dots \, \Bbrk{z_n}_{\,\lscw{z}{\, 2}{n}}
\, . \qquad
\eeq

\Vspa

Let us introduce the following multiplicative systems in
\(\C \Brk{z_1,\, \dots ,\, z_n}\,\):
\beqa\label{3.1.4}
L_n \, \hspace{-3pt} := && \!\!\!
\left\{\mgvspc{16pt}\!\right.
\mathop{\prod}\limits_{k \, = \, 1}^{N} \,
\left(\mgvspc{16pt}\!\right.
\Su_{l \, = \, 1}^{n} \, \lambda_{k,\, l} \, z_l
\left.\mgvspc{16pt}\!\!\right)^{\hspace{-3pt} 2}
\hspace{-1pt}
\, : \,
N \in \N \hspace{1pt} ,\,
\left( \lambda_{k,\, 1},\, \dots ,\, \lambda_{k,\, n} \right)
\in \C^n \,\backslash \left\{ 0 \right\} \
\nn && \!\!\! \hspace{105pt}
\text{for} \ k = 1,\, \dots ,\, N
\left.\mgvspc{16pt}\!\right\}
\hspace{-1pt} , \qquad
\\ \label{3.1.5}
R_n \, \hspace{-3pt} := && \!\!\!
\left\{\mgvspc{16pt}\!\right.
\left(\mgvspc{16pt}\!\right.
\mathop{\prod}\limits_{k \, = \, 1}^{N} \,
z_k^{\, 2}
\left.\mgvspc{16pt}\!\!\right)^{\hspace{-4pt} N}
\left(\mgvspc{16pt}\!\right.
\mathop{\prod}\limits_{1 \, \leqslant \, l \, < \, m \, \leqslant \, n} \,
z_{lm}^{\, 2}
\left.\mgvspc{16pt}\!\!\right)^{\hspace{-4pt} N}
\hspace{-1pt}
\, : \,
N \in \N
\left.\mgvspc{16pt}\!\right\}
\, \qquad
\eeqa
(\(z_{lm} = z_l - z_m \in \C \Brk{z_l,\, z_m}\)). Clearly,
\beq\label{3.1.6}
R_n \, \varsubsetneqq \, L_n
\quad \text{and} \quad
\Lin \Bbrk{z_1,\, \dots ,\, z_n}_{\, \raisebox{-2pt}{\small $R_n$}}
\, \varsubsetneqq \,
\Lin \Bbrk{z_1,\, \dots ,\, z_n}_{\, \raisebox{-2pt}{\small $L_n$}}
\, \qquad
\eeq
for every vector space $\Lin$ as the localized modules in~(\ref{3.1.6})
have induced derivations $\di_{z^{\mu}}$ (\(\mu = 1,\, \dots,\, D\)).

For every linear automorphism \(A : \C^n \to \C^n\) with matrix
\(\left( A_{kl} \right)\) and inverse matrix
\(\left( {\mathop{A}\limits^{\text{\tiny $-1$}}}_{kl}  \right)\),
and a vector space $\Lin$
we define an induced automorphism
\beq\label{add3.1.7}
r \left( A \right) \, : \,
\Lin \Bbrk{z_1,\, \dots ,\, z_n}_{\, \raisebox{-2pt}{\small $L_n$}}
\, \longrightarrow \,
\Lin \Bbrk{z_1,\, \dots ,\, z_n}_{\, \raisebox{-2pt}{\small $L_n$}}
\, , \quad
\eeq
the ``linear change of variables
\(z_k \mapsto
z_k' = \Su_{l \, = \, 1}^n \, A_{kl} \, z_l\,\)'',
replacing
\(z_k \mapsto
\Su_{l \, = \, 1}^n \, {\mathop{A}\limits^{\text{\tiny $-1$}}}_{kl}
\, z_l\gvspc{-5pt}\)
(\(k = 1,\, \dots ,\, n\,\))
(note that the multiplicative system $L_n$~(\ref{3.1.4}) is invariant
under this replacement while $R_n$ is not).
There is also a natural action of the symmetric group $\Ss_n$ on
\(\Lin \Bbrk{z_1,\, \dots ,\, z_n}_{\, \raisebox{-2pt}{\small $L_n$}}\)
and
\(\Lin \Bbrk{z_1,\, \dots ,\, z_n}_{\, \raisebox{-2pt}{\small $R_n$}}\)
induced by the permutation of variables
\(\left( z_1,\, \dots ,\, z_n \right)\,\), since the multiplicative
systems $L_n$ and $R_n$ are invariant under this action.

\Vspa

Now we will introduce a homomorphism,
commuting with the derivations $\di_{z^{\mu}}$,
\beq\label{3.1.7}
\iota_{z_1,\, \dots ,\, z_n}  \, : \,
\C \Bbrk{z_1,\, \dots ,\, z_n}_{\,
\raisebox{-2pt}{\small $L_n$}}
\ \longrightarrow \
\C \Bbrk{z_1}_{\,\lscw{z}{\, 2}{1}}
\dots \, \Bbrk{z_n}_{\,\lscw{z}{\, 2}{n}}
\, \qquad
\eeq
that
will be the expansion ``in the domain
\(\left|\hspace{1pt}z_1^{\, 2}\hspace{1pt}\right| > \dots >
\left|\hspace{1pt}z_n^{\, 2}\hspace{1pt}\right|\,\)''.
We first set
\beq\label{3.1.8}
\iota_{z_1,\, \dots ,\, z_n}
\restr{\C \Bbrk{z_1,\, \dots ,\, z_n}}
\, = \,
\ID_{\,
\raisebox{1pt}{\small \(\C \Bbrk{z_1,\, \dots ,\, z_n}\)}}
\, . \qquad
\eeq
Next, consider for every \(N \in \Z\) and constants
\(\left( \lambda_1,\, \dots ,\,  \lambda_n \right) \in \C^n\,\),
the Taylor expansions in the $D$--dimensional variables
\(z_1,\, \dots ,\, z_n\,\):
\beqa\label{3.1.9}
& \hspace{-10pt}
\iota
\left(\mgvspc{16pt}\!\right.
1 \hspace{-1pt} + 2 \hspace{-3pt}
\mathop{\raisebox{0pt}{\Large $\sum$}}\limits_{
2 \, \leqslant \, k \, < \, l \, \leqslant \, n_{}^{}}
\hspace{-6pt}
\lambda_1^2 \,
\lambda_k \lambda_l \
z_1^{\, 2} \
z_k \spr z_l + \hspace{-2pt}
\mathop{\raisebox{0pt}{\Large $\sum$}}\limits_{m \, = \, 2}^{n}
\,
\left(\!\vspe\right.
2 \, \lambda_1 \lambda_m \,
z_1 \spr z_m
+
\lambda_1^2 \lambda_m^2 \,
z_1^{\, 2} z_m^{\, 2}
\left.\vspe\!\right)
\left.\mgvspc{16pt}\!\!\right)^{\hspace{-3pt} -N}
\hspace{-3pt}  =
& \hspace{-10pt} \nn & \hspace{-10pt}
= \hspace{-3pt}
\mathop{\raisebox{0pt}{\Large $\sum$}}\limits_{
k_1,\, \dots ,\, k_n \, = \, 0}^{\infty_{}^{}}
\hspace{-6pt}
\lambda_1^{k_1} \dots \hspace{1pt} \lambda_n^{k_n} \,
f^N_{k_1 \dots \, k_n}
\left( z_1,\, \dots ,\, z_n \right)
\, \ \in \, \
\C \Brk{z_1} \Bbrk{z_2,\, \dots ,\, z_n}
\, \subset \,
& \hspace{-10pt} \nn & \hspace{-10pt}
\subset \,
\C \Bbrk{z_1,\, \dots ,\, z_n}
\, , \ & \hspace{-10pt}
\eeqa
where
\(f^N_{k_1 \dots \, k_n} \left( z_1,\, \dots ,\, z_n \right)\)
are separately homogeneous polynomials in
\(z_1, \dots , z_n\) of degrees \(k_1,\, \dots ,\, k_n\,\),
respectively,
and the coefficient (formal) series in $z_1$ for every monomial
in \(z_2,\, \dots ,\, z_n\) is actually a polynomial.
(The last is true because the polynomials
$f^N_{k_1 \, \dots \, k_n} \left( z_1,\, \dots ,\, z_n \right)\mgvspc{10pt}$
are zero if \(k_1 > k_2 + \dots + k_n\,\).)
Thus we can replace
$z_1$ by
\(\frac{\raisebox{1pt}{$z_1$}}{\raisebox{-3pt}{$z_1^{\, 2}$}}\)
in the formal series~(\ref{3.1.9})
and define for every \(N \in \Z\) and
constants \(\left( \lambda_1,\, \dots ,\, \lambda_n \right)
\in \C^n\,\), \(\lambda_1 \neq 0\,\):
\beqa\label{3.1.10}
& \hspace{-10pt}
\iota_{z_1,\, \dots ,\, z_n} \hspace{-1pt}
\left(\mgvspc{17pt}\!\right. \hspace{-4pt}
\left(\mgvspc{16pt}\!\right.
\mathop{\raisebox{0pt}{\Large $\sum$}}\limits_{
k \, = \, 1}^{n_{}^{}} \, \lambda_{k} \, z_l
\left.\mgvspc{16pt}\!\!\right)^{\hspace{-3pt} 2}
\left.\mgvspc{17pt}\!\right)^{\hspace{-3pt} -N}
\hspace{-3pt} := \hspace{-1pt}
\left( \lambda_1^2 \, z_1^{\, 2} \right)^{-N} \hspace{-1pt}
\left[\mgvspc{17pt}\!\right. \hspace{-1pt}
\iota
\left(\mgvspc{16pt}\!\right.
\!\!
1 \hspace{-1pt} + 2 \hspace{-3pt}
\mathop{\raisebox{0pt}{\Large $\sum$}}\limits_{
2 \, \leqslant \, k \, < \, l \, \leqslant \, n_{}^{}}
\hspace{-6pt}
\lambda_1^{-2} \,
\lambda_k \lambda_l \,
z_1^{\, 2} \,
z_k \spr z_l \, +
& \hspace{-10pt} \nn & \hspace{-10pt} + \hspace{-1pt}
\mathop{\raisebox{0pt}{\Large $\sum$}}\limits_{m \, = \, 2}^{n}
\left(\!\vspe\right.
2 \, \lambda_1^{-1} \lambda_m \,
z_1 \spr z_m
+
\lambda_1^{-2} \lambda_m^2 \,
z_1^{\, 2} z_m^{\, 2}
\left.\vspe\!\right)
\left.\mgvspc{16pt}\!\!\right)^{\hspace{-3pt} -N}
\vrestr{18.5pt}{\,
z_1 \mapsto \frac{\lscw{z}{}{1}}{
\raisebox{-1pt}{$\lscw{z}{\, 2^{\raisebox{3pt}{\small \({}\)}}_{}}{1}$}}}
\left.\mgvspc{19pt}\hspace{1pt}\right] \in
& \hspace{-10pt} \nn & \hspace{-10pt}
\in \, \C \Brk{z_1,\, \frc{1}{z^{\, 2}_1}} \Bbrk{z_2,\, \dots ,\, z_n}
\, \subset \, \C \Bbrk{z_1}_{\,\lscw{z}{\, 2}{1}}
\dots \, \Bbrk{z_n}_{\,\lscw{z}{\, 2}{n}}
\, . \hspace{-31pt}
& \hspace{-10pt}
\eeqa
For general constants \(\left( \lambda_1,\, \dots ,\,  \lambda_n \right)
\in \C^n \,\backslash \left\{ 0 \right\}\) we set
\beq\label{3.1.11}
\iota_{z_1,\, \dots ,\, z_n} \,
\left(\mgvspc{17pt}\!\right. \hspace{-2pt}
\left(\mgvspc{16pt}\!\right.
\Su_{k \, = \, 1}^{n} \, \lambda_{k} \, z_l
\left.\mgvspc{16pt}\!\!\right)^{\hspace{-3pt} 2}
\left.\mgvspc{17pt}\!\right)^{\hspace{-3pt} -N}
\, := \,
\iota_{z_m,\, \dots ,\, z_n} \,
\left(\mgvspc{17pt}\!\right. \hspace{-2pt}
\left(\mgvspc{16pt}\!\right.
\Su_{k \, = \, m}^{n} \, \lambda_{k} \, z_l
\left.\mgvspc{16pt}\!\!\right)^{\hspace{-3pt} 2}
\left.\mgvspc{17pt}\!\right)^{\hspace{-3pt} -N}
\eeq
(\(\in \C \Bbrk{z_1}_{\,\lscw{z}{\, 2}{1}}
\dots \, \Bbrk{z_n}_{\,\lscw{z}{\, 2}{n}}\))
if \(\lambda_1 = \dots = \lambda_{m-1} = 0\,\), \(\lambda_m \neq 0\,\).
Since
the Taylor expansions~(\ref{3.1.9})
have a multiplicative property, then
\beqa\label{3.1.12}
&
\iota_{z_1,\, \dots ,\, z_n} \,
\left(\mgvspc{17pt}\!\right. \hspace{-2pt}
\left(\mgvspc{16pt}\!\right.
\mathop{\raisebox{0pt}{\Large $\sum$}}
\limits_{k \, = \, 1}^{n} \, \lambda_{k} \, z_k
\left.\mgvspc{16pt}\!\!\right)^{\hspace{-3pt} 2}
\left.\mgvspc{17pt}\!\right)^{\hspace{-3pt} -N_1}
\iota_{z_1,\, \dots ,\, z_n} \,
\left(\mgvspc{17pt}\!\right. \hspace{-2pt}
\left(\mgvspc{16pt}\!\right.
\mathop{\raisebox{0pt}{\Large $\sum$}}
\limits_{k \, = \, 1}^{n} \, \lambda_{k} \, z_k
\left.\mgvspc{16pt}\!\!\right)^{\hspace{-3pt} 2}
\left.\mgvspc{17pt}\!\right)^{\hspace{-3pt} -N_2}
\hspace{-2pt} =
& \nn & = \,
\iota_{z_1,\, \dots ,\, z_n} \,
\left(\mgvspc{17pt}\!\right. \hspace{-2pt}
\left(\mgvspc{16pt}\!\right.
\mathop{\raisebox{0pt}{\Large $\sum$}}
\limits_{k \, = \, 1}^{n} \, \lambda_{k} \, z_k
\left.\mgvspc{16pt}\!\!\right)^{\hspace{-3pt} 2}
\left.\mgvspc{17pt}\!\right)^{\hspace{-3pt} -N_1-N_2}
\,
\eeqa
for
\(N_1,\, N_2 \in \Z\,\).
Finally,
the homomorphism $\iota_{z_1,\, \dots ,\, z_n}$~(\ref{3.1.7})
is uniquely determined by Eqs.~(\ref{3.1.8})--(\ref{3.1.12}).

\begin{mremark}\label{new-rem}
The operation $\iota_{z_1,\dots,z_n}$ applied to a rational function
$R(z_1,\dots,z_n)$, regular for \(z_2=0, \dots, z_n=0$, should give the 
Taylor expansion of $R$ in \(z_2,\dots,z_n\) around \(\left(0,\dots,0\right)\). 
Its coefficients are rational functions in $z_1$. 
In the case of the left hand side of (\ref{3.1.10}) these coefficients 
are polynomials in $z_1$ and $\lfrc{1}{z^{\, 2}_1}$, 
which follows by induction in the total order of \(z_2,\dots,z_n\).
(The author thanks A.~Retakh for his interest in this work and for
asking a question answered in this Remark.)
\end{mremark}

Note that $\iota_{z_1,\, \dots ,\, z_n}$
is a \(\C \Brk{z_1, \lfrc{1}{z^{\, 2}_1}
\hspace{1pt} ; \dots ; \hspace{1pt}
z_n, \lfrc{1}{z^{\, 2}_n}}\)--linear map
and commutes with the derivations $\di_{z_k^{\mu}}$
(\(k = 1,\, \dots ,\, n\,\), \(\mu = 1,\, \dots ,\, D\,\)).
We can also define
a \(\C \Brkl{z_1, \hspace{-1pt} \lfrc{1}{z^{\, 2}_1}}
\hspace{1pt} ; \dots ;\)
\(\Brkr{z_n, \lfrc{1}{z^{\, 2}_n}}\)--linear map
\\ ${}$ \hfill
\(\iota_{z_1,\, \dots ,\, z_n} :
\Lin \Bbrk{z_1,\, \dots ,\, z_n}_{\raisebox{-2pt}{\small $L_n$}}
\hspace{1pt} \longrightarrow \
\Lin \Bbrk{z_1}_{\,\lscw{z}{\, 2}{1}}
\dots \, \Bbrk{z_n}_{\,\lscw{z}{\, 2}{n}}
\mgvspc{14pt}\mgvspc{-8pt}\),
\hfill ${}$ \\
so that
\(\iota_{z_1,\, \dots ,\, z_n}
\restr{\Lin \Bbrk{z_1,\, \dots ,\, z_n}}
\, = \,
\ID_{\,
\raisebox{1pt}{\small \(\Lin \Bbrk{z_1,\, \dots ,\, z_n}\)}}
\mgvspc{-12pt}\)
and
$\iota_{z_1,\, \dots ,\, z_n} \left( f \, u \right)$ $=$
$\iota_{z_1,\, \dots ,\, z_n} \left( f \right)$
$\iota_{z_1,\, \dots ,\, z_n} \left( u \right)$ for
\(f \in \C \Bbrk{z_1,\, \dots ,\, z_n}_{\raisebox{-2pt}{\small $L_n$}}\)
and
\(u \in \Lin \Bbrk{z_1,\, \dots ,\, z_n}_{\raisebox{-2pt}{\small $L_n$}}\,\).

\Vspa

The map
$\iota_{z_1, \dots , z_n}$
has zero kernel in
\(\Lin \Bbrk{z_1, \dots , z_n}_{\raisebox{-2pt}{\small $L_n$}}\).
Indeed, if
\(\iota_{z_1, \dots , z_n} u = 0\)
for some \(u \in
\Lin \Bbrk{z_1, \dots , z_n}_{\raisebox{-2pt}{\small $L_n$}}\)
then
\(u \hspace{-1pt} = \hspace{-1pt} f^{-1} \, v\),
where
\(f \hspace{-1pt} := \hspace{-1pt}
\mathop{\prod}\limits_{k \, = \, 1}^{N}
\left(\raisebox{14pt}{\hspace{-2pt}}\right.
\Su_{l \, = \, 1}^{n} \, \lambda_{k,\, l} \, z_l
\left.\raisebox{14pt}{\hspace{-2pt}}\right)^{\hspace{-3pt} 2}
\),
\(v \in \Lin \Bbrk{z_1,\, \dots ,\, z_n}\mgvspc{10pt}\).
But
$\iota_{z_1,\, \dots ,\, z_n} u$ $=$
$\iota_{z_1,\, \dots ,\, z_n} \left( f^{-1} \right)$
$\iota_{z_1,\, \dots ,\, z_n} v$ and
\(\iota_{z_1,\, \dots ,\, z_n} \, v \equiv v \neq 0\),
\(\iota_{z_1,\, \dots ,\, z_n} \, \left( f^{-1} \right)
\neq 0\) (since
$\iota_{z_1,\, \dots ,\, z_n} \, \left( f^{-1} \right)$\gvspc{9pt}
is the inverse of $f$),
which contradicts the fact that in
\(\Lin \Bbrk{z_1}_{\,\lscw{z}{\, 2}{1}}
\dots \, \Bbrk{z_n}_{\,\lscw{z}{\, 2}{n}}\)\gvspc{9pt}
there are no zero divisors.

\begin{mproposition}\label{pr:2.10}
In any vertex algebra $\VA$ and
\(a_1,\, \dots ,\, a_n,\, b \in \VA\) one has
\beqa\label{3.2.1}
&
Y \left( a_1,\, z_1 \right) \dots \hspace{1pt}
Y \left( a_n,\, z_n \right) \, b
\, \in \,
\iota_{z_1,\, \dots ,\, z_n} \hspace{1pt}
\left(
\VA \Bbrk{z_1,\, \dots ,\, z_n}_{\raisebox{-2pt}{\small $R_n$}}
\right)
\, \subset \,
& \nn & \subset \,
\VA \Bbrk{z_1}_{\,\lscw{z}{\, 2}{1}}
\dots \, \Bbrk{z_n}_{\,\lscw{z}{\, 2}{n}}
\,
&
\eeqa
(see Eq.~(\ref{3.1.5})).
Moreover, the inverse image
\beqa\label{new2.17.2}
&
\mathcal{Y}_n \left( a_1,\, z_1;\, \dots ;\, a_n,\, z_n;\, b \right)
:=
\iota_{z_1,\, \dots ,\, z_n}^{-1} \hspace{1pt}
\lb\!\vspe\right.
Y \left( a_1,\, z_1 \right) \dots \hspace{1pt}
Y \left( a_n,\, z_n \right) \, b
\left.\vspe\!\rb
\, \in \,
& \nn & \in \,
\VA \Bbrk{z_1,\, \dots ,\, z_n}_{\raisebox{-2pt}{\small $R_n$}}
&
\eeqa
is $\Z_2$--symmetric in the sense that if
\(a_1,\, \dots a_n\) have fixed parities
\(p_1,\, \dots p_n\) (resp.)
then for any permutation \(\sigma \in \Ss_n\,\):
\beq\label{new2.18.3}
\hspace{2pt}
\mathcal{Y}_n \left( a_{\sigma \left( 1 \right)},\hspace{1pt}
z_{\sigma \left( 1 \right)};\hspace{1pt} \dots ;\hspace{1pt}
a_{\sigma \left( n \right)},\hspace{1pt}
z_{\sigma \left( n \right)};\hspace{1pt} b \right)
=
\left( -1 \right)^{\varepsilon \left( \sigma \right)}
\mathcal{Y}_n \left( a_1,\hspace{1pt} z_1
;\hspace{1pt} \dots ;\hspace{1pt} a_n,\hspace{1pt} z_n;\hspace{1pt} b \right)
,
\hspace{-20pt}
\eeq
where
\(\varepsilon \left( \sigma \right) :=
\Su_{\left( ij \right)} \, p_{a_i} p_{a_j} \, \text{\textit{mod}} \, 2\)
(the sum is taken over all transpositions
$\left( ij \right)$ in some representation
\(\sigma = \prod \, \left( ij \right)\gvspc{10pt}\)).
\end{mproposition}

\begin{proof}
Locality (Definition~\ref{def:2.1}~(\textit{b})) implies that
\\ ${}$ \hfill
\(\rho_n^N \, Y \left( a_1,\, z_1 \right) \dots \hspace{1pt}
Y \left( a_n,\, z_n \right) \, b \in
\VA \Bbrk{z_1,\, \dots ,\, z_n}\)
\hspace{0.3cm} for \hspace{0.3cm}
\(N \mgrt 0\mgvspc{14pt}\mgvspc{-8pt}\),
\hfill ${}$ \\
where
\(\rho_n :=
\left( \prod_{\, k} \, \lscw{z}{\, 2}{k} \right) \,
\left( \prod_{\, l \, < \, m}
\, \lscw{z}{\, 2}{lm} \right)\,\).
On the other hand,
\\ ${}$ \hfill
\(
Y \left( a_1,\, z_1 \right) \dots \hspace{1pt}
Y \left( a_n,\, z_n \right) \hspace{1pt} b
\hspace{-1pt} \in \hspace{-1pt}
\VA \Bbrk{z_1}_{\,\lscw{z}{\, 2}{1}}
\dots \, \Bbrk{z_n}_{\,\lscw{z}{\, 2}{n}}
\mgvspc{14pt}\mgvspc{-8pt}\),
\hfill ${}$ \\
because of Definition~\ref{def:2.1}~(\textit{a}).
Then (\ref{3.2.1}) follows from the fact that
\(\iota_{z_1,\, \dots ,\, z_n} \, \rho_n^{-N}\mgvspc{11pt}\)
is an inverse element
of $\rho_n^N$ in
\(\C
\Bbrk{z_1}_{\,\lscw{z}{\, 2}{1}}
\dots \, \Bbrk{z_n}_{\,\lscw{z}{\, 2}{n}}\mgvspc{10pt}\).
To prove Eq.~(\ref{new2.18.3}) we
note that for \(N \mgrt 0\,\):
\(\rho_n^N \,
\mathcal{Y}_n \left( a_1,\, z_1;\, \dots ;\, a_n,\, z_n;\, b \right)\)
is $\Z_2$--symmetric, while $\rho_n^N$ is symmetric
in \(z_1 ,\,  \dots ,\, z_n\,\).\qed
\end{proof}

\begin{mtheorem}\label{th:2.11}
In any vertex algebra $\VA$ and
\(a,\,  b,\, c \in \VA\) it follows that
\beq\label{3.3.1}
Y \left(
Y \left( a, z \hspace{-1pt} - \hspace{-1pt} w \right)
b,\hspace{1pt} w \right) \hspace{1pt} c
=
\iota_{w,\hspace{1pt} z - w}
\left( \hspace{1pt}
r_{z,\hspace{1pt} w}^{w,\hspace{1pt} z-w}
\left(
\iota_{z,\hspace{1pt} w}^{\, - \hspace{0.5pt} 1}
\left(\!\vspe\right.
Y \left( a,\hspace{1pt} z \right)
Y \left( b,\hspace{1pt} w \right) \hspace{1pt} c
\left.\vspe\!\right)
\right)
\right)
,
\eeq
where
\(Y \left( Y \left( a, z \hspace{-1pt} - \hspace{-1pt} w \right)
b, w \right) c\)
is viewed as a series belonging to
\(\VA \Bbrk{w}_{\lscw{w}{\hspace{1pt} 2}{}}
\Bbrk{z \hspace{-1pt} - \hspace{-1pt} w}_{\lscw{\left(
z \hspace{-1pt} - \hspace{-1pt} w \right)}{2}{}}\),
$\iota_{z,\, w}^{\, - \hspace{0.5pt} 1}$
is the inverse of $\iota_{z,\, w}$ on its image
and
\(r_{z,\, w}^{w,\, z-w} :
\VA \Bbrk{z,\, w}_{\raisebox{-2pt}{\small $L_n$}} \to
\VA \Bbrk{w,\, z \hspace{-1pt} - \hspace{-1pt} w}_{
\raisebox{-2pt}{\small $L_n$}}\mgvspc{10pt}\)
is the map of type~(\ref{add3.1.7})
induced by the change of variables
\(\left( z,\, w \right) \mapsto \left( w,\, z-w \right)\mgvspc{10pt}\).
\end{mtheorem}

\begin{proof}
The theorem follows from Theorem~\ref{th:2.1} (Eq.~(\ref{2.4})) and
Eq.~(\ref{n2.8}).
More precisely, we obtain the following equalities in
\(\VA \Bbrk{z,\, w} \cong \VA \Bbrk{w,\, z-w}\) for \(N \mgrt 0\hspace{1pt}\):
\\ ${}$ \hfill \(
\left(\raisebox{9pt}{\hspace{-2pt}}\right. z^{\, 2} \, w^{\, 2}
\left( z \hspace{-1pt} - \hspace{-1pt} w \right)^{\, 2}
\left.\raisebox{9pt}{\hspace{-2pt}}\right)^N
Y \left( a,\hspace{1pt} z \right)
Y \left( b,\hspace{1pt} w \right) \hspace{1pt} c
=
\mgvspc{18pt}\mgvspc{-6pt}\) \hfill ${}$ \\ ${}$ \hfill \(
=
\left(\raisebox{9pt}{\hspace{-2pt}}\right. z^{\, 2} \, w^{\, 2}
\left.\raisebox{9pt}{\hspace{-2pt}}\right)^N
\hspace{-4pt}
\mathop{\raisebox{0pt}{\large $\sum$}}\limits_{\raisebox{-1pt}{\scriptsize \(
n, m = 0\)}}^{\raisebox{1pt}{\scriptsize \(\infty\)}}
\hspace{2pt}
\mathop{\raisebox{0pt}{\large $\sum$}}\limits_{\sigma \, = \, 1}^{\har_m}
\hspace{2pt}
Y \left( a, w \right)_{\left\{ n-N, m, \sigma \right\}}
\hspace{-2pt}
Y \left( b, w \right) c \hspace{2pt}
\left(\raisebox{9pt}{\hspace{-2pt}}\right.
\left( z \hspace{-1pt} - \hspace{-1pt} w \right)^{2}
\left.\raisebox{9pt}{\hspace{-2pt}}\right)^{n}
h^{\left( m \right)}_{\sigma} \left( z \hspace{-1pt} - \hspace{-1pt} w \right)
=
\) \hfill ${}$ \\ ${}$ \hfill \(
=
\left(\raisebox{9pt}{\hspace{-2pt}}\right. z^{\, 2} \, w^{\, 2}
\left.\raisebox{9pt}{\hspace{-2pt}}\right)^N
\hspace{-1pt}
\mathop{\raisebox{0pt}{\large $\sum$}}\limits_{\raisebox{-1pt}{\scriptsize \(
n, m = 0\)}}^{\raisebox{1pt}{\scriptsize \(\infty\)}}
\hspace{2pt}
\mathop{\raisebox{0pt}{\large $\sum$}}\limits_{\sigma \, = \, 1}^{\har_m}
\hspace{3pt}
Y \left( a_{\, \left\{ n-N,\, m,\, \sigma \right\}} b,\, w \right)\, c \hspace{2pt}
\left(\raisebox{9pt}{\hspace{-2pt}}\right.
\left( z \hspace{-1pt} - \hspace{-1pt} w \right)^{2}
\left.\raisebox{9pt}{\hspace{-2pt}}\right)^{n}
h^{\left( m \right)}_{\sigma} \left( z \hspace{-1pt} - \hspace{-1pt} w \right)
=
\mgvspc{-8pt}\) \hfill ${}$ \\ ${}$ \hfill \(
=
\left(\raisebox{9pt}{\hspace{-2pt}}\right. z_1^{\, 2} \, z_2^{\, 2}
\left( z_1 \hspace{-1pt} + \hspace{-1pt} z_2 \right)^{\, 2}
\left.\raisebox{9pt}{\hspace{-2pt}}\right)^N \,
Y \left(
Y \left( a, z_1 \right)
b,\hspace{1pt} z_2 \right) \hspace{1pt} c \,
\vrestr{9pt}{z_1 = z \hspace{-1pt} - \hspace{-1pt} w,\,
z_2 = w}
\, ;
\mgvspc{-13pt}\) \hfill ${}$ \\
then the prefactors can be cancelled after applying to both sides
the corresponding $\iota^{-1}$ operators.\qed
\end{proof}

\msection{Free Field Examples. Lie Superalgebras of Formal Distributions}{sec:5}

Let us consider a central extension of the free commutative Lie superalgebra
\beq\label{eq5.1}
\mathit{Span}_{\C} \, \left\{ u_{\left\{ n,\, m,\, \sigma \right\}}^{\alpha} :
\alpha \in A,\, n \in \Z,\, m \in N \cup \left\{ 0 \right\},\,
\sigma = 1,\, \dots ,\, \har_m \right\}
\, , \qquad
\eeq
where $A$ is some index set and all generators
$u_{\left\{ n,\, m,\, \sigma \right\}}^{\alpha}$ have parities
$p_{\alpha}$ which do not depend on $n$, $m$ and $\sigma$.
The commutation relations are
presented by the following generating functions:
\beqa\label{eq5.2}
&
\left[ u^{\alpha} \left( z \right),\, u^{\beta} \left( w \right) \right]
\, = \,
\left(\raisebox{15pt}{\hspace{-3pt}}\left.
\iota_{z,\, w} \,
\frac{\raisebox{2pt}{\(Q_{\alpha\beta} \left( z-w \right)\)}}{
\raisebox{-6pt}{
\(\left[\raisebox{9pt}{\hspace{-3pt}}\right. \left( z-w \right)^2
\left.\raisebox{9pt}{\hspace{-3pt}}\right]^{\mu_{\alpha\beta}}\)}}
\, - \,
\iota_{w,\, z} \,
\frac{\raisebox{2pt}{\(Q_{\alpha\beta} \left( z-w \right)\)}}{
\raisebox{-6pt}{
\(\left[\raisebox{9pt}{\hspace{-3pt}}\right. \left( z-w \right)^2
\left.\raisebox{9pt}{\hspace{-3pt}}\right]^{\mu_{\alpha\beta}}\)}}
\right.\raisebox{15pt}{\hspace{-3pt}}\right) K
\, ,
&
\qquad
\\ \label{eq5.3n}
&
u^{\alpha} \left( z \right) \, = \,
\Su_{n \, \in \, \Z} \,
\Su_{m \, = \, 0}^{\infty} \,
\Su_{\sigma \, = \, 1}^{\har_m} \,
u_{\,\left\{ n,\, m,\, \sigma \right\}}^{\alpha} \,
\left( z^{\, 2} \right)^n \, h^{\left( m \right)}_{\sigma}
\left( z \right)
&
\eeqa
for \(\alpha,\, \beta \in A\),
where \(\mu_{\alpha\beta}\) are positive integers,
$K$ is the central element
and $Q_{\alpha\beta} \left( z \right)$
are polynomials such that
\(Q_{\alpha\beta} \left( -z \right) =
\left( -1 \right)^{p_{\alpha}p_{\beta}} Q_{\beta\alpha} \left( z \right)\).
Without lost of generality we can suppose that the leading (i.~e. harmonic)
term in the harmonic decomposition (\ref{harmdec}) for every
\(Q_{\alpha\beta} \left( z \right) =: p \left( z \right)\)
is nonzero if \(Q_{\alpha\beta} \left( z \right) \neq 0\).
Then the right hand side of Eq.~(\ref{eq5.2}) uniquely determines
the polynomials $Q_{\alpha\beta} \left( z \right)\,$.
As a consequence of Eq.~(\ref{eq5.2}) (and (\ref{3.1.12})) we have
\beq\label{eq5.3}
\left[\raisebox{9pt}{\hspace{-3pt}}\right. \left( z-w \right)^2
\left.\raisebox{9pt}{\hspace{-3pt}}\right]^{\mu_{\alpha\beta}}
\left[ u^{\alpha} \left( z \right),\, u^{\beta} \left( w \right) \right]
\, = \, 0
\, . \qquad
\eeq
The Lie super algebra $\mathfrak{H}$ so obtained has a decomposition
\beqa\label{eq5.4}
\mathfrak{H} = && \hspace{-3pt}
\mathfrak{H}_{\left\{ + \right\}}
\, \oplus \, \C \, K \, \oplus \,
\mathfrak{H}_{\left\{ - \right\}}
\, , \qquad
\\ \label{eq5.6n}
\mathfrak{H}_{\left\{ + \right\}} = && \hspace{-3pt}
\mathit{Span}_{\C} \, \left\{ u_{\left\{ n,\, m,\, \sigma \right\}}^{\alpha} :
\alpha \in A,\, n \geqslant 0,\, m \in N \cup \left\{ 0 \right\},\,
\sigma = 1,\, \dots ,\, \har_m \right\}
, \qquad
\\ \label{eq5.7n}
\mathfrak{H}_{\left\{ - \right\}} = && \hspace{-3pt}
\mathit{Span}_{\C} \, \left\{ u_{\left\{ n,\, m,\, \sigma \right\}}^{\alpha} :
\alpha \in A,\, n < 0,\, m \in N \cup \left\{ 0 \right\},\,
\sigma = 1,\, \dots ,\, \har_m \right\}
. \qquad
\eeqa
Let $\mathcal{F}$ be the Fock representation space of $\mathfrak{H}$
determined by \(\mathfrak{H}_{\left\{ - \right\}} \, \rvac = 0\)
and \(K\vrestr{9pt}{\mathcal{F}} = k \, \ID\),
where \(\rvac \in \mathcal{F}\) is the Fock vacuum.
Then the formal series~(\ref{eq5.3n}) is represented on $\mathcal{F}$
as a field for every \(\alpha \in A\).
In fact we will prove more a general statement:

\begin{mproposition}\label{pr:5.1}
Let a system of formal series~(\ref{eq5.3n}) be given with coefficients
$u^{\alpha}_{\left\{ n,\, m,\, \sigma \right\}}$
generating a Lie superalgebra $\mathfrak{L}$ and such that
the equation~(\ref{eq5.3}) is satisfied for some positive integers
$\mu_{\alpha\beta}$ and all \(\alpha,\, \beta \in A\)
($u^{\alpha}_{\left\{ n,\, m,\, \sigma \right\}}$ are supposed to
have a parity independent on $n$, $m$ and $\sigma$).
Let $\mathfrak{L}_{\left\{ + \right\}}$ and
$\mathfrak{L}_{\left\{ - \right\}}$ be the subalgebras of
$\mathfrak{L}$ which are generated by the right hand sides of
Eqs.~(\ref{eq5.6n}) and~(\ref{eq5.7n}), respectively.
Let $\mathcal{U}$ be the representation of $\mathfrak{L}$ obtained
by factorization of the universal enveloping algebra
$U \left( \mathfrak{L} \right)$ of $\mathfrak{L}$ by the left ideal
$U \left( \mathfrak{L} \right) \mathfrak{L}_{\left\{ - \right\}}$
where \(\mathfrak{L}\) is assumed to act by left multiplication.
Then the formal series~(\ref{eq5.3n}) is represented on $\mathcal{U}$ as
a field for every \(\alpha \in A\).
\end{mproposition}

\begin{proof}
Let us denote the class of \(a \in U \left( \mathfrak{L} \right)\)
in $\mathcal{U}$ by \(\left[ a \right]\).
Thus we have to prove that
\beq\label{eq5.8}
u^{\alpha} \left( z \right)
\left[
u^{\alpha_1}_{\left\{ n_1,\, m_1,\, \sigma_1 \right\}}
\dots
u^{\alpha_k}_{\left\{ n_k,\, m_k,\, \sigma_k \right\}} \ID
\right] \, \in \, \mathcal{U} \Bbrk{z}_{\,\lscw{z}{\, 2}{}}
\, \qquad
\eeq
for all \(k = 0,\, 1,\, \dots\) and all values of the indices.
We will make the proof by induction in $k$:
for \(k = 0\) Eq.~(\ref{eq5.8}) follows from the factorization by
\(\mathfrak{L}_{\left\{ - \right\}}\).
Suppose that (\ref{eq5.8}) is satisfied for \(k \geqslant 0\)
and all values of the indices.
Let us set
\\ ${}$ \hfill
\(v_k :=
\left[
u^{\alpha_1}_{\left\{ n_1,\, m_1,\, \sigma_1 \right\}}
\dots
u^{\alpha_k}_{\left\{ n_k,\, m_k,\, \sigma_k \right\}} \ID
\right]
\in \mathcal{U} .
\mgvspc{15pt}\mgvspc{-11pt}\) \hfill ${}$ \\
Then we have to prove that
\(
\left( z^{\, 2} \right)^N
u^{\alpha} \left( z \right)
u^{\beta}_{\left\{ n,\, m,\, \sigma \right\}} v_k
\in \mathcal{U} \Bbrk{z}
\)
for \(N \mgrt 0\).
By the inductive assumption
\(u^{\beta}_{\left\{ n,\, m,\, \sigma \right\}} v_k
= P_{\left\{ n+M,\, m,\, \sigma \right\}} \left( \di_{w} \right)
\left( w^{\, 2} \right)^M u^{\beta} \left( w \right) v_k
\!\left.\vspe\right|_{\,
\raisebox{1pt}{\small \(w = 0\)}}\)
for \(M \mgrt 0\)
(recall the definition (\ref{ad1.11}) of
$P_{\left\{ n,\, m,\, \sigma \right\}} \left( z \right)$).
Then by Lemma~\ref{lm:2.3}, for every \(L \in \N\) there exist
\(N \in \N\) and a polynomial
\(Q \left( z,w \right) \in \C \Brk{z,w}\) such that
\\ ${}$ \hfill \(
\left( z^{\, 2} \right)^N
u^{\alpha} \left( z \right)
u^{\beta}_{\left\{ n,\, m,\, \sigma \right\}} v_k
=
\mgvspc{15pt}\) \hfill ${}$ \\ ${}$ \hfill \(
=
\left[\raisebox{9pt}{\hspace{-3pt}}\right. \left( z-w \right)^2
\left.\raisebox{9pt}{\hspace{-3pt}}\right]^N
u^{\alpha} \left( z \right)
P_{\left\{ n+M,\, m,\, \sigma \right\}} \left( \di_{w} \right)
\left( w^{\, 2} \right)^M u^{\beta} \left( w \right) v_k
\!\left.\vspe\right|_{\,
\raisebox{1pt}{\small \(w = 0\)}}
=
\mgvspc{15pt}\) \hfill ${}$ \\ ${}$ \hfill \(
=
Q \left( z-w,\, \di_{w} \right)
\left[\raisebox{9pt}{\hspace{-3pt}}\right. \left( z-w \right)^2
\left.\raisebox{9pt}{\hspace{-3pt}}\right]^L
\left( w^{\, 2} \right)^M
u^{\alpha} \left( z \right) u^{\beta} \left( w \right) v_k
\!\left.\vspe\right|_{\,
\raisebox{1pt}{\small \(w = 0\)}}
\mgvspc{15pt}\mgvspc{-11pt}\) \hfill ${}$ \\
for \(M \mgrt 0\).
On the other hand, it follows from (\ref{eq5.3}) that
\(\left[\raisebox{9pt}{\hspace{-3pt}}\right. \left( z-w \right)^2
\left.\raisebox{9pt}{\hspace{-3pt}}\right]^L
u^{\alpha} \left( z \right) u^{\beta} \left( w \right) v_k\)
$\in$
\(\mathcal{U} \Bbrk{z, w}\) for \(L \mgrt 0\)
(as in the case of vertex algebras, after Def.~\ref{def:2.3}).
Consequently,
\(\left( z^{\, 2} \right)^N
u^{\alpha} \left( z \right)
u^{\beta}_{\left\{ n,\, m,\, \sigma \right\}} v_k \in \mathcal{U} \Bbrk{z}\)
for \(N \mgrt 0\).\qed
\end{proof}

To obtain a vertex algebra structure we need additional assumptions.

\begin{mproposition}\label{pr:5.2}
In the assumptions of Proposition~\ref{pr:5.1}
let us suppose that there exist even derivations
\(\mathrm{T}_1,\, \dots,\, \mathrm{T}_D\) of \(\mathfrak{L}\) such that
\beq\label{eq5.9}
\mathrm{T}_{\mu} \left( u^{\alpha} \left( z \right) \right)
\, = \,
\di_{z^{\mu}} \left( u^{\alpha} \left( z \right) \right)
\, \qquad
\eeq
for \(\mu = 1,\, \dots ,\, D\) and \(\alpha \in A\).
Then \(\mathfrak{L}_{\left\{ - \right\}}\) is $\mathrm{T}$--invariant
and hence $\mathrm{T}_{\mu}$ are represented on $\mathcal{U}\,$.
Suppose also that
\(
\VA =
\raisebox{3pt}{\(\mathcal{U}\)}
\hspace{-3pt}\left/\raisebox{8pt}{${}$}\right.\hspace{-4pt}
\raisebox{-4pt}{\(\mathcal{J}\)}
\)
is a quotient
representation of \(\mathfrak{L}\) by a $\mathrm{T}$--invariant
subrepresentation $\mathcal{J}$ such that the class
\(\varvac := \left[ \ID \right] \in \VA\) of
\(\ID \in U \left( \mathfrak{L} \right)\) is nonzero.
Then if $\mathrm{T}_{\mu}$ are represented on $\VA$ by
\(\Trn_{\mu} \in \mathit{End} \, \VA\) it follows that
the representation of the formal series~(\ref{eq5.3n}) on $\VA$
satisfy all the assumptions of the existence
Theorem~\ref{th:2.8} and hence $\VA$ has the structure of a
vertex algebra.
\end{mproposition}

\begin{proof}
Using Eqs.~(\ref{actions}) one can prove that Eq.~(\ref{eq5.9})
is equivalent to
\beqa\label{eq5.10}
\mathrm{T}_{\mu} \left( u^{\alpha}_{\,\left\{ n,\, m,\, \sigma \right\}} \right)
\, := \,
&& \hspace{-2pt}
\frac{
n + m +
\raisebox{-0.5pt}{{\large \(\frac{\raisebox{0pt}{{\footnotesize \(D\)}}}{\raisebox{-1pt}{{\footnotesize \(2\)}}}\)}}
\,
}{\raisebox{-3.5pt}{\(
m +
\raisebox{-0.5pt}{{\large \(\frac{\raisebox{0pt}{{\footnotesize \(D\)}}}{\raisebox{-1pt}{{\footnotesize \(2\)}}}\)}}
\,
\)}}
\ \,
\Su_{\sigma_1 \, = \, 1}^{\har_{m+1}} \,
A_{\mu\,\sigma_1\,\sigma}^{\left( m+1 \right)} \
u^{\alpha}_{\,\left\{ n,\, m+1,\, \sigma_1 \right\}}
+
\nn && \hspace{-2pt}
+
\Su_{\sigma_1 \, = \, 1}^{\har_{m-1}} \,
2 \left( n+1 \right) \,
B_{\mu\,\sigma_1\,\sigma}^{\left( m-1 \right)} \
u^{\alpha}_{\,\left\{ n+1,\, m-1,\, \sigma_1 \right\}}
\, .
\eeqa
Therefore, $\mathfrak{L}_{\left\{ - \right\}}$ is $\mathrm{T}$--invariant.
By Proposition~\ref{pr:5.1} $u^{\alpha} \left( z \right)$
acts as a field on $\VA$ for every \(\alpha \in A\).
The verifications of the other assumptions of
Theorem~\ref{th:2.8} are straightforward.\qed
\end{proof}

\begin{mcorollary}\label{cr:5.3}
The Fock space $\mathcal{F}$ defined above has the structure of a vertex
algebra which is generated by the fields~(\ref{eq5.3n}) satisfying
the relations~(\ref{eq5.2}).
\end{mcorollary}

\begin{proof}
Eq.~(\ref{eq5.10}) and \(\mathrm{T}_{\mu} \left( K \right) = 0\)
define an even derivation of the algebra
$\mathfrak{H}$~(\ref{eq5.4}) since the relations~(\ref{eq5.2}) are
$\di$--invariant.
To apply Propositions~\ref{pr:5.1} and \ref{pr:5.2} we extend
the system of formal series~(\ref{eq5.3n}) with the constant series
\(K \left( z \right) = K\).
Then the role of $\mathfrak{L}_{\left\{ - \right\}}$ is played by
$\mathfrak{H}_{\left\{ - \right\}}$ and $\mathcal{F}$ is
obtained by additional factorization of $\mathcal{U}$ by the subrepresentation
generated by $K - k\ID$ which is $\mathrm{T}$--invariant.
Finally, $\mathcal{F}$ is isomorphic to the symmetric superalgebra generated
by $\mathfrak{H}_{\left\{ + \right\}}$~(\ref{eq5.6n}) so that the class
$\varvac$ is nonzero.\qed
\end{proof}

The vertex algebra obtained in Corollary~\ref{cr:5.3} is called
a \textbf{free field} vertex algebra.
A Lie superalgebra $\mathfrak{L}$ and a system of series~(\ref{eq5.3n})
satisfying the assumptions of
Proposition~\ref{pr:5.1} and possessing a system of even derivations
\(\mathrm{T}_1,\, \dots ,\, \mathrm{T}_D\,\), such that Eq.~(\ref{eq5.9})
holds is called a \textbf{Lie superalgebra of formal distributions}.

\msection{Categorical Properties of Vertex Algebras. Representations}{sec:6}

We begin with some basic categorical notions for vertex algebras
in higher dimensions which are straightforward generalizations of
the corresponding one from the chiral CFT \cite{Kac}.
A \textbf{morphism} $f$ of vertex algebras $\VA$ and $\VA'$ is
called a parity preserving linear map \(f : \VA \to \VA'\)
such that
\beqa\label{eq6.1}
f \left( a_{\,\left\{ n,\, m,\, \sigma \right\}}b \right)
\, = && \hspace{-2pt}
f \left( a \right)_{\,\left\{ n,\, m,\, \sigma \right\}}
f \left( b \right)
\, , \quad
\\ \label{eq6.2}
f \left( \Trn_{\mu} \left( a \right) \right) \, = && \hspace{-2pt}
\Trn'_{\mu} \left( f \left( a \right) \right)
\, , \quad
\\ \label{eq6.3}
f \left(\raisebox{9pt}{\hspace{-2pt}}\right.
\varvac \left.\raisebox{9pt}{\hspace{-2pt}}\right)
\, = && \hspace{-2pt} f \left(\raisebox{9pt}{\hspace{-2pt}}\right.
\varvac' \left.\raisebox{9pt}{\hspace{-2pt}}\right)
\, \quad
\eeqa
for all \(a,\, b \in \VA\,\),
\(n \in \Z\,\),
\(m = 0,\, 1,\, \dots\,\),
\(\sigma = 1,\, \dots ,\, \har_m\)
and \(\mu = 1,\, \dots ,\, D\,\),
where $\Trn_{\mu}\,$, $\varvac$ and $\Trn'_{\mu}\,$, $\varvac'\,$, are
the translation endomorphisms and the vacuum, correspondingly in
$\VA$ and $\VA'$.
An isomorphism of vertex algebras is a morphism which is an isomorphism
as a linear map.
An injective or surjecvtive morphism $f$ is such that the map
$f$ is injective or surjective as a linear map, respectively.
The image $g \left( \mathcal{U} \right)$ and the kernel $\mathit{Ker} \, g$
of a morphism \(g : \mathcal{U} \to \VA\) are called
a \textbf{vertex subalgebra} and \textbf{ideal} of $\VA\,$, respectively.
Note that the image $g \left( \mathcal{U} \right)$ is itself
a vertex algebra.
If \(f : \VA \to \VA'\) is a surjective morphism and $\mathcal{J}$ is its
kernel then the quotient space
\(\text{\raisebox{3.5pt}{\(\VA\)}
\hspace{-4pt}\raisebox{0pt}{$\left/\raisebox{9pt}{}\right.$}\hspace{-6pt}
\raisebox{-4pt}{\(\mathcal{J}\)}}\) possesses the structure of a
vertex algebra isomorphic to $\VA'$.
It is called a \textbf{quotient} vertex algebra.

\begin{mproposition}\label{pr:6.1}
Let $\VA$ be a vertex algebra.
\begin{plist}
\item
A super-subspace $\mathcal{U}$ of $\VA$ has the structure of a
vertex subalgebra of $\VA$ under the inclusion
\(\mathcal{U} \hookrightarrow \VA\) iff
\(\varvac \in \mathcal{U}\) and
\(a_{\,\left\{ n,\, m,\, \sigma \right\}}b \in \mathcal{U}\)
for all \(a,\, b \in \mathcal{U}\) and
\(n \in \Z\), \(m = 0,1, \dots\), \(\sigma = 1, \dots \har_m\).
\item
A super-subspace $\mathcal{J}$ of $\VA$ is an ideal iff
$\mathcal{J}$ is $\Trn_{\mu}$--invariant (\(\mu = 1, \dots , D\)),
\(\varvac \notin \mathcal{J}\) and
\(a_{\,\left\{ n,\, m,\, \sigma \right\}}b \in \mathcal{J}\)
for all \(a \in \VA\), \(b \in \mathcal{J}\) and
\(n \in \Z\), \(m = 0,1, \dots\), \(\sigma = 1, \dots \har_m\).
\end{plist}
\end{mproposition}

\begin{proof}
To prove the statement~(a)
first observe that $\mathcal{U}$ is
$\Trn_{\mu}$--invariant (\(\mu = 1, \dots , D\)), since
for every \(a \in \mathcal{U}\):
\(a = Y \left( a,\, z \right) \varvac \vrestr{9pt}{z = 0}\)
and then \(\Trn_{\mu} \, a =
\di_{z^{\mu}} \left(\raisebox{9pt}{\hspace{-2pt}}\right.
Y \left( a,\, z \right) \varvac
\left.\raisebox{9pt}{\hspace{-2pt}}\right) \vrestr{9pt}{z = 0}\).
Then~(a) follows directly by the above definitions.
For the proof of part (b), as in the chiral CFT (\cite{Kac}),
we need to show that
\(a_{\,\left\{ n,\, m,\, \sigma \right\}}b \in \mathcal{J}\)
for all \(a \in \mathcal{J}\), \(b \in \VA\),
\(n \in \Z\), \(m = 0,1, \dots\), \(\sigma = 1, \dots \har_m\,\).
However the last property is a consequence of the \textit{quasisymmetry}
relation:
\beq\label{eq6qs}
Y \left( a,\, z \right) \, b \, = \,
\left( -1 \right)^{\, p_a p_b}
\, e^{\, z \spr \Trn} \, Y \left( b,\, -z \right) \, a
\,
\eeq
(here \(z \spr \Trn = \mathop{\sum}\limits_{\mu \, = \, 1}^D
a^{\mu} \, \Trn_{\mu}\)) for all \(a,\, b \in \VA\)
(the right hand side of Eq.~(\ref{eq6qs}) is understood
as an action of the series
\(e^{\, z \spr \Trn} \in \left( \mathit{End} \, \VA \right) \Bbrk{z}\)
on a series belonging to
$\VA \Bbrk{z}_{\,\lscw{z}{\, 2}{}}$).

Here is a sketch of the proof of Eq.~(\ref{eq6qs}):
using the series $\mathcal{Y}_n$ introduced in
Proposition~\ref{pr:2.10},
we first derive that
\(\mathcal{Y}_2 \left(\raisebox{9pt}{\hspace{-2pt}}\right.
a,\, z,\, b,\, w;\, \varvac \left.\raisebox{9pt}{\hspace{-2pt}}\right)
\in \VA \Bbrk{z,\, w}_{\,\lscw{\left( z-w \right)}{\, 2}{}}\)
and
\beqa\label{eq6qs1}
&
\mathcal{Y}_2 \left(\raisebox{9pt}{\hspace{-2pt}}\right.
a,\, z;\, b,\, w;\, \varvac \left.\raisebox{9pt}{\hspace{-2pt}}\right)
\vrestr{10pt}{w \, = \, 0} \, = \,
\mathcal{Y}_1 \left(\raisebox{9pt}{\hspace{-2pt}}\right.
a,\, z;\, b \left.\raisebox{9pt}{\hspace{-2pt}}\right)
\, = \, Y \left( a,\, z \right) \, b
,
& \\ \label{eq6qs2} &
e^{\, w \spr \Trn} \,
\mathcal{Y}_2 \left(\raisebox{9pt}{\hspace{-2pt}}\right.
a,\, z_1;\, b,\, z_2;\, \varvac \left.\raisebox{9pt}{\hspace{-2pt}}\right)
\, = \,
\mathcal{Y}_2 \left(\raisebox{9pt}{\hspace{-2pt}}\right.
a,\, z_1+w;\, b,\, z_2+w;\, \varvac \left.\raisebox{9pt}{\hspace{-2pt}}\right)
& \mgvspc{10pt}
\eeqa
(using the argument of the proof of Proposition~\ref{pr:2.5}
\((a) \Rightarrow (b)\));
we then apply to the left hand side of Eq.~(\ref{eq6qs2})
the $\Z_2$--symmetry~(\ref{new2.17.2}) and set
\(z_1 := 0\), \(w = -z_2 := z\).\qed
\end{proof}

Let $\VA$ be a vertex algebra over $\C^D$ and
let \(A : \C^{D'} \to \C^D\) be a linear orthogonal map
(\(\left( A \, x \right) \spr \left( A \, x \right) = \, x \spr x\))
with a matrix $\left( A_{\nu}^{\mu} \right)$:  \(A \, e_{\nu}' =
\mathop{\sum}\limits_{\nu = 1}^D \, A_{\nu}^{\mu} \, e_{\mu}\)
in the standard bases
\(\left\{ e_{\mu} \right\}_{\mu \, = \, 1}^D\) and
\(\left\{ e_{\nu}' \right\}_{\nu \, = \, 1}^{D'}\)
(\(D' \leqslant D\)).
Then the formal series \(Y' \left( a,\, x \right) \, := \,
Y \left( a,\, z \right) \vrestr{9pt}{z \, = \, A \, x}\) for
\(x = \left(\raisebox{9pt}{\hspace{-2pt}}\right.
x^1,\, \dots ,\, x^{D'} \left.\raisebox{9pt}{\hspace{-2pt}}\right)\)
and \(\left( A \, x \right)^{\mu} =
\mathop{\sum}\limits_{\nu = 1}^{D'} \, A_{\nu}^{\mu} \, x^{\nu}\,\),
are correctly defined for every \(a \in \VA\) as series belonging to
\(\left( \mathit{End} \, \VA \right) \Bbrk{x,\, \frc{1}{x^{\, 2}}}\,\).
They generate, combined with the maps
\(\Trn_{\nu}' := \mathop{\sum}\limits_{\mu = 1}^D \, A_{\nu}^{\mu} \,
\Trn_{\mu}\) for \(\nu = 1,\, \dots ,\, D'\,\),
a structure of vertex algebra on $\VA$ over $\C^{D'}$
with the same $\Z_2$--grading and vacuum \(\varvac \in \VA\).
We denote this vertex algebra by $A^* \VA$ and call it
a \textbf{restriction of} $\VA$ over $\C^{D'}$.

Let
$\VA$ and $\VA'$ be vertex algebras over $\C^D$
and let the corresponding state--field correspondence,
translation operators and vacua be:
$Y \left( a,\, z \right)$,
$\Trn_{\mu}$, $\varvac$ (in $\VA$) and
$Y' \left( b,\, x \right)$,
$\Trn_{\nu}'$, $\varvac'$
(in $\VA'$).
Then for every \(a \in \VA\) and \(b \in \VA'\) the formal series
\(Y \left( a,\, z_1 \right) \hspace{1pt} \otimes \hspace{1pt}
Y' \left( b,\, z_2 \right)\)
is a field on the superspace $\VA \otimes \VA'$ and
consequently one can define the field
\(Y'' \left( a \otimes b,\, z \right) :=
Y \left( a,\, z_1 \right) \hspace{1pt} \otimes \hspace{1pt}
Y' \left( b,\, z_2 \right) \vrestr{9pt}{z_1 = z_2 = z}\).
The fields $Y'' \left( a \otimes b,\, z \right)$
together with the operators
\(\Trn_{\mu}'':=\Trn_{\mu}+\Trn_{\mu}'\) generate
a vertex algebra structure over $\C^D$ on $\VA \otimes \VA'$
with a vacuum \(\varvac \otimes \varvac'\).
This vertex algebra is called a \textbf{tensor product}
of $\VA$ and $\VA'$ and we will denote it by $\VA \otimes \VA'$.

A \textbf{representation} of the vertex algebra $\VA$ is called a
super space \(\REP\) together with
a parity preserving linear map
\(\VA \rightarrow \left( \mathit{End} \, \REP \right)
\Bbrk{z,\, \frc{1}{z^{\, 2}}} :
a \mapsto Y_{\REP} \left( a,\, z \right)\)\gvspc{-6pt}
and even endomorphisms
\(\Trn_{\mu} \in \mathit{End} \, \REP\)
for \(\mu = 1,\, \dots ,\, D\) called again
\textbf{translation endomorphisms}
such that:
\begin{mlist}
\item
$Y_{\REP} \left( a,\, z \right)$ is a field and
$Y_{\REP} \left( a,\, z \right)\,$, $Y_{\REP} \left( b,\, z \right)$
are mutually local for all $a$ and \(b \in \VA\,\);
\item
\(Y_{\REP} \left( a,\, z \right)_{\,\left\{ n,\, m,\, \sigma \right\}}
Y_{\REP} \left( b,\, z \right)
= Y_{\REP} \left( a_{\,\left\{ n,\, m,\, \sigma \right\}}b,\, z \right)\)
for all \(a,\, b \in \VA\) and all
\(n \in \Z\,\), \(m = 0,\, 1,\, \dots\,\),
\(\sigma = 1,\, \dots ,\, \har_m\,\),
where the field $\left\{ n, m, \sigma \right\}$--products
are defined in accord with Theorem~\ref{th:2.1};
\item
\(\left[ \hspace{0.6pt} \Trn_{\mu} \hspace{0.6pt} ,\,
Y_{\REP} \left( a,\, z \right) \right] = \di_{z^{\mu}}
\, Y_{\REP} \left( a,\, z \right)\) for \(\mu = 1,\, \dots ,\, D\,\).
\end{mlist}
An example of a representation of a vertex algebra $\VA$ is provided by the
vertex algebra itself and it is called the \textbf{vacuum representation}
of~$\VA\,$.

\msection{Conformal Vertex Algebras}{sec:7}

We begin by recalling some basic facts about the conformal group
and its action for higher space dimensions $D$.
We will use the \textit{complex} conformal group \(\mathfrak{C}_{\C}\)
which is convenient to choose to be the \textit{connected complex spinor} group
\(\mathit{Spin}_0 \left( D+2;\, \C \right)
=: \mathfrak{C}_{\C}\) (the last is in fact a covering of the
``geometrical'' conformal group in space dimensions \(D \geqslant 3\)).
The geometrical action of \(\mathfrak{C}_{\C}\) on $\C^D$ will
be denoted as
\beq\label{eq7.1}
\C^D \, \ni \, z \, \mapsto \, g \, \left( z \right) \, \in \, \C^D
\quad (g \in \mathfrak{C}_{\C})
\, \qquad
\eeq
and it generally has \textit{singularities}: we will denote the
regularity of $z$ for $g$ as ``\(g \left( z \right) \in \C^D\)''.
The Lie algebra \(\mathfrak{c}_{\C}\) of \(\mathfrak{C}_{\C}\) is
isomorphic to \(\mathit{so} \left( D+2;\, \C \right)\) and it has
generators $\Trn_1$, $\dots$, $\Trn_D$, $\SCT_1$, $\dots$, $\SCT_D$,
$H$ and $\Omega_{\mu\nu}$ for \(1 \leqslant \mu < \nu \leqslant D\)
(\(\Omega_{\nu\mu} := - \Omega_{\mu\nu}\)),
with the following commutation relations:
\beqa\label{eq6.4o}
\left[ \, H \, , \, \Omega_{\mu\nu} \, \right]
\ = && \! 0
\, = \,
\left[ \, \Trn_{\mu} \, , \,
\Trn_{\nu} \, \right]
\, = \,
\left[ \, \SCT_{\mu} \, , \,
\SCT_{\nu} \, \right]
\, ,
\nn
\left[ \, \Omega_{\mu_1\nu_1} \, , \,
\Omega_{\mu_2\nu_2} \, \right]
\ = && \!
\delta_{\mu_1\mu_2}\,\Omega_{\nu_1 \nu_2}
\, + \,
\delta_{\nu_1\nu_2}\,\Omega_{\mu_1 \mu_2}
\, - \, \delta_{\mu_1\nu_2}\,\Omega_{\nu_1 \mu_2}
\, - \, \delta_{\nu_1\mu_2}\,\Omega_{\mu_1 \nu_2}
\, ,
\mgvspc{10pt} \nn &&
\hspace{-65pt}
\begin{array}{rlrl}
\left[ \, H \, , \, \Trn_{\mu} \, \right]
\, = & \,
\Trn_{\mu}
\, , \
&
\left[ \, H \, , \, \SCT_{\mu} \, \right]
\, = & \,
- \, \SCT_{\mu}
\, , \quad
\mgvspc{10pt} \\
\left[ \, \Omega_{\mu\nu} \, , \,
\Trn_{\rho} \, \right]
\, = & \,
\delta_{\mu\rho}\, \Trn_{\nu}
\, - \,
\delta_{\nu\rho}\, \Trn_{\mu}
\, , \
&
\left[ \, \Omega_{\mu\nu} \, , \,
\SCT_{\rho} \, \right]
\, = & \,
\delta_{\mu\rho}\, \SCT_{\nu}
\, - \,
\delta_{\nu\rho}\, \SCT_{\mu}
\, , \quad
\mgvspc{12pt}
\end{array}
\mgvspc{14pt} \nn
\left[ \, \SCT_{\mu} \, , \,
\Trn_{\nu} \, \right]
\, = && \!
2 \, \delta_{\mu\nu}\, H
\, + \, 2 \, \Omega_{\mu\nu}
\, .
\eeqa
Thus: $\Trn_1$, $\dots$, $\Trn_D$ are the generators of the translations
on $\C^D\,$, \(t_a \left( z \right) := z+a\,\),
\(t_a = e^{a \spr \Trn}\), \(a \spr \Trn \equiv
\mathop{\sum}\limits_{\mu \, = \, 1}^D
a^{\mu} \Trn_{\mu}\);
$H$ is the generator of the dilations
\(e^{\hspace{1pt} \lambda \hspace{1pt} H} \left( z \right) =
e^{\hspace{1pt} \lambda} z\);
\(\Omega_{\mu\nu}\)~are the generators of the orthogonal group of \(\C^D\)
($e^{\hspace{1pt} \hspace{1pt} \vartheta \hspace{1pt} \Omega_{\mu\nu}}$
being the rotation on angle
$\vartheta$ in the plane $\left( \mu,\, \nu \right)$ of $\R^D\hspace{1pt}$;
and finally, \(\SCT_{\mu}\) are generator of the special conformal
transformations:
\beq\label{eq7.3}
\sct_a \, := \, e^{a \spr \SCT} \, , \quad
\sct_a \left( z \right) \, = \, \frac{z+z^2 \, a}{
1 + 2 \, a \spr z + a^{\, 2} \, z^{\, 2}}
\quad (a,\, z \in \C^D)
\, . \qquad
\eeq
(For the explicit expression of the generators
$\Trn_1$, $\dots$, $\Trn_D$, $\SCT_1$, $\dots$, $\SCT_D$,
$H$ and $\Omega_{\mu\nu}$ in terms of the standard generators of
\(\mathit{so} \left( D+2;\, \C \right)\)~--~see for example~\cite{Dub}.)
We will call the Lie subalgebra of $\mathfrak{c}_{\C}$ generated
by $\Omega_{\mu\nu}$ (\(1 \leqslant \mu < \nu \leqslant D\)) the
\textbf{rotation subalgebra} (\(\cong
\mathit{so} \left( D;\, \C \right)\))
and its corresponding subgroup in $\mathfrak{C}_{\C}$~--~the
\textbf{spinor rotation subgroup} (\(\cong
\mathit{Spin}_0 \left( D;\, \C \right)\)).

An important element of the group $\mathfrak{C}_{\C}$ is the
\textbf{Weyl inversion} $j_{W}$:
\beq\label{eq7.4o1}
j_W \left( z \right) \, := \, \frac{R_D \left( z \right)}{z^{\, 2}}
\, , \qquad
\eeq
where $R_{\mu} \left( z \right)$ for
\(z = \left( z^1,\, \dots ,\, z^D \right)\) and \(\mu = 1,\, \dots ,\, D\,\), is
the reflection
\beq\label{eq7.20}
R_{\mu} \left( z^1,\, \dots ,\, z^D \right) \, = \,
\left( z^1,\, \dots ,\, -z^{\mu},\, \dots ,\, z^D \right)
\, . \qquad
\eeq
$j_{W}$ is represented in $\mathit{SO} \left( D+1,\, 1;\, \R \right)$
as a rotation on $\pi$ in the plane $\left( D , D+1 \right)\hspace{1pt}$,
i.~e., \(j_W = e^{\pi \left( \Trn_D + \SCT_D \right)}\).
Note that \(j_W^2 \left( z \right) = z\) for all $z$ but nevertheless
\(j_W^2\) as an element of
$\mathit{Spin}_0 \left( D+2,\, \C \right)$
is the nonunit central element \(C = - \ID_{\mathit{Cliff}}\,\),
\beq\label{eq7.6o11}
j_W^2 \, = \, C \, , \quad
\eeq
where $\ID_{\mathit{Cliff}}$ is the Clifford algebra unit.

The passage from the vertex algebras
to the globally conformal invariant QFT
needs first an additional symmetry structure for our vertex algebras.
For this purpose we will extend the abelian Lie algebra
of the translations $\Trn_1$, $\dots$, $\Trn_D$ to
the conformal one
\(\mathfrak{c}_{\C} \cong \mathit{so}(D+2;\, \C)\).

\begin{mdefinition}\label{def:7.1o3}
A \textbf{conformal vertex algebra} is called a vertex algebra $\VA$
endowed with an action of $\mathfrak{c}_{\C}$ by even linear automorphisms
such that
\beqa\label{eq6.5o}
\left[ \hspace{1pt} H \, , \,
Y \left( a,\, z \right) \, \right]
\, = && \hspace{-2pt}
z \spr \di_{z} \, Y \left( a,\, z \right) \, + \,
Y \left( H\hspace{1pt} a,\, z \right)
\, , \quad
\\ \label{eq6.6o}
\left[ \hspace{1pt} \Omega_{\mu\nu} \, , \,
Y \left( a,\, z \right) \, \right]
\, = && \hspace{-2pt}
z^{\mu} \, \di_{z^{\nu}} Y \left( a,\, z \right) \, - \,
z^{\nu} \, \di_{z^{\mu}} Y \left( a,\, z \right) \, + \,
Y \left( \Omega_{\mu\nu}\hspace{1pt} a,\, z \right)
\, , \quad \mgvspc{12pt}
\\ \label{eq6.7o}
\left[ \hspace{1pt} \SCT_{\mu} \, , \,
Y \left( a,\, z \right) \, \right]
\, = && \hspace{-2pt}
\left( -z^{\, 2} \, \di_{z^{\mu}}  + 2 \, z^{\mu} \, z \spr \di_{z} \right)
Y \left( a,\, z \right) \, + \,
2 \, z^{\mu} \, Y \left( H\hspace{1pt} a,\, z \right) \, + \,
\mgvspc{12pt}
\nn && \hspace{-2pt}
+ \, 2 \mathop{\sum}\limits_{\nu \, = \, 1}^D
z^{\nu} \, Y \left( \Omega_{\nu\mu} \hspace{1pt} a,\, z \right) \, + \,
Y \left( \SCT_{\mu}\hspace{1pt} a,\, z \right)
\, \qquad
\eeqa
(\(z \spr \di_{z} \equiv \Su_{\mu \, = \, 1}^D
z^{\mu} \, \di_{z^{\mu}}\),
\(z^{\, 2} \equiv z \spr z \equiv \mathop{\sum}\limits_{\mu \, = \, 1}^D
z^{\mu} z^{\mu}\)).
The compatibility of the commutation relations~(\ref{eq6.4o})
with Eqs.~(\ref{eq6.5o})--(\ref{eq6.7o}) is obtained by
a straightforward computation.
We require also
that:
\begin{mlist}
\item
the enodomorphism $H$
is diagonalizable with nonnegative eigenvalues
(the \textbf{energy positivity} condition).
\item
The representation of the rotation subalgebra
\(\mathit{so} \left( D;\, \C \right) \subset \mathfrak{c}_{\C}\) on $\VA$
decomposes into a direct sum of finite dimensional irreducible
subrepresentations.
Then this representation can admit integration to an action of the
spinor rotation
subgroup
\(\mathit{Spin}_0 \left( D;\, \C \right)\).
\item
Let $C$ be the central element~(\ref{eq7.6o11})
(\(C^2 = \ID\))
then \(H + \frac{\raisebox{0pt}{$1$}}{\raisebox{-2pt}{$4$}} \,
\left( \ID - C \right)\) has an integer spectrum.
In particular, $H$ has only integer or half-integer eigenvalues.
\item
The vacuum $\varvac$ is the only one $\mathfrak{c}_{\C}$--invariant element of $\VA$
up to multiplication,
i.~e., \(X \, a = 0\) for every \(X \in \mathfrak{c}_{\C}\) 
\ $\Leftrightarrow$ \ \(a \sim \varvac\).
\end{mlist}
\end{mdefinition}

If \(a \in \VA\) is an eigenvector of $H$ we will denote its eigenvalue
by $\mathit{wt}_H \left( a \right)$ and call it \textbf{weight} of $a$:
\beq\label{eq6.8o2}
H \, a \, = \, \mathit{wt}_H \left( a \right) a
\, . \qquad
\eeq
Then if \(a,\, b \in \VA\) have fixed weights:
\beq\label{eq6.9o2}
\mathit{wt}_H \left( a_{\,\left\{ n,\, m,\, \sigma \right\}}b \right)
\, = \,
\mathit{wt}_H \left( a \right) \, + \,
\mathit{wt}_H \left( b \right) \, + \, 2 \, n \, + \, m
\, , \qquad
\eeq
which follows from the equation
\\ ${}$ \hfill
\(a_{\,\left\{ n,\, m,\, \sigma \right\}}b
= P_{\left\{ n+N,\, m,\, \sigma \right\}} \left( \di_{z} \right)
\left( z^{\, 2} \right)^N Y \left( a,\, z \right) b
\vrestr{9pt}{\, z \, = \, 0}\)\gvspc{15pt}\gvspc{-12pt}
\hfill ${}$ \\
for \(N \mgrt 0\) (see Eq.~(\ref{ad1.11})) and
the relation~(\ref{eq6.5o}).

As a consequence of Definition~\ref{def:7.1o3}~(\textit{a})
and (\textit{d}), and the commutation
relations (\ref{eq6.4o}) the endomorphisms
$\SCT_{\mu}$ will have a nilpotent action on
a conformal vertex algebra~$\VA$.
Thus the representation of $\SCT_1$, $\dots$, $\SCT_D$,
$H$ and $\Omega_{\mu\nu}$ (\(1 \leqslant \mu < \nu \leqslant D\,\))
on $\VA$ can be integrated to a group action.
Recall that
the last generators
span a Lie subalgebra \(=: \mathfrak{c}_{\C, 0}\) of $\mathfrak{c}_{\C}$
which corresponds to the subgroup $\mathfrak{C}_{\C, 0}$ of
\(\mathfrak{C}_{\C}\)~--~the \textit{connected part of the stabilizer}
of \(0 \in \C^D\).
Note that $\mathfrak{C}_{\C, 0}$ is isomorphic to the
inhomogeneous connected spinor group of $\C^D$ with dilations
(i.~e. the complex Euclidean spinor group with dilations)
and it is simply connected.
Denote the obtained action by
\beq\label{eq6.8o1}
\pi_0 \, : \, \mathfrak{C}_{\C, 0} \to \mathit{Aut}_0 \, \VA
\, \qquad
\eeq
and define
\beq\label{eq6.9o1}
\pi_z \left( g \right) \, := \,
\pi_0 \left( t_{g \left( z \right)}^{-1} \, g \, t_z \right)
\, \qquad
\eeq
for all
pairs $\left( z,\, g \right)$ in some neighbourhood of
\(\left( 0,\, \ID \right) \in \C^D \times \mathfrak{C}_{\C}\) such that
\(g \left( z \right) \in \C^D\).
Note that \(t_{g \left( z \right)}^{-1} \, g \, t_z \left( 0 \right) = 0\)
if $z$ is regular for $g$
(i.~e., \(g \left( z \right) \in \C^D\)) so that
\(t_{g \left( z \right)}^{-1} \, g \, t_z \left( 0 \right) \in
\mathfrak{C}_{\C, 0}\) for small $z$ and $g$.

\begin{mproposition}\label{pr:7.1}
\begin{plist}
\item
The function $\pi_z \left( g \right)$~(\ref{eq6.9o1}) is rational
in $z$ for every \(g \in \mathfrak{C}_{\C}\) with values in
$\left( \mathit{End} \, \VA \right)_{0}$, i.~e. it can be expressed as
a ratio of a polynomial belonging to
$\left( \mathit{End} \, \VA \right)_{0} \Brk{z}$ and a
polynomial belonging to
$\C \Brk{z}$.
It has the {\rm cocycle} property
\beq\label{eq6.10o1}
\pi_z \left( g_1 g_2 \right) \, = \,
\pi_{g_2 \left( z \right)} \left( g_1 \right)
\pi_z \left( g_2 \right)
\,  \quad
\mathit{iff} \quad
g_1 g_2 \left( z \right),\, g_2 \left( z \right) \, \in \, \C^D
\, \qquad
\eeq
and satisfies \(\pi_z \left( t_a \right) = \ID\).
\item
Let the assumptions of Definition~\ref{def:7.1o3} are supposed
except the condition~(\textit{c}).
Then if the cocycle $\pi_z \left( g \right)$~(\ref{eq6.9o1})
possesses a continuation to a rational function in $z$ it follows that
the condition~(\textit{c}) of Definition~\ref{def:7.1o3} is also satisfied.
\end{plist}
\end{mproposition}

\begin{proof}
(a)
Eq.~(\ref{eq6.10o1}) is a straightforward consequence of (\ref{eq6.9o1})
for small $z$, $g_1$ and $g_2$.
If $g$ belongs to the (spinor) Euclidean group of $\C^D$ with dilations
then \(t_{g \left( z \right)}^{-1} \, g \, t_z = g_1\) does not
depend on $z$ and it is just the homogeneous
part of $g$ (i.~e. the projection on the spinor and dilation group).
Thus if we prove the rationality of $\pi_z \left( j_W \right)$
it will follow for the general $\pi_z \left( g \right)$
because of Eq.~(\ref{eq6.10o1}) and the fact the conformal group
$\mathfrak{C}_{\C}$ is generated by $j_W$ and the spinor
Euclidean group with dilations.
To compute
$\pi_z \left( j_W \right)$ first we observe that
\beq\label{eq7.22o1}
\hspace{3pt}
t_{j_W \left( z \right)}^{-1} \, j_W \, t_z =
\sct_{-R_D \left( z \right)} \, O
\, , \quad
O_z \left( w \right) =
\frac{1}{z^{\, 2}} \,
\mathop{\sum}\limits_{\mu \, = \, 1}^{D} \hspace{-1pt}
\frac{\left( R_D \left( z \right) \right)^{\mu}}{
\sqrt{z^{\, 2}}} \ R_{\mu} \left( R_D \left( w \right) \right)
\, ,
\hspace{-20pt}
\eeq
where \(\sct_{-R_D \left( z \right)} ,\, O_z \in \mathfrak{C}_{\C,0}\)
so that
\beq\label{eq7.15o2}
\pi_z \hspace{-2pt} \left( j_W \right) \, = \,
\pi_0 \left(\raisebox{9pt}{\hspace{-2pt}}\right.
\sct_{-R_D \left( z \right)} \left.\raisebox{9pt}{\hspace{-2pt}}\right) \,
\pi_0 \left( O_z \right)
\, .
\eeq
Then $\pi_0 \left( O_z \right)$ is rational due to
Definition \ref{def:7.1o3}~(\textit{c}):
indeed, if $\pi_0$ is a subrepresentation of the spinor representation
of the Clifford algebra
$\mathit{Cliff} \left( D;\, \R \right)$
(for even $D$ it is uniquely determined and for odd $D$,
there are two irreducible representations, up to equivalence),
then $O_z$ is represented by
\(\pm \left( z^{\, 2} \right)^{-H -
\raisebox{-0.5pt}{$\frac{1}{2}$}} \,
\mathop{\sum}\limits_{\mu \, = \, 1}^D
z^{\mu} \, \gamma_D \, \gamma_{\mu}\hspace{1pt}\)),
where $\gamma_{\mu}$ are the Clifford algebra generators;
in the general case $\pi_0$ is a direct sum of subrepresentations
of some tensor power of the above one.

To prove the second part~(b) we observe that
the last argument is invertible: the rationality of $\pi_0 \left( O_z \right)$
implies the condition~(\textit{c}) of Definition~\ref{def:7.1o3}.\qed
\end{proof}

A standard consequence of the commutation relations~(\ref{eq6.4o})
is that the full eigen{\-}spa{\-}ces of the action of $H$ on a conformal
vertex algebra are invariant for the action of the
rotation subalgebra $\mathit{so} \left( D;\, \C \right)$.
Therefore, the irreducible subrepresentations of
$\mathit{so} \left( D;\, \C \right)$ are eigenspaces for $H$.

\begin{mlemma}\label{lm:7:2o12}
For every element $a$ in a conformal vertex algebra $\VA$ there exists
a finite dimensional subspace $\mathcal{U}$ of $\VA$ which contains $a$
and is invariant with respect to the representation $\pi_0$~(\ref{eq6.8o1}).
Therefore, $\mathcal{U}$ is also invariant for the cocycle
$\pi_z \left( g \right)$~(\ref{eq6.9o1}).
\end{mlemma}

\begin{proof}
Because of Definition~\ref{def:7.1o3} (\textit{b}) it is sufficient
to prove the lemma in the case when $a$ belongs to some irreducible
subrepresentation $\mathcal{U}_0$ of the rotation group.
Then \(H\vrestr{9pt}{\mathcal{U}_0} = d \, \ID\)\gvspc{-5pt} for \(2 d \in
\left\{ 0 \right\} \cup \N\).
Therefore, \(\SCT_{\mu_1} \dots \SCT_{\mu_{2d}} \, \mathcal{U}_0 = 0\)
for all  \(\mu_k =  1,\, \dots ,\, D\) (\(k = 1,\, \dots ,\, 2d\))
and we can set then the space $\mathcal{U}$ to be the linear span of all
vectors belonging to
\(\SCT_{\mu_1} \dots \SCT_{\mu_k} \, \mathcal{U}_0\)
(\(k = 0,\, \dots ,\, 2d-1\,\)).\qed
\end{proof}

Let $f \left( \lambda \right)$ and $f' \left( \lambda,\, \lambda' \right)$
be functions with values in a finite dimensional vector space $\Lin$
which are holomorphic in $\lambda$ and $\lambda'$ in a neighebouhood
of \(0 \in \C\).
Then we set
\(\iota_{\lambda} \, f \left( \lambda \right) \in \Lin \Bbrk{\lambda}\) and
\(\iota_{\lambda,\, \lambda'} \,
f' \left( \lambda,\, \lambda' \right) \in \Lin \Bbrk{\lambda,\, \lambda'}\)
to be just the Taylor series of $f$ and $f'$ around \(0 \in \C\).
This definition is applicable to the functions
$\pi_z \left( e^{\, \lambda \hspace{1pt} X} \right)$ and
\(\pi_{e^{\, \lambda' \hspace{1pt} X'}
\left( z \right)} \left( e^{\, \lambda \hspace{1pt} X} \right)\)
for \(X,\, X' \in \mathfrak{c}_{\C}\) because of
Lemma~\ref{lm:7:2o12}.
Then Eq.~(\ref{eq6.10o1}) implies
\beq\label{eq7.16o7}
\iota_{\lambda,\, \lambda'} \left(\raisebox{9pt}{\hspace{-2pt}}\right.
\pi_z \left(\raisebox{9pt}{\hspace{-2pt}}\right.
e^{\, \lambda \hspace{1pt} X} \, e^{\, \lambda' \hspace{1pt} X'}
\left.\raisebox{9pt}{\hspace{-2pt}}\right)
\left.\raisebox{9pt}{\hspace{-2pt}}\right)
\, = \,
\iota_{\lambda,\, \lambda'} \left(\raisebox{9pt}{\hspace{-2pt}}\right.
\pi_{e^{\, \lambda \hspace{1pt} X} \left( z \right)}
\left(\raisebox{9pt}{\hspace{-2pt}}\right.
e^{\, \lambda' \hspace{1pt} X'}
\left.\raisebox{9pt}{\hspace{-2pt}}\right)
\left.\raisebox{9pt}{\hspace{-2pt}}\right) \,
\iota_{\lambda,\, \lambda'} \left(\raisebox{9pt}{\hspace{-2pt}}\right.
\pi_z \left(\raisebox{9pt}{\hspace{-2pt}}\right.
e^{\, \lambda \hspace{1pt} X}
\left.\raisebox{9pt}{\hspace{-2pt}}\right)
\left.\raisebox{9pt}{\hspace{-2pt}}\right)
\, . \
\eeq
We will distinct the above notations $\iota_{\lambda}$ and
$\iota_{\lambda,\, \lambda'}$ from the similar
$\iota_{z_1,\, \dots,\, z_n}$ from Sect.~\ref{sec:4} by the
type of the arguments $\lambda$, $\lambda'$ and $z_k$.

It follows from the above constructions that the following equation is valid
for any conformal vertex algebra $\VA\,$:
\beq\label{eq7.17o7}
e^{\, \lambda \hspace{1pt} X} \,
Y \left( a,\, z \right) \,
e^{\, -\lambda \hspace{1pt} X} \, b
\, = \,
Y \left(\raisebox{9pt}{\hspace{-2pt}}\right.
\iota_{\lambda} \,
\pi_z \left(\raisebox{9pt}{\hspace{-2pt}}\right.
e^{\, \lambda \hspace{1pt} X}
\left.\raisebox{9pt}{\hspace{-2pt}}\right)
a,\, e^{\, \lambda \hspace{1pt} X} \left( z \right)
\left.\raisebox{9pt}{\hspace{-2pt}}\right)
\, b
\, \
\eeq
as series belonging to
$\VA \Bbrk{z}_{\,\lscw{z}{\, 2}{}} \Bbrk{\lambda}$
for all \(a,\, b \in \VA\) and \(X \in \mathfrak{c}_{\C}\).
In the case of \(X \in \mathfrak{c}_{\C, 0}\) Eq.~(\ref{eq7.17o7})
follows from the construction of $\pi_0$~(\ref{eq6.8o1})
and relations~(\ref{eq6.5o})--(\ref{eq6.7o}).
For \(X = \Trn_{\mu}\,\), (\ref{eq6.8o1}) is just a consequence of
Eq.~(\ref{add2.14.1}).

\msection{Hermitean Structure in Conformal Vertex Algebras}{sec:8}

Besides the conformal structure the passage from the vertex algebras to
the QFT requires a \textit{Hermitean} structure.
In this section we will use besides the variables $z$, $w$, etc., their
$D$--dimensional conjugate variables
\(\overline{z} \, \equiv \, \overline{\left( z^1,\, \dots,\, z^D \right)} =
\left( \overline{z_1},\, \dots,\, \overline{z_D} \right)\), etc.
If $z$ is considered as a formal variable then $\overline{z}$ will be
treated as an independent formal variable and if $z$ takes values in $\C^D$
then $\overline{z}$ will be its complex conjugate.
We set also
\\ ${}$ \hfill \(
\overline{u \left( z \right)} \, = \,
\overline{u} \left( \overline{z} \right)
\, , \quad
\overline{\overline{z}} \, = \, z
\
\mgvspc{15pt}\mgvspc{-8pt}\) \hfill ${}$ \\
for a series \(u \left( z \right) \in \C \Bbrk{z,\, \frc{1}{z^{\, 2}}}\),
where $\overline{u} \left( z \right)$ stands for the series with
complex conjugate coefficients.

Define the following \textbf{conjugation}
\beq\label{eq7.23o3}
z \, \mapsto
\, z^* \, := \, \frac{\overline{z}}{\overline{z}^{\, 2}}
\, . \quad
\eeq
It can be written also by the Weyl reflection~(\ref{eq7.4o1}) as:
\beq\label{eqn-new}
z^* \, = \, j_W \left( R_D \left( \overline{z} \right) \right) \, \equiv \,
j_W^{-1} \left( R_D \left( \overline{z} \right) \right)
\, . \
\eeq
The corresponding conjugations on the complex conformal
Lie algebra and group
are introduced by:
\beqa\label{eq7.26o3}
&
\left( \Trn_{\mu} \right)^* \, := \, - \SCT_{\mu}
\quad (1 \leqslant \mu \leqslant D)
\, , \
& \nn &
H^{\hspace{1pt} *} \, = \, - \, H
\, , \quad
\Omega_{\mu\nu}^{\hspace{1pt} *} \, = \, \Omega_{\mu\nu}
\quad (1 \leqslant \mu < \nu \leqslant D)
& \nn &
\left( e^{\, \lambda \hspace{1pt} X} \right)^* \, = \,
e^{\, \overline{\lambda} \hspace{1pt} X^*}
\quad \mathrm{for} \quad
X \, \in \, \mathfrak{c}_{\C} \quad \mathrm{and} \quad
\lambda \, \in \, \C
\,
&
\eeqa
as \(\left[ X_1,\, X_2 \right]^* = \left[ X_1^*,\, X_2^* \right]\) and
\(g_1^* \, g_2^* = \left( g_1 \, g_2 \right)^*\) for
\(X_1,\, X_2 \in \mathfrak{c}_{\C}\) and
\(g_1,\, g_2 \in \mathfrak{C}_{\C}\).
Since the conjugation $*$ uses the \textit{conformal inversion}
\(z \mapsto \frac{\textstyle z}{\textstyle z^{\, 2}}\)\gvspc{10pt}\gvspc{-8pt}
under which the generators $\Trn_{\mu}$ and $\SCT_{\mu}$ are conjugated,
we have the consistency relation:
\beq\label{eq7.27o3}
g \left( z \right)^* \, = \, g^* \left( z^* \right)
\, \qquad
\eeq
for all \(g \in \mathfrak{C}_{\C}\) and \(z \in \C^D\) such that
\(z^*,\, g \left( z \right) \in \C^D\).

For a series
\(a \left( z \right) \in \Lin \Bbrk{z,\, \frc{1}{z^{\, 2}}}\),
$\Lin$ being a complex vector space,
the substitution $a \left( z^* \right)$ is correctly defined
by Eq.~(\ref{1.22}) as the series
\beq\label{eq8.2o8}
a \left( z^* \right) \, := \,
J \left[ a \left( w \right) \right]
\vrestr{9pt}{w = \overline{z}} \, \in \,
\Lin \Bbrk{\overline{z},\, \frc{1}{\overline{z}^{\, 2}}}
\, . \
\eeq
If $\Lin$ is
endowed with a Hermitean form
\(\La a \Vl b \Ra \in \C\) for \(a,\, b \in \Lin\,\), then we will
use the convention:
\beq\label{eq6.8o}
\La a \Vl u \left( z \right) \, b \Ra
\, = \,
\La \overline{u} \left( \overline{z} \right) \, a \Vl b \Ra
\, = \,
u \left( z \right) \La a \Vl b \Ra
\, \quad
(\La a \Vl b \Ra \, = \, \overline{\La b \Vl a \Ra})
\, \
\eeq
for a series \(u \left( z \right) \in \C \Bbrk{z,\, \frc{1}{z^{\, 2}}}\).

\begin{mdefinition}\label{def7.2o4}
A conformal vertex algebra $\VA$ with \textbf{Hermitean structure}
is a conformal vertex algebra equipped with a \textit{nondegenerate}
Hermitean form
\(\La a \Vl b \Ra \in \C\) for \(a,\, b \in \VA\),
compatible with the $Z_2$--grading of $\VA$
(\(\La a \Vl b \Ra = 0\) if \(a \in \VA_0\) and \(b \in \VA_1\))
and possessing an antilinear even involution
\(a \mapsto \cnj{a}\) (\(a \in \VA\); \(\cnj{\left( \lambda\, a \right)} =
\overline{\lambda}\, a\), \(\cnjj{a} = a\),
\(\cnj{\left( \VA_{0,1} \right)} = \VA_{0,1}\))
satisfying the following conditions
\beqa\label{eq8.4o8}
&
\La a \Vl X \, b \Ra \, = \, - \La X^* \, a \Vl b \Ra
\, , \
& \\ \label{eq7.19} &
\La a \Vl Y \left( \cnj{c},\, z \right) b \Ra \, = \,
\overline{
\La b \Vl Y \left(\raisebox{9pt}{\hspace{-2pt}}\right.
\pi_{z^*} \hspace{-2pt} \left( j_W \right)^{-1} c,\, z^*
\left.\raisebox{9pt}{\hspace{-2pt}}\right) a  \Ra}
\, \ & \mgvspc{12pt}
\eeqa
for all \(a,\, b,\, c \in \VA\) and \(X \in \mathfrak{c}_{\C}\,\).
Here the last equality is understood in the sense of rational functions in $z$
and it is correct
in view of the following remark.
\end{mdefinition}

\begin{mremark}\label{rm:7.1o5}
\begin{plist}
\item
As a consequence of Eq.~(\ref{eq6.9o2}) and the orhogonality
of the different eigen{\-}spaces of $H$
(because of (\ref{eq8.4o8}) and (\ref{eq7.26o3})) we have
\beq\label{eq6.10o}
\La a \Vl Y \left( c,\, z \right) b \Ra \, \in \,
\C \Brk{z, \frc{1}{z^{\, 2}}} \, \equiv \,
\C \Brk{z}_{\,\lscw{z}{\, 2}{}}
\, , \
\eeq
(note that in accord with Definition~\ref{def:2.1} we only have
\(\La a \Vl Y \left( c,\, z \right) b \Ra \in
\C \Bbrk{z}_{\,\lscw{z}{\, 2}{}}\)).
Note also that due to the decomposition~(\ref{eq7.15o2})
we will also have
\beq\label{eq8.10o12}
\pi_z \! \left( j_W^{\hspace{1pt} \pm 1} \right) \, a \, \in \,
\VA \Brk{z,\, \frc{1}{z^{\, 2}}}
\, . \
\eeq
\item
The conjugation in Eq.~(\ref{eq8.4o8}) is a combination of the
``Minkowski'' conjugation \(z \mapsto R_D \left( \overline{z} \right)\)
and the Weyl reflection
since it can be rewritten (using Eq.~(\ref{eqn-new})) also as
\beq\label{eqn-new2}
\La a \Vl Y \left( \cnj{c},\, z \right) b \Ra \, = \,
\La Y \left(\raisebox{9pt}{\hspace{-2pt}}\right.
\pi_{R_D \left( \overline{z} \right)}
\hspace{-2pt} \left( j_W^{-1} \right) c,\, j_W^{-1}
\left( R_D \left( \overline{z} \right) \right)
\left.\raisebox{9pt}{\hspace{-2pt}}\right) a \Vl b \Ra
\, . \
\eeq
This conjugation law is also idempotent as a consequence of
Eq.~(\ref{eqn-new3}) below.
\end{plist}
\end{mremark}

\begin{mproposition}\label{pr:6.1o}
In any conformal vertex algebra $\VA$ endowed with a Hermitean form
\(\La a \Vl b \Ra \in \C\) for \(a,\, b \in \VA\), satisfying
Eqs.~(\ref{eq8.4o8}) and (\ref{eq6.10o}), the {\rm correlation} functions
\(\La b \Vl
Y \left( a_1,\, z_1 \right) \dots Y \left( a_n,\, z_n \right) c \Ra\)
are rational for all $a_1$, $\dots$, $a_n$, $b$, \(c \in \VA\)
in the sense that they belong to
\(\iota_{z_1,\, \dots ,\, z_n}
\left( \C \Brk{z_1,\, \dots ,\, z_n}_{\,
\raisebox{-2pt}{\small $R_n$}} \right)\)
(see Sect.~\ref{sec:4}).
In particular, they are regular for \(z_k^{\, 2} \neq 0\) and
\(\left( z_k - z_l \right)^{\, 2} \neq 0\) (\(k,\, l\) $=$
\(1,\, \dots ,\, n\)).
The vacuum correlation functions
\beq\label{eq6.14o1}
\Wf_n \left( a_1,\, \dots,\, a_n \hspace{1pt};\, z_1,\, \dots,\, z_n \right)
\, := \,
\iota_{z_1,\, \dots ,\, z_n}^{-1} \,
\La \varvac \Vl
Y \left( a_1,\, z_1 \right) \dots Y \left( a_n,\, z_n \right) \varvac \Ra
\eeq
are {\rm globally conformal invariant} in the sense that
\beqa\label{eq6.15o1}
&
\Wf_n \left(\raisebox{9pt}{\hspace{-2pt}}\right.
\pi_{z_1} \hspace{-2pt} \left( g \right) a_1,\,
\dots, \,
\pi_{z_n} \hspace{-2pt} \left( g \right) a_n \hspace{1pt};\,
g \left( z_1 \right),\, \dots,\,
g \left( z_n \right)
\left.\raisebox{9pt}{\hspace{-2pt}}\right) \, = \,
& \nn & \, = \,
\Wf_n \left( a_1,\, \dots,\, a_n \hspace{1pt};\, z_1,\, \dots,\, z_n \right)
&
\eeqa
(for all \(g \in \mathfrak{C}_{\C}\))
as rational functions in $z_1$, $\dots$, $z_n$.
\end{mproposition}

\begin{proof}
It follows from locality and Eq.~(\ref{eq6.10o}) that
\\ ${}$ \hfill \(
\left(\raisebox{11pt}{\hspace{-2pt}}\right.
\mathop{\prod}\limits_{k \, = \, 1}^n z_k^{\, 2}
\left.\raisebox{11pt}{\hspace{-2pt}}\right)^N
\left(\raisebox{11pt}{\hspace{-2pt}}\right.
\mathop{\prod}\limits_{1 \, \leqslant \, l \, < \, m \, \leqslant \, n}
z_{lm}^{\, 2}
\left.\raisebox{11pt}{\hspace{-2pt}}\right)^N \,
\La b \Vl
Y \left( a_1,\, z_1 \right) \dots Y \left( a_n,\, z_n \right) c \Ra
\, \in \, \C
\Brk{z_1,\dots ,z_n}
\mgvspc{17pt}\mgvspc{-14pt}\) \hfill ${}$ \\
for \(N \mgrt 0\) (\(z_{lm} := z_l - z_m\)).
Then we multiply by
\\ ${}$ \hfill \(
\iota_{z_1,\, \dots ,\, z_n} \,
\left(\raisebox{11pt}{\hspace{-2pt}}\right.
\mathop{\prod}\limits_{k \, = \, 1}^n z_k^{\, 2}
\left.\raisebox{11pt}{\hspace{-2pt}}\right)^{-N}
\left(\raisebox{11pt}{\hspace{-2pt}}\right.
\mathop{\prod}\limits_{1 \, \leqslant \, l \, < \, m \, \leqslant \, n}
z_{lm}^{\, 2}
\left.\raisebox{11pt}{\hspace{-2pt}}\right)^{-N} \,
\mgvspc{17pt}\mgvspc{-14pt}\) \hfill ${}$ \\
as in the proof of Proposition~\ref{pr:2.10}.
To prove Eq.~(\ref{eq6.15o1}) we first obtain by Eq.~(\ref{eq7.17o7})
and by the conformal invariance of the vacuum
(Def.~\ref{def:7.1o3} (\textit{d})) that
\beqa\label{eq6.16o1}
&
e^{\, \lambda \hspace{1pt} X} \,
Y \left( a_1,\, z_1 \right) \, \dots \, Y \left( a_n,\, z_n \right) \, \varvac
\, = \, & \nn & \, = \,
Y \left(\raisebox{9pt}{\hspace{-2pt}}\right.
\iota_{\lambda} \,
\pi_{z_1} \left(\raisebox{9pt}{\hspace{-2pt}}\right.
e^{\, \lambda \hspace{1pt} X}
\left.\raisebox{9pt}{\hspace{-2pt}}\right)
a_1,\, e^{\, \lambda \hspace{1pt} X} \left( z_1 \right)
\left.\raisebox{9pt}{\hspace{-2pt}}\right)
\, \dots \,
Y \left(\raisebox{9pt}{\hspace{-2pt}}\right.
\iota_{\lambda} \,
\pi_{z_n} \left(\raisebox{9pt}{\hspace{-2pt}}\right.
e^{\, \lambda \hspace{1pt} X}
\left.\raisebox{9pt}{\hspace{-2pt}}\right)
a_n,\, e^{\, \lambda \hspace{1pt} X} \left( z_n \right)
\left.\raisebox{9pt}{\hspace{-2pt}}\right)
\, \varvac &
\qquad
\eeqa
for all \(X \in \mathfrak{c}_{\C}\) as an equality in
\(\VA \Bbrk{z_1}_{\,\lscw{z}{\, 2}{1}}
\dots \, \Bbrk{z_n}_{\,\lscw{z}{\, 2}{n}} \, \Bbrk{\lambda}\).
Then in view of Eq.~(\ref{eq8.4o8}) and the conformal invariance
of the vacuum we find
\beqa\label{eq6.17o1}
&
\iota_{\lambda} \,
\Wf_n \left(\raisebox{9pt}{\hspace{-2pt}}\right.
\pi_{z_1} \hspace{-2pt} \left(\raisebox{9pt}{\hspace{-2pt}}\right.
e^{\hspace{1pt} \lambda \hspace{1pt} X}
\left.\raisebox{9pt}{\hspace{-2pt}}\right)
a_1,\,
\dots, \,
\pi_{z_n} \hspace{-2pt} \left(\raisebox{9pt}{\hspace{-2pt}}\right.
e^{\hspace{1pt} \lambda \hspace{1pt} X}
\left.\raisebox{9pt}{\hspace{-2pt}}\right)
a_n \hspace{1pt};\,
e^{\hspace{1pt} \lambda \hspace{1pt} X} \left( z_1 \right),\, \dots,\,
e^{\hspace{1pt} \lambda \hspace{1pt} X} \left( z_n \right)
\left.\raisebox{9pt}{\hspace{-2pt}}\right)
\vrestr{10pt}{\lambda \, = \, 0}
\, = \, & \nn & \, = \,
\Wf_n \left( a_1,\, \dots,\, a_n \hspace{1pt};\, z_1,\, \dots,\, z_n \right)
&
\eeqa
for all \(X \in \mathfrak{c}_{\C}\)
which actually proves~(\ref{eq6.15o1}).\qed
\end{proof}

A special case of the GCI~(\ref{eq6.15o1}) is the
\textit{translation invariance}:
\beq\label{eq7.24o5}
\Wf_n \left( a_1,\, \dots ,\, a_n ;\, z_1,\, \dots ,\, z_n \right)
\, = \,
W_n \left( a_1,\, \dots ,\, a_n ;\,
z_{12},\, \dots ,\,  z_{n-1 \hspace{1pt} n} \right)
\
\eeq
(\(z_{kl} = z_k - z_l\),
\(a_1,\, \dots ,\, a_n \in \VA\)),
since \(\pi_z \left( t_a \right) = \ID\).
A consequence of locality for the correlation
functions~(\ref{eq6.14o1}) is the $\Z_2$--symmetry:
\beq\label{eq7.25o5}
\hspace{2pt}
\Wf_n \hspace{-2pt} \left( a_1, \dots , a_n ; z_1, \dots , z_n \right)
\hspace{-1pt} = \hspace{-1pt}
\left( -1 \right)^{\epsilon \left( \sigma \right)} \hspace{-2pt}
\Wf_n \hspace{-2pt} \left( a_{\sigma \left( 1 \right)}, \dots ,
a_{\sigma \left( n \right)} ;
z_{\sigma \left( 1 \right)}, \dots ,
z_{\sigma \left( n \right)} \right)
\hspace{-20pt}
\eeq
for every permutation \(\sigma \in \Ss_n\), where
$\epsilon \left( \sigma \right)$ is the $\Z_2$--parity of
$\sigma$ introduced in Proposition~\ref{pr:2.10}.

\begin{mproposition}\label{pr:8.2o10}
Let $\VA$ be a conformal vertex algebra with Hermitean structure.
Then
\beqa\label{eq8.16o10}
& \hspace{-20pt}
\cnj{\left( \pi_z \! \left( g \right) \, a \right)} =
\pi_{R_D \left( \overline{z} \right)} \! \left( j_W \, g^* j_W^{-1} \right)
\, \cnj{a}
\quad \mathit{or} \quad
\cnj{\pi_z \! \left( g \right)} =
\pi_{R_D \left( \overline{z} \right)} \! \left( j_W \, g^* j_W^{-1} \right)
& \\ \label{eqn-new3} & \hspace{-20pt}
\pi_{z^{**}} \left( j_W \right)^{-1}
\left(\raisebox{9pt}{\hspace{-2pt}}\right.
\pi_{z^*} \left( j_W \right)^{-1} a
\left.\raisebox{9pt}{\hspace{-2pt}}\right)^+ \, = \, a
\, . &
\eeqa
\end{mproposition}

\begin{proof}
As a consequence of Eqs.~(\ref{eq7.19}), (\ref{eq8.4o8})
and (\ref{eq7.17o7}) we have the following equalities in
$\C \Bbrk{z,\, \frc{1}{z^{\, 2}}} \Bbrk{\lambda}$
($\lambda$ being one component variable)
for all \(a,\, b,\, c \in \VA\) and \(X \in \mathfrak{c}_{\C}\):
\beqa\label{eq8.18o12}
&&
\La b \Vl
Y \left(\raisebox{9pt}{\hspace{-2pt}}\right.
\pi_z \left(\raisebox{9pt}{\hspace{-2pt}}\right.
e^{\, \lambda \hspace{1pt} X}
\left.\raisebox{9pt}{\hspace{-2pt}}\right)
a,\, e^{\, \lambda \hspace{1pt} X} \left( z \right)
\left.\raisebox{9pt}{\hspace{-2pt}}\right)
\, c \Ra
\, = \,
\La b \Vl e^{\, \lambda \hspace{1pt} X} \,
Y \left( a,\, z \right) \, e^{\, - \lambda \hspace{1pt} X}
\, c \Ra
\, = \,
\nn && \
\, = \,
\La e^{\, \overline{\lambda} \hspace{1pt} X^*} \,
Y \left(\raisebox{9pt}{\hspace{-2pt}}\right.
\pi_{z^*} \left( j_W \right)^{-1} \cnj{a},\, z^*
\left.\raisebox{9pt}{\hspace{-2pt}}\right) \,
e^{\, - \overline{\lambda} \hspace{1pt} X^*} \, b \Vl c \Ra
\, = \,
\nn && \
\, = \,
\La Y \left(\raisebox{9pt}{\hspace{-2pt}}\right.
\pi_{z^*} \left(\raisebox{9pt}{\hspace{-2pt}}\right.
e^{\, \overline{\lambda} \hspace{1pt} X^*}
\left.\raisebox{9pt}{\hspace{-2pt}}\right)
\pi_{z^*} \left( j_W \right)^{-1} \, \cnj{a},\,
e^{\, \overline{\lambda} \hspace{1pt} X^*} \! \left( z^* \right)
\left.\raisebox{9pt}{\hspace{-2pt}}\right) \, b \Vl c \Ra
\, . \
\eeqa
On the other hand:
\beqa\label{eq8.19o12}
&
\La b \Vl
Y \left(\raisebox{9pt}{\hspace{-2pt}}\right.
\pi_z \! \left(\raisebox{9pt}{\hspace{-2pt}}\right.
e^{\, \lambda \hspace{1pt} X}
\left.\raisebox{9pt}{\hspace{-2pt}}\right)
a,\, e^{\, \lambda \hspace{1pt} X} \! \left( z \right)
\left.\raisebox{9pt}{\hspace{-2pt}}\right)
\, c \Ra
\, = \, &
\nn  & \, = \,
\La Y \left(\raisebox{9pt}{\hspace{-2pt}}\right.
\pi_{e^{\, \overline{\lambda} \hspace{1pt} X^*} \left( z^* \right)}
\left( j_W \right)^{-1}
\cnj{\left(\raisebox{9pt}{\hspace{-2pt}}\right.
\pi_z \left(\raisebox{9pt}{\hspace{-2pt}}\right.
e^{\, \lambda \hspace{1pt} X}
\left.\raisebox{9pt}{\hspace{-2pt}}\right) \, a
\left.\raisebox{9pt}{\hspace{-2pt}}\right)},\,
e^{\, \overline{\lambda} \hspace{1pt} X^*} \! \left( z^* \right)
\left.\raisebox{9pt}{\hspace{-2pt}}\right) \, b \Vl c \Ra
\, . \
&
\eeqa
Since the Hermitean form is nondegenerate, comparing the right hand sides
of the last equations and replacing
\(g = e^{\, \lambda \hspace{1pt} X}\) we obtain
\beq\label{eq8.20o12}
\cnj{\left( \pi_z \left( g \right) \, a \right)} =
\pi_{g \left( z \right)^*} \left( j_W \right) \,
\pi_{z^*} \left( g^* \right) \,
\pi_{z^*} \left( j_W \right)^{-1} \, \cnj{a}
\, , \
\eeq
which actually proves (\ref{eq8.16o10}) if we further use the cocycle
property~(\ref{eq6.10o1}) and Eq.~(\ref{eqn-new}).

Eq.(\ref{eqn-new3}) is a corollary of Eq.~(\ref{eq8.16o10}),
the conjugation law \(j_W^* = j_W^{-1}\)
(see Appendix~A: Eq.~(\ref{eqA.4}) and the relation between $h$ and $j_W$) and
the cocycle property~(\ref{eq6.10o1}).\qed
\end{proof}

\msection{Connection with Globally Conformal Invariant QFT}{sec:9}

In this section we will construct a connection between
vertex algebras and Wightman QFT with GCI in higher dimensions.
For this purpose we start by introducing some additional notations.
We set $D$--dimensional Minkowski space $M$ to have a signature
$\left( D-1,1 \right)$ so that its scalar product is:
\beq\label{eq9.1}
x \spr y \, := \, - x^0 \, y^0 + \vec{x} \spr \vec{y} \, \equiv \,
- x^0 \, y^0 + x^1 \, y^1 + \cdots + x^{D-1} \, y^{D-1}
\
\eeq
for
\(x = \left( x^0,\, \vec{x} \right) \equiv
\left( x^0,\, x^1,\, \dots,\, x^{D-1} \right) \in \R^D\),
etc. and \(x^{\, 2} \equiv x \spr x\).
The complex Minkowski space $M_{\C}$ is just \(M + i M \cong \C^D\) with
the complexified scalar product~(\ref{eq9.1}).
For simplicity of notation we will use the same sign ``$\cdot$''
for two different scalar products (metrics) in $\C^D$:
the first is Euclidean one~(\ref{eqnew1.1}) used with variables
labeled by letters as $z$ and $w$; the second is the above Minkowski
metric used with variables labeled by letters as $x$, $y$
(in the real case) and $\zeta$ (in the complex case).
We will identify this metrics via isomorphism:
\beq\label{eq9.2a}
\isom \, : \, \left( \vec{z},\, z^D \right) \, \longmapsto \,
\left( -i\,z^D,\, \vec{z} \right)
\, , \quad
\vec{z} \, := \, \left( z^1,\, \dots z^{D-1} \right)
\, . \
\eeq

\renewcommand{\thesection}{\arabic{section}}
\subsection{Analytic picture on the compactified Minkowski space}
\renewcommand{\thesection}{\arabic{section}.}\label{subsec:9.1}
The connection between the structures of the vertex algebras and QFT
is based on the following coordinate transformation:
\beq\label{eq9.2}
\C^D \, \in \, z \, = \, \left( \vec{z},\, z^D \right)
\, \longleftrightarrow \,
\zeta \, = \, \left( \zeta^0,\, \vec{\zeta} \right)
\, \in \, \C^D
\eeq
where
\(\vec{\zeta} := \left( \zeta^1,\, \dots,\, \zeta^{D-1} \right)\) and
\beqa\label{eq9.3}
&&
\zeta^0 \, = \,
\frac{i}{2} \ \, \frac{1-z^2}{\frac{\textstyle 1+z^2}{\textstyle 2} + z^D}
\ , \quad \ \,
\vec{\zeta} \, = \, \frac{1}{\frac{\textstyle 1+z^2}{\textstyle 2} + z^D} \ \,
\vec{z}
\ , \quad \ \,
\\ \label{eq9.4} &&
\vec{z} \, = \,
\frac{1}{\frac{\textstyle 1+\zeta^2}{\textstyle 2} - i\,\zeta^0} \ \, \vec{\zeta}
\ , \quad \ \,
z^D \, = \,
\frac{1}{2} \ \, \frac{1-\zeta^2}{\frac{\textstyle 1+\zeta^2}{\textstyle 2} - i\,\zeta^0}
\ . \quad \ \,
\eeqa
We will regard the coordinates $\zeta$ and $z$ as two different
charts on the complex \textit{compactified Minkowski space} $\M_{\C}$.
They are particular cases of the system of charts considered in
Appendix~A.
The map
\beq\label{eq9.5}
h \, :\, \zeta \, \mapsto \, z ,\quad
\, \
\eeq
is invertible rational transformation of $\C^D$.
We will use the same letter $h$ also for the transformation
$\isom \circ h$
which can be
represented by an element of the connected complex conformal group
\(\mathfrak{C}_{\C} \equiv \mathit{Spin}_0 \left( D+2;\,\C \right)\) of $\C^D$
as it is derived in Appendix~A.
Other properties of the above transformations are:

{\it \begin{plist}
\item \hspace{-1pt}
\(h^2 = j_W^{-1}\) (where $j_W$ is defined by Eq.~(\ref{eq7.4o1})).
\item \hspace{-1pt}
If \(\zeta \hspace{-1pt}\leftrightarrow\hspace{-1pt} z\)
under the correspondences~(\ref{eq9.3})
and~(\ref{eq9.4}) then
\(\overline{\zeta} \hspace{-1pt}\leftrightarrow\hspace{-1pt} z^*\),
where $\overline{\zeta}$ is the standard (coordinate) complex conjugation
and $z^*$ is the conjugation given by Eq.~(\ref{eq7.23o3}).
\item \hspace{-1pt}
The transformation $h$ is regular on $M$ and maps it on a precompact
subset of $\C^D$:
\beqa\label{eq9.6}
&
\overline{h \left( M \right)} \, = \,
\M \, := \,
\left\{ z \in \C^D : z^* = z \right\} \, = \,
\left\{\raisebox{10pt}{\hspace{-2pt}}\right.
z \in \C^D : z = e^{i \, \vartheta} u,\,
& \nn &
\vartheta \in \R,\, u \in \R^D,\, u^{\, 2} = 1
\left.\raisebox{10pt}{\hspace{-2pt}}\right\}
\, \cong \,
\Sr^1 \hspace{-1pt} \times \Sr^{D-1}
\hspace{-2pt}\raisebox{-1pt}{$\left/\raisebox{9pt}{}\right.$}
\raisebox{-4pt}{$\Z_2$}
\, . & \qquad
\eeqa
\item \hspace{-1pt}
Let
\beq\label{eq9.7}
\mathfrak{T}_{\pm}
\, = \,
\left\{\raisebox{12pt}{\hspace{-2pt}}\right.
\mzeta \, = \, x + i \, y \, \in \, \C^D \, : \
\pm y^0 > \left| \vec{y} \right| :=
\left(\raisebox{11pt}{\hspace{-2pt}}\right.
\Su_{i \, = \, 1}^{D-1} \left( y^i \right)^2
\left.\raisebox{11pt}{\hspace{-2pt}}\right)^{\frac{1}{2}}
\left.\raisebox{12pt}{\hspace{-2pt}}\right\}
\,
\eeq
be the {\rm forward} and {\rm backward tube} in $\M_{\C}$. Then $h$ is regular on
$\mathfrak{T}_+$ and
\beq\label{eq9.8}
h \left( \mathfrak{T}_+ \right) \hspace{-1pt} = \hspace{-1pt}
T_+ \hspace{-1pt} := \hspace{-1pt}
\left\{\raisebox{12pt}{\hspace{-3pt}}\right. z \in \mathbb{C}^D
\hspace{-2pt} : \hspace{2pt}
\left| \hspace{1pt} z^{\, 2} \right| \hspace{-1pt} < \hspace{-1pt} 1,\
z \cdot \overline{z} =
\left| \hspace{1pt} z^1 \right|^2 \hspace{-2pt} + \dots +
\left| \hspace{1pt} z^D	 \right|^2
\hspace{-2pt} < \hspace{-1pt}
\frac{1 + \left| \hspace{1pt} z^{\, 2} \right|^2}{2}
\left.\raisebox{12pt}{\hspace{-3pt}}\right\}
\hspace{-2pt} .
\eeq
(Note that
\(\mathcal{B}_{\frac{1}{2}} \varsubsetneqq T_+ \varsubsetneqq \mathcal{B}_1\),
where
$\mathcal{B}_{\lambda}$ is the Hermitean ball
\(\left\{ z : z \spr \overline{z} < \lambda \right\}\).)
\end{plist}}
These statements follow by the considerations made in Appendix~A
and by a straightforward computation.
We will only make a comment about the derivation of Eq.~(\ref{eq9.8}).
It can be obtained using the transformation properties of the
interval \(\left( \zeta - \overline{\zeta} \right)^2\)
under the conformal transformation $h$
(note that \(\zeta \in \mathfrak{T}_+ \cup \mathfrak{T}_-\) iff
\(\left( \zeta - \overline{\zeta} \right)^2 > 0\),
see also \cite{NT 01} Eq. (A.6)).
The regularity of $h$ on $\mathfrak{T}_{\pm}$ follows by
the boundedness of the transformed interval
\(\left( h \left( \zeta \right) - h \left( \overline{\zeta} \right) \right)^2\).

We will use the \textit{active} point of view for the conformal group action.
This means that the group
\(\mathfrak{C}_{\C}\)
will be assumed to act on the points of the compactified Minkowski space $\M_{\C}$,
i.~e. the conformal transformations will not be considered as coordinate
changes.
Then every conformal transformation will have different
\textit{coordinate expressions} in different charts of $\M_{\C}$.
In particular, for the above two charts
the corresponding coordinate expressions for an element \(g \in \mathfrak{C}_{\C}\)
will be:
\beq\label{eq9.7a2}
\zeta \, \mapsto \, g^{\left( M \right)} \left( \zeta \right)
\quad \mathrm{and} \quad
z \, \mapsto \, g \left( z \right)
\, \quad (g \, \in \, \mathfrak{C}_{\C})
\, . \
\eeq
These two coordinate actions are connected by the conjugation
\(g \hspace{-1pt} \left( z \right) \hspace{-1pt} \text{\small $=$} \hspace{-2pt}
\mathop{h}\limits^{-1} \hspace{-3pt}
\left( g^{\left( \hspace{-1pt} M \hspace{-1pt} \right)} \hspace{-2pt}
\left( h \hspace{-1pt} \left( z \right) \right)
\hspace{-1pt} \right)\hspace{-1pt}\).
When there is no danger of confusion we will just write
$g \left( \zeta \right)$ for $g^{\left( M \right)} \left( \zeta \right)$.
Under these conventions we have one more property:
{\it \begin{plist}
\setcounter{tmpc}{4}
\item \hspace{-1pt}
There are two natural real forms
for the complex conformal group $\mathit{Spin} \left( D+2;\, \C \right)$
defined by the conjugations
\beqa\label{eq9.9}
&&
g \mapsto \overline{g} \, , \quad
\overline{g} \left( z \right) \, := \, \overline{g \left( \overline{z} \right)}
\, , \quad
\\ \label{eq9.10} &&
g \mapsto g^* \, , \quad
g^* \left( z \right) \, := \, g \left( z^* \right)^*
\, \quad
\eeqa
(\(\overline{g_1\, g_2} = \overline{g_1} \, \overline{g_2}\) and
\(\left( g_1 \, g_2 \right)^* = g_1^* \, g_2^*\)) and the corresponding
real subgroups are
the Euclidean conformal group $\mathit{Spin} \left( D+1,\, 1 \right)$
and the Minkowski conformal group $\mathit{Spin} \left( D,\, 2 \right)$,
respectively.
\end{plist}}

\renewcommand{\thesection}{\arabic{section}}
\subsection{From GCI~QFT to vertex algebras}
\renewcommand{\thesection}{\arabic{section}.}\label{subsec:9.2}
We continue with the construction of the passage from the Wightman QFT with GCI
to vertex algebras with Hermitean positive definite structure.

It is convenient to define the axiomatic QFT with GCI as a bilinear
map \(\left( f,\, a \right) \mapsto \phi \left[ f,\, a \right]\)
which gives an operator for every complex Schwartz test function
\(f \in \mathfrak{S} \left( M \right)\) on Minkowski space
and a vector $a$ from a complex vector space $F$.
The operator $\phi \left[ f,\, a \right]$ is supposed to act on
a dense invariant domain $\DOM$ in a Hilbert space $\HS$.
The axiomatic assumptions which we impose to the fields
$\phi \left[ f,\, a \right]$ are the Wightman axioms
(\cite{Jost 65}, Chapt.~III)
in which
Lorentz covariance
is replaced
by the GCI condition for the correlation functions.
In more details, $\phi \left[ f,\, a\right]$ is assumed to be
\textit{nonzero} operator valued distribution for any fixed nonzero \(a \in F\)
formally written as:
\beq\label{eq9.11}
\phi \left[ f,\, a \right] \, = \,
\mathop{\int}\limits_{\hspace{-7pt} M} \hspace{-1pt}
\phi \left( x,\, a \right) \, f \left( x \right) \, d^D x
\, . \
\eeq
The Hermitean conjugation of the fields requires the existence of a conjugation
\(a \mapsto a^+\) on $F$ such that for all \(a \in F\) and
\(\Psi_1,\, \Psi_2 \in \DOM\):
\beq\label{eq9.13a}
\La \Psi_1 \Vl \phi \left[ f,\, a \right] \Psi_2 \Ra \, = \,
\La \phi \left[ \overline{f},\, a^+ \right] \Psi_1 \Vl \Psi_2 \Ra
\, . \
\eeq
We
demand
the existence of a unitary representation of
the Poincar\'e translations which leave invariant the domain $\DOM$
and a fixed element \(\rvac \in \DOM\) called \textit{vacuum}
so that they satisfy the covariance axiom, the spectral condition
and the uniqueness of the vacuum.
The locality axiom requires that the space $F$ is $\Z_2$--graded,
\(F = F_0 \oplus F_1\),
so that \(F_{0,\, 1}^+ = F_{0,\, 1}\) and
\beq\label{eq9.12}
\phi \left( x_1,\, a_1 \right) \phi \left( x_2,\, a_2 \right)
- \left( -1 \right)^{p_1 \hspace{1pt} p_2} \,
\phi \left( x_2,\, a_2 \right) \phi \left( x_1,\, a_1 \right) \, = \, 0
\, \
\eeq
if \(x_{12}^{\, 2} \equiv \left( x_1 - x_2 \right)^2 > 0\) and
\(a_k \in F_{p_k}\) (\(k = 0,\, 1\)).
The condition of GCI is imposed on the correlation functions
\(\lvac \phi \left( x_1,\, a_1 \right) \dots \phi \left( x_n,\, a_n \right)
\rvac\) and it supposes first that there exists a cocycle
$\pi_x^M \left( g \right)$:
a rational function in \(x \in M\) for fixed
\(g \in
\mathfrak{C}_{\C}\)
with values in  $\left( \mathit{End} \, F\right)_0$
(i.~e. a ratio of a polynomial in $x$
whose coefficients are even endomorphisms of $F$ and a complex polynomial),
regular in the domain of $g$ and satisfying the cocycle
property~(\ref{eq6.10o1}) and triviality for Poincar\'{e} translations,
\(\pi_x^M \left( \tau_a \right) = \ID_F\)
(\(\tau_a \left( x \right) := x+a\));
then the correlation functions should be invariant under the substitution
\beq\label{eq9.13}
\phi \left( x,\, a \right) \, \mapsto \,
\phi \left( g \left( x \right),\, \pi_x^M \left( g \right) \, a \right)
\, \
\eeq
in the sense of \cite{NT 01}, Sect.~2.
The cocycle should be consistent with the conjugation on $F$
in the sense that:
\beq\label{eq9.15a}
\left[\raisebox{9pt}{\hspace{-2pt}}\right.
\pi_{\zeta} \left( g \right) a
\left.\raisebox{9pt}{\hspace{-2pt}}\right]^+
\, = \,
\pi_{\overline{\zeta}}^M \left( g^* \right) a^+
\, . \
\eeq
Note that from the triviality of $\pi_x^M \left( \tau_a \right)$
it follows that $\pi_x^M \left( g \right)$ does not depend
on $x$ for the transformations $g$ belonging to
the complex \textit{Weyl} group, the complex Poincar\'{e} group
with dilations, and this is an action of this group on $F$.
Since the linear space $F$ can be infinite dimensional we impose
an additional condition on the cocycle $\pi^M$: the above representation
of the complex Weyl group is supposed to be decomposable into a
direct sum of finite dimensional irreducible representations.
Finally, the axiom of completeness is naturally generalized.
This completes our characterization of axiomatic QFT with GCI.

Under this assumptions the result about the analytic continuation
of the vector--valued distribution
\(\phi \left( x_1,\, a_1 \right) \dots \phi \left( x_n,\, a_n \right)
\rvac\) (\cite{Jost 65} IV.2) comes true:
they are boundary value of the functions
\beq\label{eq9.14}
\Phi_n \left( \zeta_1,\, a_1; \dots;\, \zeta_n,\, a_n \right) \, \in \,
\DOM
\, , \
\eeq
analytic in the \textit{tube} domain
\beq\label{eq9.15}
\mathcal{T}_n \, := \,
\left\{
\left( \zeta_1,\, \dots,\, \zeta_n \right) \, : \,
\zeta_{k+1 \hspace{1pt} k} \in \mathfrak{T}_+
\ \mathrm{for} \ k \, = \, 1,\, \dots,\, n-1
,\, \zeta_n \in \mathfrak{T}_+
\right\}
\, \
\eeq
(\(\zeta_{k+1 \hspace{1pt} k} \, := \, \zeta_{k+1} - \zeta_k\)).

Direct generalizations of the Theorems~3.1 and~4.1
of~\cite{NT 01} lead to rationality of the
Wightman functions:
\beq\label{eq9.16}
\lvac \, \phi \left( x_1,\, a_1 \right) \dots \phi \left( x_n,\, a_n \right)
\, \rvac \, = \,
\frac{\mathcal{P} \left( x_1,\, \dots,\, x_n;\, a_1,\, \dots,\, a_n \right)}{
\mathop{\prod}\limits_{1 \, \leqslant \, k \, < \, l \, \leqslant \, n}
\left( x_{kl}^{\, 2} + i \, 0 \, x_{kl}^0 \right)^{\mu \left( a_k,\, a_l \right)}}
\ , \
\eeq
where $\mathcal{P} \left( x_1,\, \dots,\, x_n;\, a_1,\, \dots,\, a_n \right)$
are polynomials and $\mu \left( a,\, b \right)$ is such a function that:
\beq\label{eq9.17}
\left[ \left( x-y \right)^2 \right]^{\mu \left( a,\, b \right)}
\left[\raisebox{10pt}{}
\phi \left( x,\, a \right) \phi \left( y,\, b \right)
- \left( -1 \right)^{p \hspace{1pt} q} \,
\phi \left( y,\, b \right) \phi \left( x,\, a \right) \right] \, = \, 0
\, \
\eeq
for all \(x,\, y \in M\) (in the sense of distributions) and all \(a \in F_p\)
and \(b \in F_q\) (\(p,\, q \, = \, 0,\, 1\)).
It is natural to expect that the rationality of the correlation functions
will imply stronger analytic properties for the analytic vector
functions (\ref{eq9.14}):

\begin{mtheorem}\label{th:9.1}
Let the system of fields \(\phi \left( x,\, a \right)\)
satisfies the above conditions.
Then the analytic vector--valued functions (\ref{eq9.14})
possess a continuation which we will denote again by
\(\Phi_n \left( \zeta_1,\, a_1; \dots;\, \zeta_n,\, a_n \right)\)
such that it is holomorphic in the domain of all sets
\(\left( \zeta_1,\, \dots ,\, \zeta_n \right)\)
of {\rm mutually nonisotropic}
points of the forward tube $\mathfrak{T}_+\,$.
The functions
\(
\left(\raisebox{10pt}{\hspace{-2pt}}\right.
\mathop{\prod}\limits_{k \, < \, l}
\left( \zeta_{kl}^{\, 2} \right)^{\mu \left( a_k,\, a_l \right)}
\left.\raisebox{10pt}{\hspace{-2pt}}\right)
\Phi_n \left( \zeta_1,\, a_1; \dots;\, \zeta_n,\, a_n \right)\)
possess analytic continuation in the domain of all
\(\zeta_1,\, \dots,\, \zeta_n \in \mathfrak{T}_+\).
\end{mtheorem}

We will prove this theorem in Appendix~B.

\Vspa

A straightforward corollary of Theorem~\ref{th:9.1} is that the real
connected conformal group $\mathit{Spin}_0 \left( D,\, 2 \right)$
has a unitary representation $U \left( g \right)$ on $\HS$ such that
\beqa\label{eq9.18}
\hspace{-20pt}
U \left( g \right)
\Phi_n \left( \zeta_1, a_1;\, \dots;\, \zeta_n, a_n \right)
\hspace{-1pt} = \hspace{-1pt}
\Phi_n \left( g \left( \zeta_1 \right),
\pi_{\zeta_1}^M \left( g \right) a_1; \dots;
g \left( \zeta_n \right),
\pi_{\zeta_n}^M \left( g \right) a_n \right)
\
\eeqa
for all (Minkowski) \textit{real} conformal transformations
\(g \in \mathfrak{C} := \mathit{Spin}_0 \left( D,\, 2 \right)\),
all \(\zeta_1,\) \(\dots ,\) \(\zeta_n \in \mathfrak{T}_+\)
for which \(\zeta_{kl}^{\, 2} \neq 0\) for \(k <l \),
and for all vectors
\(a_1 \dots a_n \in F\).
Indeed, Eq.~(\ref{eq9.18}) determines $U \left( g\right)$ as a preserving
norm map on a dense subspace of $\HS$, in accord with the GCI.
Note also that $\pi_{\zeta} \left( g \right)$ is defined (regular) for
\(g \in \mathit{Spin}_0 \left( D,\, 2 \right)\)
and \(g \left( \zeta \right) \in \mathfrak{T}_+\) since $\mathfrak{T}_+$
is a \textit{homogeneous space} for the real conformal group~\cite{Uhl 63}.

As a corollary of Theorem~\ref{th:9.1} the operator--valued
generalized functions
\\ ${}$ \hfill
\(
\left(\raisebox{10pt}{\hspace{-2pt}}\right.
\mathop{\prod}\limits_{1 \, \leqslant k \, < \, l \, \leqslant n}
\left( \zeta_{kl}^{\, 2} \right)^{\mu \left( a_k,\, a_l \right)}
\left.\raisebox{10pt}{\hspace{-2pt}}\right)
\mathop{\prod}\limits_{m \, = \, 1}^n
\phi \left( x_m,\, a_m \right)
\mgvspc{15pt}\mgvspc{-3pt}\) \hfill ${}$ \\
together with all their derivatives in \(x_1,\, \dots,\, x_n\)
are defined (and regular) for coinciding arguments \(x_1 = \dots = x_n =: x\)
and are again operator distributions in $x$ acting on a
common dense invariant domain
\(\widetilde{\DOM} \supseteq \DOM\).
Denote the linear span of all these operator functions $\psi \left( x \right)$
together with the constant function
\(\ID \left( x \right) \equiv \ID_{\HS}\)
by $\mathcal{A}$.
The space $\mathcal{A}$ has natural $\Z_2$--grading induced by those of
the fields $\phi \left( a,\, x \right)$ (i.~e. the $\Z_2$--grading of $F$).
The construction of the operator distributions
\(\psi \left( x \right) \in \mathcal{A}\)
is done via the correlation functions~(\ref{eq9.16}) so we conclude
that every correlator
\(\lvac \psi_1 \left( x_1 \right) \dots \psi_n \left( x_n \right) \rvac\),
for \(\psi_1,\, \dots,\, \psi_n \in \mathcal{A}\),
is rational and $\Z_2$--symmetric in the sense of Eq.~(\ref{eq9.16})
and the vector distribution
\(\psi_1 \left( x_1 \right) \dots \psi_n \left( x_n \right) \rvac\)
is a boundary value of an analytic function
\beq\label{eq9.22a2}
\Yy^M_n \left( \psi_1,\, \zeta_1;\, \dots;\, \psi_n,\, \zeta_n \right) \,
(\, \equiv \,
\psi_1 \left( \zeta_1 \right) \dots \psi_n \left( \zeta_n \right) \rvac)
\, \
\eeq
satisfying Theorem~\ref{th:9.1}.
Note that
\beqa\label{eq9.25a1}
&
\La \Yy^M_n \left( \psi_1^+,\, \overline{\zeta_1};\, \dots;\,
\psi_n^+,\, \overline{\zeta_n} \right)
\Vl
\Yy^M_m \left( \psi_1',\, \zeta_1';\, \dots;\, \psi_m',\, \zeta_m' \right)
\Ra
\, = & \nn & = \,
\lvac
\Yy^M_{n+m} \left( \psi_1,\, \zeta_1;\, \dots;\, \psi_n,\, \zeta_n;\,
\psi_1',\, \zeta_1';\, \dots;\, \psi_m',\, \zeta_m' \right)
\Ra
&
\eeqa
(the scalar product of the vacuum with $\Yy^M_{n+m}$)
where \(\psi \mapsto \psi^+\) is the conjugation induced on $\mathcal{A}$
by the Hermitean conjugation of \(\psi \left( x \right) \in \mathcal{A}\).

Next observe that the cocycle $\pi^M$ on $F$ gives rise to a cocycle on $\mathcal{A}$,
which we denote again by $\pi^M$, such that it is a continuation of the initial one
under the natural inclusion of $F$ in~$\mathcal{A}$ and
\beq\label{eq9.22a1}
U \left( g \right) \psi \left( x \right) U \left( g \right)^{-1}
\, = \,
\left( \pi^M_x \left( g \right) \psi \right) \left( x \right)
\, \
\eeq
for \(g \in \mathfrak{C}\) and \(x \in M\), regular for $g$
(i.~e., \(g \left( x \right) \in M\)).
In particular,
\beq\label{eq9.23a1}
U \left( g \right) \phi \left( x,\, a \right) U \left( g \right)^{-1}
\, = \,
\phi \left( x,\, \pi^M_x \left( g \right) \hspace{1pt} a \right)
\, \quad (a \, \in \, F,\quad g \left( x \right) \, \in \, M)
\, . \
\eeq
Eqs.~(\ref{eq9.22a1}) and~(\ref{eq9.23a1}) are understood in the sense of distributions
in the domain \(\left\{\raisebox{9pt}{\hspace{-2pt}}\right. x:\)
\(g \left( x \right) \in M \left.\raisebox{9pt}{\hspace{-2pt}}\right\}\).
The sketch of the arguments is the following:
first observe that the action of the Weyl group on $\mathcal{A}$,
which is naturally generated by those on $F$,
is again decomposable into a direct sum of
finite dimensional irreducible representations
(since it is representable by sum of tensor products of such representations).
The energy positivity (see the argument before Proposition~\ref{pr:7.1})
allows then to define an extension of the cocycle $\pi^M$ on $\mathcal{A}$
so that every element \(\psi \in \mathcal{A}\) belongs to
a finite dimensional $\pi^M$--invariant subspace (see Lemma~\ref{lm:7:2o12}).
This makes sensible the action of $\pi_x^M \left( g \right)$ in
Eqs.~(\ref{eq9.22a1}) and~(\ref{eq9.23a1}) as an action of multiplicators.
After this precision we can prove~(\ref{eq9.23a1})
from~(\ref{eq9.18}) and~(\ref{eq9.22a1}) is then obtained.
Note also that
for all \(g \in \mathfrak{C}\) we have
a similar equation as~(\ref{eq9.17}) for
\(\Yy^M_n \left( \psi_1,\, \zeta_1;\, \dots;\, \psi_n,\, \zeta_n \right)\).

The heuristic connection between the system of fields $\mathcal{A}$
and the vertex operators \(Y \left( a,\, z \right)\) which we are
going to introduce is
\beq\label{eq9.27a1}
Y \left( a,\, z \right) \, = \,
\left( \pi_{e \left( z \right)}^M \left( h \right) \psi \right)
\left( h^{-1} \left( z \right) \right)
\, \
\eeq
where $h$ is the complex conformal transformation introduced in Sect.~9.1.
In more details, let us first define the transformed vector functions
\beqa\label{eq9.28a1}
& \hspace{-5pt}
\Yy_n \left( \psi_1,\, z_1;\, \dots;\, \psi_n,\, z_n \right)
:= &
\nn & \hspace{-5pt}  :=
\Yy^M_n \left(\raisebox{9pt}{\hspace{-2pt}}\right.
\pi_{e \left( z \right)}^M
\left( h^{-1} \right) \hspace{1pt} \psi_1,\, h^{-1} \left( z_1 \right);\, \dots;\,
\pi_{e \left( z \right)}^M
\left( h^{-1} \right) \hspace{1pt} \psi_n,\, h^{-1} \left( z_n \right)
\left.\raisebox{9pt}{\hspace{-2pt}}\right)
\, , & \qquad
\eeqa
such
that the products
\(
\left(\raisebox{10pt}{\hspace{-2pt}}\right.
\mathop{\prod}\limits_{k \, < \, l}
\left( z_{kl}^{\, 2} \right)^{\mu_{kl}}
\left.\raisebox{10pt}{\hspace{-2pt}}\right)
\Yy_n \left( z_1,\, a_1; \dots;\, z_n,\, a_n \right)\)
are analytic in
\(\left(\raisebox{8pt}{\hspace{-2pt}}\right. z_1,\)
$\dots,$  \(z_n \left.\raisebox{8pt}{\hspace{-2pt}}\right)\) $\in$ $T_+^n$
(see Eq.~(\ref{eq9.8})) for sufficiently large \(\mu_{kl} \in \N\)
(\(1 \leqslant k < l \leqslant n\)) and they are covariant in the sense:
\beqa\label{eq9.29a1}
\hspace{-16pt}
U \left( g \right)
\Yy_n \left( \psi_1, z_1; \dots; \psi_n, z_n \right)
\hspace{-1pt} = \hspace{-1pt}
\Yy_n
\left( \pi_{z_1} \hspace{-2pt} \left( g \right) \psi_1,
g \left( z_1 \right); \dots;
\pi_{z_1} \hspace{-2pt} \left( g \right) \psi_n, g \left( z_n \right)
\right)
\hspace{-1pt} , \
\eeqa
where
\beq\label{eq9.30a1}
\pi_z \left( g \right) \, := \,
\pi_{e \left( z \right)}^M \left( h \, g \, h^{-1} \right)
\, \
\eeq
is a rational cocycle on $\mathcal{A}$ in the $z$--chart
as those of Sect.~\ref{sec:7}.
Note that both cocycles $\pi^M_{\zeta} \left( g \right)$ and
$\pi_z \left( g \right)$ are constant
(not depending on $\zeta$ and $z$, resp.) for different subgroups
of $\mathfrak{C}_{\C}$:
the complex Weyl group of Minkowski space and the complex Euclidean
group with dilations for $z$--chart, respectively.
This is because we consider the conformal group action in an active sense.
In particular, the translations $\tau_a$ and $t_a$ for both charts
(\(\tau_a^{\left( M \right)} \left( \zeta \right) := \zeta+a\) and
\(t_a \left( z \right) := z+a\))
are different, as in the first case these are the Poincar\'{e} translations
while in the second case these translations are generated by the $\Trn$'s
of Sect.~\ref{sec:7}.

Now we are ready to define the vertex algebra structure that arises
from the considered GCI~QFT.
Set the vertex algebra space $\VA$ to be the image of the linear map:
\beq\label{eq9.31a1}
\mathcal{A} \, \ni \,\psi \, \mapsto \,
\Yy_1 \left( \psi,\, z \right) \vrestr{10pt}{z \, = \, 0} \, \in \, \HS
\, . \
\eeq
In fact, this map is injective as a corollary
of the separating property of the vacuum
(indeed, the translation covariance in~(\ref{eq9.29a1})
will lead to cancelation of
the whole function $\Yy_1 \left( \psi,\, z \right)$).
So we will further \textbf{identify} the space $\mathcal{A}$ by $\VA$
under this injection, thus transferring the $\Z_2$--grading and the above
cocycle $\pi_z \left( g \right)$ on $\VA$.
Note that the space $\VA$ is dense in the Hilbert space $\HS$
which is a consequence of the completeness axiom.
The vertex algebra vacuum $\varvac$ we set to be the QFT vacuum $\rvac$
(note that \(\Yy_1 \left( \ID,\, z \right) \equiv \rvac\)).
The action of the conformal Lie algebra $\mathfrak{c}_{\C}$
(which include the action of the translation endomorphisms $\Trn_{\mu}$)
is the derivative action of that of~(\ref{eq9.29a1})
(since $\mathcal{A}$ include all the derivative fields
we will not thus leave the space $\VA$).
The Hermitean form on $\VA$ is the restricted Hilbert scalar product.
Finally, the state--field correspondence is defined as
\beq\label{eq9.32a1}
Y \left( a,\, z \right) \, b \, := \,
\left( z^{\, 2} \right)^{-N}
\iota
\left(\raisebox{10pt}{\vspace{-4pt}}\right.
\left[\raisebox{9pt}{\vspace{-1pt}}\right.
\left( z-w \right)^2 \left.\raisebox{9pt}{\vspace{-1pt}}\right]^N
\, \Yy_2 \left( a,\, z;\,b,\, w \right)
\vrestr{10pt}{w \, = \, 0}
\left.\raisebox{10pt}{\vspace{-4pt}}\right)
\, \
\eeq
for \(N \mgrt 0\), where $\iota$ stands for the Taylor expansion at $0$ of
the resulting analytic function in $z$.
The coefficients in the formal series~(\ref{eq9.32a1}) belong to $\VA$
since the derivatives of the regularized products
\(\left( x_{12}^{\, 2} \right)^N
\psi_1 \left( x_1 \right) \psi_2 \left( x_2 \right)\) (for \(N \mgrt 0\))
of fields \(\psi_1,\, \psi_2 \in \mathcal{A}\), at coinciding arguments
\(x_1=x_2\), again belong to $\mathcal{A}$.

\begin{mtheorem}\label{th:9.2}
The above defined $\VA$ is a vertex algebra with Hermitean structure.
The space $\VA$ is dense in the Hilbert space $\HS$ and
coincides with the {\rm finite conformal energy} space:
the linear span of all eigenvectors of the conformal Hamiltonian $H$
(having a discrete spectrum).
\end{mtheorem}

We will sketch the \textit{proof} only for
some of the properties which one has to verify.
First, one can find by a straightforward computation
that under the transformation~(\ref{eq9.28a1})
the conversed Eq.~(\ref{eq9.25a1}) reads as
\beqa\label{eq9.33a2}
&
\La \Yy_n \left(\raisebox{9pt}{\hspace{-2pt}}\right.
\pi_{z_1^*} \hspace{-2pt} \left( j_W \right)^{-1} a_1^+,\, z_1^*
;\, \dots;\,
\pi_{z_n^*} \hspace{-2pt} \left( j_W \right)^{-1} a_n^+,\, z_n^*
\left.\raisebox{9pt}{\hspace{-2pt}}\right)
\Vl
\Yy_m \left( b_1,\, w_1;\, \dots;\, b_m,\, w_m \right)
\Ra
\, = & \nn & = \,
\La  \varvac \Vl
\Yy_{n+m} \left( a_1,\, z_1;\, \dots;\, a_n,\, z_n;\,
b_1,\, w_1;\, \dots;\, b_m,\, w_m \right)
\Ra
\,
&
\eeqa
(one uses Eqs.~(\ref{eq9.15a}) and~(\ref{eqA.4}), the cocycle property
as well as the properties (a) and (b) of the previous subsection).
As a corollary of this equation and Eq.~(\ref{eq9.32a1}) we then obtain
that Eq.~(\ref{eq7.19}) from the definition of the
Hermitean structure on vertex algebra is satisfied.

Next one can derive the equality
\beq\label{eq9.33a1}
\hspace{0pt}
Y \left( a_1, z_1 \right) \dots Y \left( a_n, z_n \right) \varvac
=
\iota_{z_1,\, \dots,\, z_n} \left( \rho_n^{-N} \right)
\, \iota
\left( \rho_n^N \, \Yy_n \left( a_1, z_1;\,\dots;\, a_n, z_n \right)
\right)
\hspace{0pt}
\eeq
for \(N \mgrt 0\),
where
\(
\rho_n :=
\mathop{\prod}\limits_{1 \, \leqslant \, k \, < \, l \, \leqslant \, n}
z_{kl}^{\, 2}
\)
and the second \(\iota\) stand for the Taylor expansion of the
resulting analytic function in \(z_1,\, \dots,\, z_n\) at
\(\left( 0,\, \dots,\, 0 \right)\).
For this purpose, first we note that the left hand side in~(\ref{eq9.33a1})
belongs to
\(\VA \Bbrk{z_1}_{\,\lscw{z}{\, 2}{1}}
\dots \, \Bbrk{z_n}_{\,\lscw{z}{\, 2}{n}}\) which follows inductively in $n$.
Then Eq.~(\ref{eq9.33a1}) is proven by induction in $n$ starting from the
definition equality~(\ref{eq9.32a1}) and for the inductive step using
Eq.~(\ref{eq9.33a2}).

From Eq.~(\ref{eq9.33a1}) it follows
the locality (Definition~\ref{def:2.1}~(\textit{b})).
The covariance properties of Definition~\ref{def:7.1o3}
(Eqs.~(\ref{eq6.5o})--(\ref{eq6.7o}) together with
Definition~\ref{def:2.1}~(\textit{c})) follows from
the conformal covariance law~(\ref{eq9.29a1}).
Finally, we point out that the condition~(\textit{d})
of Definition~\ref{def:7.1o3} is a
consequence of the rationality of the cocycle $\pi_z \left( g \right)$.
This condition also imply the second statement in the Theorem~\ref{th:9.2}.

\renewcommand{\thesection}{\arabic{section}}
\subsection{The converse passage}
\renewcommand{\thesection}{\arabic{section}.}\label{subsec:9.3}
Here we start by a vertex algebra $\VA$ with a strongly positive
Hermitean structure.
Thus $\VA$ is a prehilbert space and denote by $\HS$ its Hilbert
completion.
We are going to construct a GCI~QFT on $\HS$.

The first step is to reconstruct the functions
$\Yy_n \left( a_1,\, z_1;\, \dots;\, a_n,\, z_n \right)$
as an analytic in the domain of
\(\left( z_1,\, \dots ,\, z_n \right) \in T_+^n\).
Here they are defined by the sum of the formal series
\(
\rho_n^N
\left(\raisebox{10pt}{\hspace{-2pt}}\right.
\mathop{\prod}\limits_{m \, = \, 1}^n
Y \left( a_m,\, z_m \right)
\left.\raisebox{10pt}{\hspace{-2pt}}\right) \varvac
\) for \(N \mgrt 0\),
where
\(\rho_n =
\mathop{\prod}\limits_{1 \, \leqslant k \, < \, l \, \leqslant n} z_{kl}^{\, 2}\),
divided by
\(\rho_n^{-N}\).
The last formal series converges due to Fact~\ref{ft:1.3}
since its Hilbert norm converges to a polynomial
in accord with Proposition~\ref{pr:6.1o}.

Next one proves Eq.~(\ref{eq9.33a2}) as a consequence of
the iterated Eq.~(\ref{eq7.19}) in Definition~\ref{def7.2o4}.

Then one can define a unitary representation
of the real conformal group $\mathfrak{C}$ on $\HS$ via
Eq.~(\ref{eq9.29a1}) as the right hand side does not
change its Hilbert norm due to the GCI~(\ref{eq6.15o1}).

Now we have to make the converse passage in Eq.~(\ref{eq9.28a1})
and obtain the functions
\(\mathcal{Y}_n^M \left( a_1,\, \zeta_1;\,\dots;\, a_n,\, \zeta_n \right)\)
analytic in the same domain as those of previous subsection.
They can be proven to satisfy again Eq.~(\ref{eq9.25a1})
and the corresponding conformal covariance law
with a cocycle $\pi^M_x \left( g \right)$ on $\VA$ given by
the converse equality of Eq.~(\ref{eq9.30a1}).

Finally we define a system of local fields
\(\left\{ \fy \left( a,\, x \right) : a \in \VA \right\}\)
using the generalized vector functions obtained by the limit
\beq\label{eq9.35a1}
\hspace{0pt}
\mathcal{Y}_n^M \left( a_1,\,  \zeta_1;\, \dots;\, a_n,\,  \zeta_n \right)
\, \to \,
\fy \left( a_1,\,  x_1 \right) \dots
\fy \left( a_n,\,  x_n \right) \rvac
\, \
\eeq
(\(\rvac := \varvac\)), for:
\(\left( \zeta_1, \dots , \zeta_n \right) \in \mathcal{T}_+\)
(Eq.~(\ref{eq9.15})),
\(\mathit{Re} \, \zeta_k = x_k\),
\(\mathit{Im} \, \zeta_{l+1 \, l} \to 0\)
(\(k = 1,\, \dots ,\, n\) and \(l = 1,\, \dots ,\, n-1\))
and \(\mathit{Im} \, \zeta_n \to 0\).
In order to express the action of $\fy \left( a,\, x \right)$
by the vector distributions \(\fy \left( a_1,\,  x_1 \right) \dots
\fy \left( a_n,\,  x_n \right) \rvac\) one should use
Eq. (\ref{eq9.25a1}).

Summarizing the above construction we have found:

\begin{mtheorem}\label{th:9.3}
The above defined system of fields
\(\left\{ \fy \left( a,\, x \right) : a \in \VA \right\}\)
with the cocycle $\pi^M_x \left( g \right)$
satisfies the axioms of the GCI~QFT.
\end{mtheorem}

\msection{Outlook}{sec:10}

In conclusion, we shall discuss some possibilities
for finding new models for the vertex algebras in higher dimensions.
They are based on the construction of vertex algebra from Lie
algebras of formal distributions explained in Sect.~\ref{sec:5}.
A~straightforward generalization of the commutation relations~(\ref{eq5.2})
is presented by the so called Lie field models.
In our notations these models should be defined by the relations:
\beqa\label{eq10.1}
\left[ \hspace{2pt}
u^{\alpha} \left( z_1 \right), u^{\beta} \left( z_2 \right)
\hspace{1pt} \right]
\, = && \hspace{-2pt}
\mathop{\sum}\limits_{\gamma \, = \, 1}^A \hspace{5pt}
\mathop{\int}\limits_{\hspace{-7pt}\overline{M}} \hspace{-1pt}
D_{\alpha\beta\gamma} \left( z_1, z_2,\, z_3 \right) \,
u^{\gamma} \left( z_3 \right) \, d^D \hspace{-1pt} z_3 \, + \,
F_{\alpha\beta} \left( z_1,\, z_2 \right) \, \ID
\nn
D_{\alpha\beta\gamma} \left( z_1, z_2,\, z_3 \right) \, = && \hspace{-2pt}
\iota_{z_3,\, z_1,\, z_2} \hspace{-2pt}
\left(\raisebox{10pt}{\hspace{-2pt}}\right.
W_{\alpha\beta\gamma}^{\left( 3 \right)} \left( z_{12}, z_{23} \right)
\left.\raisebox{10pt}{\hspace{-2pt}}\right)  \, - \,
\iota_{z_3,\, z_2,\, z_1} \hspace{-2pt}
\left(\raisebox{10pt}{\hspace{-2pt}}\right.
W_{\alpha\beta\gamma}^{\left( 3 \right)} \left( z_{12}, z_{23} \right)
\left.\raisebox{10pt}{\hspace{-2pt}}\right)
\nn
F_{\alpha\beta} \left( z_1, z_2 \right) \, = && \hspace{-2pt}
\iota_{z_1,\, z_2} \hspace{-2pt}
\left(\raisebox{10pt}{\hspace{-2pt}}\right.
W_{\alpha\beta}^{\left( 2 \right)} \left( z_{12} \right)
\left.\raisebox{10pt}{\hspace{-2pt}}\right) \, - \,
\iota_{z_2,\, z_1} \hspace{-2pt}
\left(\raisebox{10pt}{\hspace{-2pt}}\right.
W_{\alpha\beta}^{\left( 2 \right)} \left( z_{12} \right)
\left.\raisebox{10pt}{\hspace{-2pt}}\right)
\, , \ \mgvspc{16pt}
\eeqa
where \(W_{\alpha\beta}^{\left( 2 \right)} \left( z,\, w \right),\,
W_{\alpha\beta\gamma}^{\left( 3 \right)} \left( z,\, w \right)
\in \C \Brk{z,w}_{\,
\raisebox{-2pt}{\small $z^{\, 2} w^{\, 2} \left( z+w \right)^2$}}\),\gvspc{-5pt}
\(z_{kl} = z_k - z_l\,\), and the (complex) measure $d^D z$ on $\M$ is provided
by the restriction of the volume form \(d z^1 \wedge \dots \wedge d z^D\) on
the real submanifold $\M$~(\ref{eq9.6}).
The integral of Eq.~(\ref{eq10.1}) can be understood by the substitution
\beq\label{eq10.2}
u^{\gamma} \left( z_3 \right) \, = \,
e^{\, z_{23} \spr \di_{z_2}} \, u^{\gamma} \left( z_2 \right)
\, \
\eeq
and by formal integration of the kernel
$D_{\alpha\beta\gamma} \left( z_{12},\, z_{23} \right)$
with \(e^{\, z_{23} \spr \di_{z_2}} \, dz_{23}\) as a series
belonging to
\(\C \Bbrk{z_{12}}_{\,\lscw{z}{\, 2}{12}} \Bbrk{\di_{z_2}}\).
Sum  of integrals of this type
should appear in any GCI~QFT
with possibly infinite number $A$ of fields.
It was shown by Robinson \cite{Rob 65} that there are no nonfree
(i.~e. with \(D_{\alpha\beta\gamma} \left( z,w \right) \neq 0\)
for some $\alpha$, $\beta$ and $\gamma$) such \textit{scalar} models
(i.~e. \(A=1\)) in space-time dimension
higher than~2.
For nonscalar but finite component models it is not clear
whether there exists a solution of the Jacobi identity for the
commutators in (\ref{eq10.1}).
From the point of view of the GCI the interesting solutions
are those for which
the functions $W_{\alpha\beta}^{\left( 2 \right)}$ and
$W_{\alpha\beta\gamma}^{\left( 3 \right)}$ are
conformally invariant (finite component) $2$- and $3$--point functions.
They would provide a conformal and Hermitean structure on the
vertex algebra of the type considered in Sects.~\ref{sec:7}
and \ref{sec:8}.
For example there are nontrivial candidates for a tree point
function of a \textit{nonabelian gauge field} $F_{\mu\nu}^a$
displayed in \cite{NST 03} and it is an interesting question
whether they yield the Jacobi identity for the corresponding commutators.

\begin{acknowledgements}
The author thanks Bojko Bakalov, Petko Nikolov,
Ventseslav Rizov and Ivan Todorov for useful discussions.
The author acknowledges the hospitality of the Erwin
Schr\"odinger International Institute for Mathematical Physics (ESI)
where this work was conceived
as well as
partial support by the Bulgarian National Council for Scientific Research under
contract F-828,
by the Research Training Network within the Framework Programme 5 of the European
Commission under contract HPRN-CT-2002-00325.
\end{acknowledgements}

\setcounter{section}{0}

\setcounter{equation}{0}
\Apsection{Affine system of charts on complex compactified Minkowski space}{app:A}
In this section we will follow the notation of \cite{NT 01}, Sect.~2 and Appendix~A
but our considerations will be basically done over the complex field $\C$.

Recall that as an algebraic variety,
the complex $D$--dimensional compactified Min{\-}kow{\-}ski space $\M_{\C}$
is defined as a complex $D$--dimensional projective nondegenerated quadric.
The real compactified Minkowski space $\M$ is then characterized by a
conjugation \(\M_{\C} \ni p \mapsto p^* \in \M_{\C}\) as
\(\M \equiv \left\{ p \in \M_{\C} : p = p^* \right\}\) such that
the corresponding real restriction of the complex quadric has a signature
\(\left( D,\, 2 \right)\).

The manifold $\M_{\C}$ is a homogeneous space of the connected complex conformal
group \(\mathfrak{C}_{\C} \equiv \mathit{Spin}_0 \left( D+2;\, \C \right)\)
which acts on it via its orthogonal action on the quadric.
The stabilizer $\mathfrak{C}_{\C,\, p}$ of a point \(p \in \M_{\C}\)
is isomorphic to the complex spinor Weyl group: the complex spinor Poincar\'{e}
group with dilations.
The action of $\mathfrak{C}_{\C,\, p}$ leave invariant two natural
subsets of $\M_{\C}$:
the set $K_{\C,\, p}$ of all mutually isotropic points of $\M_{\C}$
with respect to $p$
(recall that the isotropy relation is conformally invariant and coincides on
the projective quadric with the orthogonality relation of the rays);
the other invariant subspace is the complement
\(M_{\C,\, p} := \M_{\C} \backslash K_{\C,\, p}\).
The set $M_{\C,\, p}$ is open and dense in $\M_{\C}$ and the action of
$\mathfrak{C}_{\C,\, p}$ induces on it a structure of affine space with
a conformal class of flat metric.
Thus we obtain a complex affine atlas
$\left\{ M_{\C,\, p} : p \in \M_{\C} \right\}$
on $\M_{\C}$ indexed by its points.

To illustrate the above construction note that
a particular case of affine chart of $\M_{\C}$ is the injection
of the complex Minkowski space $M_{\C}$ in $\M_{\C}$,
\(M_{\C} \hookrightarrow \M_{\C}\).
In this case \(M_{\C} \equiv M_{\C,\, p_{\infty}}\) for a \textit{special} point
\(p_{\infty} \in \M_{\C}\): the tip of the cone
\(K_{\infty} \equiv K_{\C,\, p_{\infty}}\) of ``infinite'' points.
The complex spinor Weyl group of the Minkowski space $M_{\C}$ coincides then with
$\mathfrak{C}_{\C,\, p_{\infty}}$.
Every other stabilizer $\mathfrak{C}_{\C,\, p}$ is conjugated to
$\mathfrak{C}_{\C,\, p_{\infty}}$:
\(\mathfrak{C}_{\C,\, p} = g \, \mathfrak{C}_{\C,\, p_{\infty}} \, g^{-1}\)
if \(g \left( p_{\infty} \right) = p\).
Thus \(M_{\C,\, p} = g \left( M_{\C} \right)\) which transfers the affine
structure from $M_{\C}$ to $M_{\C,\, p}$.

Fixing a point \(q \in M_{\C,\, p}\) it becomes a vector space (with center~$q$).
In the case of the (complex) Minkowski space $M_{\C}$ the center
is denoted by $p_0$ (note that $p_0$ and $p_{\infty}$ are real points).
Thus every pair \(p,\, q \in \M_{\C}\) of mutually nonisotropic points
determines a vector space included in $\M_{\C}$ as a dense open subset.
Since there always exists a transformation \(g \in \mathfrak{C}_{\C}\)
such that \(g \left( p_{\infty} \right) = p\) and \(g \left( p_0 \right) = q\)
(\cite{NT 01} Proposition~1.1) then the map
\(g : M_{\C} \to g \left( M_{\C} \right) \equiv M_{\C,\, p}\) will be an
isomorphism of the corresponding vector spaces.

In the projective description of $\M_{\C}$ a vector chart $M_{\C,\, p}$
with center $q$ is determined by two representatives
\(\VEC{\eta}_{\infty},\, \VEC{\eta}_0 \in \C^{D+2}\),
\(p = \left\{\raisebox{9pt}{\hspace{-3pt}}\right.
\lambda \, \VEC{\eta}_{\infty}
\left.\raisebox{9pt}{\hspace{-3pt}}\right\}\),
\(q = \left\{\raisebox{9pt}{\hspace{-3pt}}\right.
\lambda \, \VEC{\eta}_0
\left.\raisebox{9pt}{\hspace{-3pt}}\right\}\)
(\(\VEC{\eta}_{\infty}^{\, 2} = \VEC{\eta}_0^{\, 2} = 0\)
as in \cite{NT 01} Eq.~(A.1))
with fixed mutual normalization \(\VEC{\eta}_{\infty} \spr \VEC{\eta}_0 = 1\).
Then the orthogonal complement
\(\left\{\raisebox{9pt}{\hspace{-2pt}}\right.
\VEC{\eta}_{\infty},\, \VEC{\eta}_0
\left.\raisebox{9pt}{\hspace{-2pt}}\right\}^{\perp} \cong \C^D\)
(\(\mathit{Span} \left\{\raisebox{9pt}{\hspace{-2pt}}\right.
\VEC{\eta}_{\infty},\, \VEC{\eta}_0
\left.\raisebox{9pt}{\hspace{-2pt}}\right\}\) has
nondegenerated metric) plays the role of vector space of the chart,
\(M_{\C,\, p} \cong \left\{\raisebox{9pt}{\hspace{-2pt}}\right.
\VEC{\eta}_{\infty},\, \VEC{\eta}_0
\left.\raisebox{9pt}{\hspace{-2pt}}\right\}^{\perp}\).
A particular case of the last correspondence is the \textit{Klein--Dirac}
compactification formulae \cite{NT 01} Eq.~(A.2) where
\(\VEC{\eta}_{\infty} \equiv \VEC{\xi}_{\infty}\) and
\(\VEC{\eta}_0 \equiv \VEC{\xi}_0\) so that
\beqa\label{eqA.1}
&
\left\{\raisebox{9pt}{\hspace{-2pt}}\right.
\VEC{\xi}_{\infty},\, \VEC{\xi}_0
\left.\raisebox{9pt}{\hspace{-2pt}}\right\}^{\perp}
\, \ni \, \mathop{\sum}\limits_{\mu \, = \, 0}^{D-1}
\, \zeta^{\mu} \, \VEC{e}_{\mu} \, \mapsto \,
\left\{\raisebox{9pt}{\hspace{-3pt}}\right.
\lambda \, \VEC{\xi}_{\zeta}
\left.\raisebox{9pt}{\hspace{-3pt}}\right\} \, \in \,
M_{\C}
\, ,
& \nn &
\VEC{\xi}_{\zeta} \, := \,
\VEC{\xi}_0
\, - \, \frac{\raisebox{2pt}{$\zeta^{\, 2}$}}{\raisebox{-2pt}{$2$}}
\ \VEC{\xi}_{\infty}
\, + \, \mathop{\sum}\limits_{\mu \, = \, 0}^{D-1}
\, \zeta^{\mu} \, \VEC{e}_{\mu}
\, , &
\eeqa
where the expressions for \(\VEC{\xi}_{\infty}\) and \(\VEC{\xi}_0\)
can be found in Appendix~A of \cite{NT 01}.
In the general case every vector
\(\VEC{v} \in \left\{\raisebox{9pt}{\hspace{-2pt}}\right.
\VEC{\eta}_{\infty},\, \VEC{\eta}_0
\left.\raisebox{9pt}{\hspace{-2pt}}\right\}^{\perp}\)
is mapped to
the point \(\left\{\raisebox{9pt}{\hspace{-3pt}}\right.
\lambda \, \VEC{\eta}_{\VEC{v}}
\left.\raisebox{9pt}{\hspace{-3pt}}\right\} \in M_{\C,\, p}\)
determined by the representative
\beq\label{eqA.2}
\VEC{\eta}_{\VEC{v}} \, := \,
\VEC{\eta}_0
\, - \, \frac{\VEC{v}^{\, 2}}{2}
\ \VEC{\eta}_{\infty}
\, + \, \VEC{v}
\, . \
\eeq

There is a special kind of affine complex charts $M_{\C,\, p}$ determined by the
condition that they cover the whole real compact Minkowski space,
i.~e. \(\M \subset M_{\C,\, p}\):

\begin{mproposition}\label{pr:A.1n}
\(\M \subset M_{\C,\, p}\) \ iff \
\(p \in \mathfrak{T}_+ \cup \mathfrak{T}_- \subset M_{\C}\) \
($\mathfrak{T}_{\pm}$ are defined in Eq.~(\ref{eq9.7})).
\end{mproposition}

\begin{proof}
First, \(q \in \M\) iff \(q = q^*\).
On the after hand, let \(q = q^*\)
and \(p = \left\{\raisebox{9pt}{\hspace{-3pt}}\right.
\lambda \, \VEC{\eta}
\left.\raisebox{9pt}{\hspace{-3pt}}\right\}\),
\(q = \left\{\raisebox{9pt}{\hspace{-3pt}}\right.
\lambda \, \VEC{\theta}
\left.\raisebox{9pt}{\hspace{-3pt}}\right\}\) as
\(\VEC{\theta} = \VEC{\theta}{\raisebox{10pt}{\hspace{1pt}}}^*\)
(\(\VEC{\theta} \in \R^{D+2}\)). Then
\(q \in \M_{\C,\, p}\) iff \(\VEC{\eta} \spr \VEC{\theta} \neq 0\),
which is also equivalent to
\(\VEC{\eta} \spr \VEC{\theta} \neq 0\) and
\(\VEC{\eta}{\raisebox{9pt}{\hspace{1pt}}}^* \spr \VEC{\theta} \neq 0\).
Thus \(q \in M_{\C,\, p}\) iff
\(\VEC{\theta} \notin \mathit{Re} \, \left\{\raisebox{9pt}{\hspace{-2pt}}\right.
\VEC{\eta},\,
\VEC{\eta}{\raisebox{9pt}{\hspace{1pt}}}^*
\left.\raisebox{9pt}{\hspace{-2pt}}\right\}^{\perp}
\equiv
\left\{\raisebox{9pt}{\hspace{-2pt}}\right.
\mathit{Re} \, \VEC{\eta},\,
\mathit{Im} \, \VEC{\eta}
\left.\raisebox{9pt}{\hspace{-2pt}}\right\}^{\perp}_{\R}
\)
(the last one stands for orthogonal complement in $\R^{D+2}$).
Therefore,
\(\M \subset M_{\C,\, p}\) iff
the space
\(\left\{\raisebox{9pt}{\hspace{-2pt}}\right.
\mathit{Re} \, \VEC{\eta},\,
\mathit{Im} \, \VEC{\eta}
\left.\raisebox{9pt}{\hspace{-2pt}}\right\}^{\perp}_{\R}\)
has definite restriction of the metric
(i.~e. it does not contain isotropic vectors).
Since we have chosen the signature of $\R^{D+2}$ to be \(\left( D,\, 2 \right)\)
we conclude that the space
\(\mathit{Span}_{\R} \, \left\{\raisebox{9pt}{\hspace{-2pt}}\right.
\mathit{Re} \, \VEC{\eta},\,
\mathit{Im} \, \VEC{\eta}
\left.\raisebox{9pt}{\hspace{-2pt}}\right\}\) should be of signature
\(\left( 0,\, 2 \right)\).
In particular,
\(\VEC{\eta} \spr \VEC{\xi}_{\infty} \neq 0\), i.~e.
\(p \in M_{\C} \equiv M_{\C,\, p_{\infty}}\), and if we set
\(\VEC{\eta} := \VEC{\xi}_{\zeta}\),
accordingly to Eq.~(\ref{eqA.1}), then we derive that
\(0 > \left(\raisebox{9pt}{\hspace{-3pt}}\right.
\mathit{Im} \, \VEC{\xi}_{\zeta}
\left.\raisebox{9pt}{\hspace{-3pt}}\right)^2 =
- 4
\left(\raisebox{9pt}{\hspace{-3pt}}\right.
\VEC{\xi}_{\zeta} \hspace{-1pt} - \hspace{-1pt}
\VEC{\xi}_{\zeta}{\raisebox{10pt}{\hspace{-2pt}}}^*
\left.\raisebox{9pt}{\hspace{-3pt}}\right)^2 =
2\,
\left(\raisebox{9pt}{\hspace{-3pt}}\right.
\mathit{Im} \, \zeta
\left.\raisebox{9pt}{\hspace{-3pt}}\right)^2\).
But the last means that \(\zeta \in \mathfrak{T}_+ \cup \mathfrak{T}_-\).\QED
\end{proof}

\begin{mremark}\label{rm:A.1}
If we start with the real form on the quadric $\M_{\C}$ of signature
\(\left( r+1, s+1 \right)\) with \(s \neq 1\) and \(r \neq 1\)
then it follows from the above proof that there does not exist
affine chart containing the whole corresponding real quadric.
\end{mremark}

Since all points of $\mathfrak{T}_+$ as well as $\mathfrak{T}_-$
lies in single orbits under the action of the real conformal group
\(\mathfrak{C}\) (\cite{Uhl 63})
then the charts in Proposition~\ref{pr:A.1n} are conjugated
one to another
in both cases of \(p \in \mathfrak{T}_+\) and \(p \in \mathfrak{T}_-\).
Besides that, we prefer the case when \(p \in \mathfrak{T}_-\) because
\(p \notin M_{\C,\, p}\).
Thus up to real conformal transformation we can choose $p$ to be
the fixed point \(-i\, e_0 = \left( -i,\, \vec{0} \right)\).
The center of the chart is convenient to be the conjugated point
\(p^* \in \mathfrak{T}_+\) since then the conjugation becomes simpler
in the coordinates of the chart.
The vector space of the chart is then isomorphic to
\(\left\{\raisebox{9pt}{\hspace{-2pt}}\right.
\VEC{\xi}_{i\, e_0},\,
\VEC{\xi}_{-i\, e_0}
\left.\raisebox{9pt}{\hspace{-2pt}}\right\}^{\perp}\) which coincides
with
\(\mathit{Span} \, \left\{\raisebox{9pt}{\hspace{-2pt}}\right.
\VEC{e}_1,\, \dots,\, \VEC{e}_D
\left.\raisebox{9pt}{\hspace{-2pt}}\right\}\)
in the notations of \cite{NT 01} Eq.~(A.1).
The real part of the last space is Euclidean so we denote this
chart by \(E_{\C} \hookrightarrow \M_{\C}\), \(E_{\C} \cong \C^D\).
The Eq.~(\ref{eqA.2}) takes then the form similar to Eq.~(\ref{eqA.1}):
\beqa\label{eqA.3}
\C^D
\, \ni \, z \, \mapsto \,
\left\{\raisebox{9pt}{\hspace{-3pt}}\right.
\lambda \, \VEC{\eta}_z
\left.\raisebox{9pt}{\hspace{-3pt}}\right\} \, \in \,
M_{\C}
\, , \quad
\VEC{\eta}_z \, := \,
\VEC{\eta}_0
\, - \, \frac{z^{\, 2}}{2}
\ \VEC{\eta}_{\infty}
\, + \, \mathop{\sum}\limits_{\mu \, = \, 1}^{D}
\, z^{\mu} \, \VEC{e}_{\mu}
\, ,
\eeqa
where \(\VEC{\eta}_0 := \frac{\textstyle 1}{\textstyle 2}
\left(\raisebox{9pt}{\hspace{-3pt}}\right.
\VEC{e}_{-1} + i \VEC{e}_0
\left.\raisebox{9pt}{\hspace{-3pt}}\right) =
\frac{\textstyle 1}{\textstyle 2} \, \VEC{\xi}_{i \, e_0}
\)
and \(\VEC{\eta}_{\infty} :=
- \VEC{e}_{-1} + i \VEC{e}_0
= - 2 \, \VEC{\eta}_0{\raisebox{10pt}{\hspace{-2pt}}}^*\).
The relations~(\ref{eq9.3}) and~(\ref{eq9.4}), giving the connection
between the coordinates in the charts $M_{\C}$ and $E_{\C}$,
are derived by the equation
\beq\label{eqnAnew1}
\VEC{\xi}_{\zeta} \, \sim \, \VEC{\eta}_{z} \, . \
\eeq
Note that the transformation $h$, which acts on
$\C^{D+2}$ as an Euclidean rotation in the plain
\(\mathit{Span} \left\{\raisebox{9pt}{\hspace{-3pt}}\right.
i \VEC{e}_0,\, \VEC{e}_D
\left.\raisebox{9pt}{\hspace{-3pt}}\right\}\) of an angle
$\frac{\textstyle \pi}{\textstyle 2}$, conjugates the charts
$M_{\C}$ and $E_{\C}$
since
\beq\label{eqnAneq2}
h \, \VEC{\eta}_z \, = \, \VEC{\xi}_{\isom \left( z \right)}
\, \
\eeq
($\isom$ is defined by Eq.~(\ref{eq9.2a})).
This is exactly the transformation~(\ref{eq9.5})
and its square acting as an Euclidean rotation in the plain
\(\mathit{Span} \left\{\raisebox{9pt}{\hspace{-3pt}}\right.
i \VEC{e}_0,\, \VEC{e}_D
\left.\raisebox{9pt}{\hspace{-3pt}}\right\}\) of an angle
$\pi$ is the Weyl reflection $j_W^{-1}$~(\ref{eq7.6o11})
(here the axis of $i \VEC{e}_0$ corresponds to the axis ``$D+1$''
in the notations of Sect.~\ref{sec:7}).
Note also that
\beq\label{eqA.4}
h^* \, = \, h^{-1}
\, . \
\eeq

We conclude this geometric review with 
a remark about the projective interpretation of the
tube domains $\mathfrak{T}_{\pm}$~(\ref{eq9.7}) of the complex
Minkowski space $M_{\C}$.
First observe that for a complex vector \(\VEC{\eta} \in \C^{D+2}\)
we have: \(\VEC{\eta}^{\, 2} = 0\) iff 
\(
\left(\raisebox{9pt}{\hspace{-3pt}}\right.
\mathit{Re} \, \VEC{\eta}
\left.\raisebox{9pt}{\hspace{-3pt}}\right)^2 =
\left(\raisebox{9pt}{\hspace{-3pt}}\right.
\mathit{Im} \, \VEC{\eta}
\left.\raisebox{9pt}{\hspace{-3pt}}\right)^2
\) and
\(
\left(\raisebox{9pt}{\hspace{-3pt}}\right.
\mathit{Re} \, \VEC{\eta}
\left.\raisebox{9pt}{\hspace{-3pt}}\right) \spr
\left(\raisebox{9pt}{\hspace{-3pt}}\right.
\mathit{Im} \, \VEC{\eta}
\left.\raisebox{9pt}{\hspace{-3pt}}\right) = 0
\).
Therefore, there are three natural subsets of rays
\(\left\{\raisebox{9pt}{\hspace{-3pt}}\right.
\lambda \, \VEC{\eta}
\left.\raisebox{9pt}{\hspace{-3pt}}\right\} \in \M_{\C}\),
invariant with respect to the real conformal group $\mathfrak{C}$:
\begin{plist}
\item[(\textit{a})]
\(\left(\raisebox{9pt}{\hspace{-3pt}}\right.
\mathit{Re} \, \VEC{\eta}
\left.\raisebox{9pt}{\hspace{-3pt}}\right)^2 =
\left(\raisebox{9pt}{\hspace{-3pt}}\right.
\mathit{Im} \, \VEC{\eta}
\left.\raisebox{9pt}{\hspace{-3pt}}\right)^2 > 0\),
\item[(\textit{b})]
\(\left(\raisebox{9pt}{\hspace{-3pt}}\right.
\mathit{Re} \, \VEC{\eta}
\left.\raisebox{9pt}{\hspace{-3pt}}\right)^2 =
\left(\raisebox{9pt}{\hspace{-3pt}}\right.
\mathit{Im} \, \VEC{\eta}
\left.\raisebox{9pt}{\hspace{-3pt}}\right)^2 < 0\),
\item[(\textit{c})]
\(\left(\raisebox{9pt}{\hspace{-3pt}}\right.
\mathit{Re} \, \VEC{\eta}
\left.\raisebox{9pt}{\hspace{-3pt}}\right)^2 =
\left(\raisebox{9pt}{\hspace{-3pt}}\right.
\mathit{Im} \, \VEC{\eta}
\left.\raisebox{9pt}{\hspace{-3pt}}\right)^2 = 0\).
\end{plist}
In the selected signature $\left( D,2 \right)$ of the real
quadric $\M$, the case (\textit{b}) corresponds to the
union \(\mathfrak{T}_+ \cup \mathfrak{T}_-\):
this follows from the equality:
\(
\left(\raisebox{9pt}{\hspace{-3pt}}\right.
\mathit{Im} \, \VEC{\xi}_{\zeta}
\left.\raisebox{9pt}{\hspace{-3pt}}\right)^2 =
2\,
\left(\raisebox{9pt}{\hspace{-3pt}}\right.
\mathit{Im} \, \zeta
\left.\raisebox{9pt}{\hspace{-3pt}}\right)^2
\), for \(\zeta \in M_{C}\), 
already used in the proof of Proposition~\ref{pr:A.1n}.
The separation between $\mathfrak{T}_+$ and $\mathfrak{T}_-$ in the subspace
(\textit{b}) of rays, corresponds to the \textit{orientation} of the
pair \(\left(\raisebox{9pt}{\hspace{-3pt}}\right.
\mathit{Re} \, \VEC{\eta},\, \mathit{Im} \, \VEC{\eta}
\left.\raisebox{9pt}{\hspace{-3pt}}\right)\).

\setcounter{equation}{0}
\Apsection{Proof of Theorem~\ref{th:9.1}}{app:B}
To prove the theorem \ref{th:9.1} we need some technical preparation.

\begin{mproposition}\label{pr:A.1}
Let $\Theta_0 \left( \zeta_1,\, \dots,\, \zeta_n \right)$
be a continuous function with values in a Hilbert
space ${\cal H}$ and defined in an open domain
$U_0$ in $M_{\C}^{n}\,$.
Let $U_0$ be contained in a connected domain
$U$ in $M_{\C}^{n}\,$, such that the scalar product
\beq\label{2.2.22}
F_0 \left( \zeta_1,\, \dots,\, \zeta_{2n} \right) \, := \,
\La
\Theta_0\left(\overline{\zeta_1},\, \dots,\,\overline{\zeta_n}\right)
\Vl \Theta_0 \left( \zeta_{n+1},\, \dots,\, \zeta_{2n} \right) \Ra
\ , \qquad
\eeq
(\(\overline{\zeta} \equiv \overline{x+iy} = x-iy\) for \(x,\, y \in M\))
has a continuation to a strongly analytic function
$F \left( \zeta_1,\, \dots,\, \zeta_{2n} \right)$
in the domain
$U^* \!\times U$ in $M_{\C}^{\times\, 2n}\,$,
where
$U^*$ is the set of complex conjugate
elements of~$U\,$.
Then there exists a single-valued continuation
$\Theta \left( \zeta_1,\, \dots,\, \zeta_n \right)$ of
$\Theta_0 \left( \zeta_1,\, \dots,\, \zeta_n \right)\,$,
to strongly analytic function
(i.~e. analytic in norm)
on the domain $U\,$,
with values in $\HS$ such that
\beq\label{2.2.23}
F \left( \zeta_1,\, \dots,\, \zeta_{2n} \right) \, = \, \La
\Theta\left(\overline{\zeta_1},\, \dots,\,\overline{\zeta_n}\right)
\Vl \Theta \left( \zeta_{n+1},\, \dots,\, \zeta_{2n} \right) \Ra
\eeq
in $U^* \!\times U\,$.
\end{mproposition}

To prove this proposition it is useful the following
simple fact:

\begin{mfact}\label{ft:1.3}
Let \(\LB \Psi_n \RB_{n \, = \, 1}^{\infty}
\subset \HS\,\), then if the sequence
$\LB\La \Psi_m \Vl \Psi_n \Ra\RB_{m,\, n \, = \, 1}^{\infty}$
is Cauchy fundamental then the sequence
$\LB \Psi_n \RB_{n \, = \, 1}^{\infty}$
is fundamental in the norm of $\HS$ too.
\end{mfact}

\noindent
This \textit{follows} from
\(\left\| \Psi_m \hspace{-1pt} - \hspace{-1pt} \Psi_n \right\|^2
\hspace{-1pt} = \hspace{-1pt}
\La \Psi_m \Vl \Theta_m \Ra \hspace{-1pt} + \hspace{-1pt} \La \Psi_n \Vl \Psi_n \Ra
\hspace{-1pt} - \hspace{-1pt} 2 \La \Psi_m \Vl \Psi_n \Ra
\hspace{-1pt} \mathop{\longrightarrow}
\limits_{m,\hspace{1pt}n \to \infty} \hspace{-1pt} 0\).
Continuing with the \textit{proof} of Proposition~\ref{pr:A.1}
we first observe that
the vector--valued function
$\Theta_0 \left( \zeta_1,\, \dots,\, \zeta_n \right)$
is strongly analytical in the domain $U_0\,$,
which follows from the analyticity of the scalar product
(\ref{2.2.22}) and the above Fact~\ref{ft:1.3}.
Let \(\left( \zeta_1,\, \dots,\, \zeta_n \right) \in U\,\).
The domain $U$ is connected and consequently there exists
piecewise linear path in $U$ connecting
$\left( \zeta_1,\, \dots,\, \zeta_n \right)$ with $U_0\,$.
Then for every interval
\(\left( \zeta_1 \left( t \right),\, \dots,\,
\zeta_n \left( t \right) \right)\,\), \( t \in
\left[ \, 0 \, , \, 1 \, \right]\) we can
make the requested continuation by the following
fact:

\vspace{0.08in}

\noindent
\textit{If} $\Psi_0 \left( t \right)$
\textit{is a strongly analytic function with values in}
$\HS$ \textit{defined in a neighborhood of} \(t = 0 \in \C\)
\textit{and such that the scalar product}
\(h_0 \left( t_1,\, t_2 \right) \hspace{-1pt} := \hspace{-1pt}
\La \hspace{-1pt}\Psi_0 \left( \hspace{1pt}\overline{t_1}\hspace{1pt} \right)
\hspace{-2pt}\Vl\hspace{-1pt}
\Psi_0 \left( t_2 \right) \hspace{-2pt}\Ra\)
\textit{possess an analytic continuation}
\(h \left( t_1,\, t_2 \right)\)
\textit{in} $\left( t_1,\, t_2 \right)$
\textit{in some neighborhood of}
\(\left[ \, 0 \, , 1 \, \right]^{\times 2} \subset \C^{\, 2}\,\),
\textit{then the function} $\Psi_0 \left( t \right)$
\textit{can be continued in a neighborhood of}
$\left[ \, 0 \, , \, 1 \, \right]$ \textit{in} $\C$
\textit{to strongly analytic function} $\Psi \left( x \right)\,$,
\textit{for which}
\(h \left( t_1,\, t_2 \right) =
\La \Psi \left( \,\overline{t_1}\, \right) \Vl
\Psi \left( t_2 \right) \Ra\,\).

\vspace{0.08in}

\noindent
To prove the last statement
we first note that there exists a positive number
$\rho$ such that if \(D_{\rho} \subset \C\) stands for the open
complex disk with centre at \(0 \in \C\) and radius $\rho\,$, then
the set
\(\left[ \, 0 \, , \, 1 \right]^{\times 2} \! +
D_{\rho}^{\times 2} =
\left\{\mgvspc{10pt}\hspace{-2pt}\right. u+v : u \in
\left[ \, 0 \, , \, 1 \right]^{\times 2},\
v \in D_{\rho}^{\times 2}
\left.\mgvspc{10pt}\hspace{-2pt}\right\}\)
is contained in the analyticity domain of $h \left( t_1,\, t_2 \right)\,$.
Thus, if for every
\(t \in \left[ \, 0 \, , \, 1 \, \right]\,\),
$J_t$ stands for the interval in $\R$ with centre $t$ and radius $\rho\,$,
then $J_t \times J_t$ lies in the convergence domain of the Taylor
expansion of the function
$h \left( t_1,\, t_2 \right)$ around $\left( t,\, t \right)\,$.
Since the length of the intervals $J_t$ is constant, then there exist
a finite number $J_{t_k}\,$,
for \(k = 0,\, \dots,\, n\) such that
\(0 = t_0 < \dots < t_k < \dots < t_n = 1\) and
\(t_{k+1} \in J_{t_k}\) (\(0 \leqslant k < n\,\)).
In such a way if the requested analytic expansion $\Psi \left( t \right)$
is done in a neighborhood of
$\left[ \, 0 \, , \, t_k \, \right]$ (\(0 \leqslant k \leqslant n \,\)),
then the Taylor expansion of $\Psi \left( t \right)$ with centre
$t_k$ will be convergent in $J_k$~--~by the construction of $J_k$
and Fact~\ref{ft:1.3}.
Thus we arrive to $t_{k+1}$ and by induction, to
\(t_n = 1\,\).

The single-vauledness of the obtained continuation
\(\Theta \left( \zeta_1,\, \dots,\, \zeta_n \right)\)
follows by Eq. (\ref{2.2.23}).
Indeed, if we denote by $\HS_0\,$,
the closed linear span of the vectors of the type
$\Theta_0 \left( \zeta_1,\, \dots,\, \zeta_n \right)\,$,
for all $\left( \zeta_1,\, \dots,\, \zeta_n \right)$ in $U_0\,$,
then in accord with the above construction the continuation
$\Theta \left( \zeta_1,\, \dots,\, \zeta_n \right)$ will take
its values again in $\HS_0\,$.
Thus the single-valuedness of
\(\Theta \left( \zeta_1,\, \dots,\, \zeta_1 \right)\) will follow
by the single-valuedness of the scalar product (\ref{2.2.23}).
The strong analyticity of
\(\Theta \left( \zeta_1,\, \dots,\, \zeta_1 \right)\) is implied
again by Eq.~(\ref{2.2.23}), as those of
$\Theta_0 \left( \zeta_1,\, \dots,\, \zeta_n \right)\,${\spc}.\qed

To \textit{prove} the Theorem~\ref{th:9.1} we just need to apply
the Proposition~\ref{pr:A.1}
for the analytic function (\ref{eq9.14}), taking $U_0$ to be
$\mathcal{T}_n$~(\ref{eq9.15}) and for $U$ using the domain of
all sets of mutually nonisotropic points of $\mathfrak{T}_+\,$.

\end{document}